\documentclass[usenatbib,fleqn,useAMS]{mnras}
\usepackage[T1]{fontenc}
\usepackage{times}
\usepackage{ae,aecompl}
\usepackage{amsmath}	
\usepackage{amssymb}	
\usepackage{graphicx}
\usepackage{subfigure}
\usepackage{bm}
\usepackage{multicol}        

\newcommand\msun{\rm{M_{\odot}}}

\def\stacksymbols #1#2#3#4{\def\theguybelow{#2}
        \def\verticalposition{\lower#3pt}
        \def\spacingwithinsymbol{\baselineskip0pt\lineskip#4pt}
        \mathrel{\mathpalette\intermediary#1}}
\def\intermediary #1#2{\verticalposition\vbox{\spacingwithinsymbol
        \everycr={}\tabskip0pt
        \halign{$\mathsurround0pt#1\hfil##\hfil$\crcr#2\crcr
                \theguybelow\crcr}}}

\def\lta{\stacksymbols{<}{\sim}{2.5}{.2}}
\def\gta{\stacksymbols{>}{\sim}{2.5}{.2}}

\title[The 3-phase condensation model in massive galaxies]{Raining on black holes and massive galaxies:\\
the top-down multiphase condensation model}
\author[M.\,Gaspari et al.]{M. Gaspari$^1$\thanks{{\it Einstein} and {\it Spitzer} Fellow.}\thanks{E-mail: mgaspari@astro.princeton.edu}, P. Temi$^2$, F. Brighenti$^3$\\
$^1$Department of Astrophysical Sciences, Princeton University, 4 Ivy Lane, Princeton, NJ 08544-1001 USA\\
$^2$Astrophysics Branch, NASA/Ames Research Center, MS 245-6, Moffett Field, CA 94035, USA\\
$^3$Astronomy Department, University of Bologna, Via Ranzani 1, 40127 Bologna, Italy}

\pubyear{2016}

\begin{document}
\label{firstpage}
\pagerange{\pageref{firstpage}--\pageref{lastpage}}
\maketitle

\begin{abstract}
The atmospheres filling massive galaxies, groups, and clusters display remarkable similarities with rainfalls. Such plasma halos are shaped by AGN heating and subsonic turbulence ($\sigma_v\sim$\,150\,km\,s$^{-1}$), as probed by {\it Hitomi}. The new 3D high-resolution simulations show the soft X-ray ($<1$\,keV) {\it hot} plasma cools rapidly via radiative emission at the high-density interface of the turbulent eddies, stimulating a {\it top-down} condensation cascade of {\it warm}, $10^4$\,K filaments. The ionized (optical/UV) filaments extend up to several kpc and form a thin skin enveloping the neutral filaments (optical/IR/21-cm). The peaks of the warm filaments further condense into {\it cold} molecular clouds ($<50$\,K; radio) with total mass up to several $10^7\,\msun$, i.e., 5/50$\times$ the neutral/ionized masses. The multiphase structures inherit the chaotic kinematics and are dynamically supported. In the inner 500\,pc, the cold/warm clouds collide in inelastic way, mixing angular momentum and leading to {\it chaotic cold accretion} (CCA). The black hole accretion rate (BHAR) can be modeled via quasi-spherical viscous accretion, $\dot M_\bullet\propto \nu_{\rm c}$, with clump collisional viscosity $\nu_{\rm c}\equiv\lambda_{\rm c}\,\sigma_v$ and mean free path $\sim$\,100 pc. Beyond the inner kpc region, slow pressure torques drive the gas angular momentum transport. In CCA, the BHAR is recurrently boosted up to 2 dex compared with the disc evolution. The disc arises via helical condensation as turbulence is subdominant; with negligible rotation too, the molecular phase is prevented via compressional heating. The CCA BHAR distribution is lognormal with pink noise power spectrum $\propto f^{-1}$ characteristic of fractal chaotic phenomena. The rapid self-similar CCA variability can explain the light curve variability of AGN and HMXBs. An improved criterium to trace thermal instability, or better, nonlinear condensation is proposed: $\sigma_v/v_{\rm cool}\lesssim1$. The 3-phase CCA reproduces crucial observations of cospatial multiphase gas in massive galaxies, as {\it Chandra} X-ray images, SOAR H$\alpha$ warm filaments images and kinematics, {\it Herschel} [C$^+$] emission, and ALMA giant molecular associations. CCA plays key role in AGN feedback, AGN unification/obscuration, the evolution of BHs, galaxies, and clusters. 
\end{abstract}

\begin{keywords}
BH accretion -- ISM, IGM, ICM -- turbulence -- 3D hydro simulations
\vspace{-0.78cm}
\end{keywords}


\section{Introduction \& observations}\label{s:intro}
In analogy to the quiet revolution of numerical weather forecasts (\citealt{Bauer:2015}), 
astrophysicists have started to value 
that the gaseous atmospheres of galaxies (ISM), groups (IGM), and clusters (ICM) are not static isothermal spheres but live halos continuously perturbed by turbulent motions. Such atmospheres are complex hydrodynamical systems displaying nonlinear multiphase structures.
\citeauthor{McKee:1977} (1977) suggested that supernovae (SNe) explosions produce a 3-phase medium: SNe mainly bursting from the cold disc of spiral galaxies heat the gas via shocks up to $10^6\,{\rm K}$ creating the volume-filling hot plasma. The strong compressive motions driven by SNe trigger nonlinear thermal instability (TI) leading to a bistable warm ($\sim\,$$10^4\,{\rm K}$) and cold ($\lta\,$$10^2\,{\rm K}$) phase in pressure equilibrium.

We propose here a model in which the multiphase halo -- in particular related to the IGM/ICM -- is formed {\it top-down}, i.e., not necessarily emerging from the cold to warm/hot phase via SN heating, but condensing from the hot plasma (a few keV) all the way down to the neutral and molecular phases.
X-ray observations have shown that hot plasma halos are ubiquitous from massive galaxy clusters, groups, and galaxies down to at least $M_\ast\simeq10^{10.8}\ \msun$ (\citealt{Anderson:2015}; for earlier works, \citealt{Trinchieri:1986,Fabbiano:1992} and refs.~within).
Such halos emerge from the virialization shocks of cosmological structures (\citealt{Kravtsov:2012}) and cool down via radiative emission (\S\ref{s:hot}). At the same time, the active galactic nucleus (AGN) feedback preserves the global atmosphere from collapsing in a pure cooling flow (\citealt{McNamara:2007} for a review), albeit locally the turbulent gas can become thermally unstable.

This work continues our systematic investigation on multiphase halos.
The initial {\it chaotic cold accretion model} (CCA) was introduced in \citeauthor{Gaspari:2013_cca} (2013; G13).
The CCA model is based on a simple concept related to our everyday experience of weather:
cold clouds condense out of a rarefied, turbulent atmosphere and rain down, hence why CCA can be understood as {\it raining onto black holes}. 
Interestingly, the density ratio between the hot and warm phase ($\sim\,$$10^{3}$) is analogous to the ratio between air and water.  
Similar to storms, 
turbulence is a key element to drive CCA in massive galaxies, groups, and clusters. 
Recently, {\it Hitomi} telescope has probed the presence of turbulence in the X-ray plasma via Fe line broadening, detecting line-of-sight velocity dispersion $164\pm10$\,km\,s$^{-1}$ in Perseus cluster core (\citealt{Hitomi:2016}).
The vortical eddies generated across scales varying by orders of magnitude induce strong nonlinear coupling between different physical processes, as plasma cooling and heating. The advancements in high-performance supercomputing have allowed us to enter a new era in astrophysics in which we can directly study the fully nonlinear 3D hydro equations with complex physics, going beyond analytic models with restrictive assumptions (\citealt{Bondi:1952,Shakura:1973}).
Although single processes are relatively easy to isolate, their nonlinear interactions can completely change the dynamics driving new phenomena.

In G13, we tested the evolution of SMBH accretion by simulating the 3D flow affected by heating, cooling, and turbulence,
from the galactic (10s kpc scales) down to a fraction of a pc. We tested both hot and warm ($10^4$\,K) mode accretion, with either isolated physics or their nonlinear interaction (\citealt{Gaspari:2016_IAU} for a review).
In hot adiabatic flows, either coherent or chaotic vorticity stifles the accretion rate by a factor of several compared with the classic Bondi rate, $\dot M_\bullet<\dot M_{\rm B}\equiv \lambda\, 4\pi (GM_\bullet)^2\rho_{\rm B}/c^3_{{\rm s,\,B}}\propto M_\bullet^2\, K_{\rm B}^{-3/2}$, where $\lambda$ is a normalization factor of order unity, $\rho_{\rm B}$, $c_{\rm s,\, B}$, and $K_{\rm B}$ are the gas density, sound speed, and entropy at large radii ($\gta\,$1 kpc), respectively.
In the turbulent and heated halo, the chaotic collisions promote angular momentum cancellation and boost the accretion rate
up to $100\times$ the Bondi rate, promoting highly efficient AGN feedback via massive outflows (\citealt{Gaspari:2011a,Gaspari:2011b,Gaspari:2012a,Gaspari:2012b} for the large-scale runs tracking the Gyr feedback loop).
In \citeauthor{Gaspari:2015_cca} (2015; G15), we tested the impact of rotation in the hot and warm flow.
As turbulent Taylor number ${\rm Ta_{\rm t}}\equiv v_{\rm rot}/\sigma_v<1$, the chaotic evolution and rain prevail. 
In the opposite regime, the evolution is more quiescent and driven by a warm disc.

The importance of warm gas condensation has been independently tested and corroborated by other groups 
(e.g., \citealt{Quataert:2000,Pizzolato:2005,Soker:2006,Barai:2012,McCourt:2012,Sharma:2012,Mathews:2012,Li:2014,Valentini:2015}). 
The relevance of chaotic collisions has been highlighted by \citet{King:2006} and \citet{Nayakshin:2007}.
Recently, Voit et al.~studied condensation and thermal instability -- renaming it `precipitation' -- by using the observed radial ICM plasma entropy profiles and $L({\rm H\alpha})-K_0$ diagrams (e.g., \citealt{Voit:2015_prof,Voit:2015_nat,Voit:2016})
and later linking them to the specific star formation rates and metallicity histories of galaxies (\citealt{Voit:2015_SFR}); other authors have discussed the limitations of such observational analysis (\citealt{Panagoulia:2014,McNamara:2016}) remarking issues tied to projection effects and to the {\it Chandra} resolution limit.

In recent years, multiwavelength observations have started to unveil the multiphase structure of massive galaxies --
mainly classified as early-type/elliptical galaxies (ETGs) and dominating the cores of groups and clusters.
Often thought to be `red and dead', massive galaxies are far from being dead,
hosting considerable amount of ionized, neutral, and molecular gas, in addition to the hot plasma halo.  
In optical, we observe extended H$\alpha$ (6564.6\,\AA) and [N$^+$] (6585.2\,\AA) filaments and clouds up to 10 kpc from the center, 
which are tightly correlated with soft X-ray and FUV emission
(\citealt{McDonald:2009,McDonald:2010,McDonald:2011a,Werner:2014,Tremblay:2015} and refs.~therein).
The H$\alpha$ luminosities in giant elliptical galaxies are of the order of a few $10^{40}\,{\rm erg\,s^{-1}}$, implying ionized gas masses $M_{\rm ion}\approx10^6$\,-\,$10^7\,\msun$. NGC 5044 is the archetypal system showing a remarkable network of chaotic H$\alpha$ filaments (\citealt{Gastaldello:2009}).
Besides filamentary structures, the warm gas is also found in another state, that of a coherent rotating structure, which is typically correlated with more quiescent systems (\citealt{Hamer:2016}). 

Massive galaxies not only harbor warm gas ($10^4\,{\rm K}$) but also neutral and molecular gas down to $\sim\,$10\,K (e.g., \citealt{Combes:2007}).
ALMA has opened up the gate to high-resolution detections of molecular gas in ETGs (for earlier studies see \citealt{Knapp:1996} and refs.~within). In NGC 5044 group we observe 24 giant molecular associations (GMA) within $r\lta4\,{\rm kpc}$ (\citealt{David:2014}) with inferred molecular mass $M_{\rm H_2}\approx5\times10^7\ \msun$ and chaotic dynamics.
In massive galaxy clusters ($40\times$ more massive than groups as NGC 5044), 
ALMA has detected $M_{\rm H_2}\approx5\times10^{10}\,\msun$ in the core of A1835 (\citealt{McNamara:2014}), which may be supported in a rotating turbulent disc or outflowing structure. In A1664 cluster (\citealt{Russell:2014}), the molecular gas shows asymmetric velocity structure, with two gas clumps flowing into the nucleus. In the massive cluster PKS 0745-191 affected by an extreme AGN outburst of $5\times10^{45}$\,erg\,s$^{-1}$ (\citealt{Russell:2016}), ALMA shows no sign of a disc with projected molecular filaments extending out towards the kpc-scale X-ray cavities.  
In A2597 brightest cluster galaxy (BCG; \citealt{Tremblay:2016}), three narrow redshifted absorption features are detected by ALMA, similar to Milky Way giant molecular clouds, which are redshifted by 240\,-\,335\,km\,s$^{-1}$ falling along line of sight toward the SMBH; the authors argue that the accreting clouds are confined within $r<100$\,pc. 

The cold phase ($\sim\,$$100\,{\rm K}$) may be traced via far-infrared [C$^+$] 157\,$\mu$m line emission (\citealt{Crawford:1985}), assuming photodissociation as main mechanism.
{\it Herschel} data showed that [C$^+$] emission is cospatial with the warm H$\alpha$+[N$^+$] emission, with ratio in the range $0.4$\,-\,$0.8$ (\citealt{Werner:2014}). Such large extension and ratio suggest [C$^+$] is tracing also the warm gas which cools from $10^4\,{\rm K}$. 
The observed [C$^+$] velocity dispersions, 100\,-\,300\,km\,s$^{-1}$, are similar to the kinematics of the
other phases (\citealt{Edge:2001,Werner:2009,Canning:2013,Sanders:2013,David:2014}), corroborating internal condensation over gas stripping from infalling galaxies. 
The mid-infrared {\it Spitzer} spectra of the [C$^+$] bright galaxies 
typically show dust and PAH emission too
(\citealt{Panuzzo:2011}), albeit ETGs dust-to-cold-gas ratios are at least 1 dex lower than in the Milky Way (\citealt{Smith:2012,Hirashita:2015}).

The multiwavelength survey ATLAS$^{\rm 3D}$ (\citealt{Cappellari:2011}) has confirmed the multiphase structure of ETGs.
Ionized gas (H$\beta$, [O$^{++}$], and [NI]) is observed in 73\% of the sample (\citealt{Davis:2011}), HI gas in 40\% (\citealt{Serra:2012}), and molecular CO gas in 22\% (\citealt{Young:2011}). The $M_{\rm HI}$ and $M_{\rm H_2}$ masses are significant, spanning a range
$10^7$\,-\,$10^9\ \msun$, with the gas-richest galaxies residing in the poorest environments. The more massive galaxies residing near the central, denser regions of the group
tend to populate the lower mass range, although the sample needs to be carefully reselected on main X-ray properties (as strong cool cores) to remove contaminations of disc-like and merging galaxies. 
Albeit atomic gas has been relatively ignored in ETG observations due to the very high sensitivity required, we will show it is an important component of the multiphase halo.
Recent studies have started to report significant [OI] 63\,$\mu$m flux from several massive galaxies (\citealt{Hamer:2014,Werner:2014}; see also \citealt{Thom:2012} for HI absorbers and \citealt{Kilborn:2009} for HI in groups). From the [C$^+$] flux, the inferred (\citealt{Hamer:2014}) neutral gas masses may be $M_{\rm HI}\sim10^7-10^8\,\msun$, in line with the ATLAS$^{\rm 3D}$ findings.

While in G13 and G15 the focus was on the condensation down to the first stable regime related to the warm gas ($T\ge10^4\,{\rm K}$, e.g., due to strong photoionization heating), we now continue to investigate the role of the multiphase condensation and CCA down to the neutral ($10^2$\,-\,$10^4\,{\rm K}$) and molecular ($<50\,{\rm K}$) gas phases, in analogy to the ISM studies of the last decade
(e.g., \citealt{Koyama:2002,Joung:2006,Kim:2013,Gatto:2015,Girichidis:2016}; \S\ref{s:cold}).
This is achieved with extremely high-resolution 3D hydrodynamic simulations, including turbulence, rotation, cooling, and heating processes (described in \S\ref{s:init}). 
As indicated by the above observations, the phenomenon is complex and entails different dynamics.
Our step by step experiments cover thus 3 major dynamical states of a system.
In \S\ref{s:cool_e03}, we analyze the condensation evolution in a quiescent rotating hot halo, 
leading to the formation of a multiphase disc.
In \S\ref{s:cool_e00}, we show the evolution of a cooling halo without rotation, inducing a compact multiphase cloud.
In \S\ref{s:cca}, we dissect the CCA evolution with full physics arising from the turbulent and chaotic environment. 
Each section focuses on the theory and examines consistency with the most recent observational data.
In particular, \S\ref{s:cca_comp} focuses on the CCA synthetic observations in the X-ray, UV, optical/IR, and radio bands, taking NGC 5044 -- one of the best studied massive ETG -- as exemplary test case. 
We remark our focus is on the accretion process; the strong AGN outburst and uplift phase will be tackled elsewhere.
In \S\ref{s:conc}, we summarize the main findings.
Key result is the formation of a 3-phase medium via top-down condensation cascade (at variance with the bottom-up model by \citealt{McKee:1977}), which reproduces the tightly correlated multiphase structures and properties observed in massive galaxies. 

\vspace{-0.41cm}
\section[]{Physics \& numerics} \label{s:init}
The core physics and numerics have been introduced in G13-G15.
Here we summarize the main features and focus on the new ingredients as neutral and molecular gas cooling.

\vspace{-0.41cm}
\subsection{Hydrodynamics and source terms} \label{s:physics}
\noindent
We use a modified version of the AMR code FLASH4 (\citealt{Fryxell:2000}) to integrate the 3D equations of hydrodynamics 
in Eulerian and conservative form:
\begin{equation}\label{cont}
\frac{\partial\rho}{\partial t} + \boldsymbol{\nabla}\cdot\left(\rho \boldsymbol{v}\right) = 0
\end{equation}
\vspace{-0.41cm}
\begin{equation}\label{mom}
\frac{\partial\rho \boldsymbol{v}}{\partial t} + \boldsymbol{\nabla}\cdot\left(\rho \boldsymbol{v} \otimes \boldsymbol{v}\right) + \boldsymbol{\nabla}{P} = \rho(\boldsymbol{g}_\bullet+\boldsymbol{g}_\ast+\boldsymbol{g}_{\rm dm}+\boldsymbol{a}_{\rm turb})
\end{equation}
\vspace{-0.41cm}
\[
\frac{\partial\rho e }{\partial t} + \boldsymbol{\nabla}\cdot\left[\left(\rho e  + P\right) \boldsymbol{v}\right] = \rho\boldsymbol{v}\cdot (\boldsymbol{g}_\bullet+\boldsymbol{g}_\ast+\boldsymbol{g}_{\rm dm}+\boldsymbol{a}_{\rm turb})\; +
\]
\vspace{-0.41cm}
\begin{equation}\label{ene}
\quad\quad\quad\quad\quad\quad\quad\\ +\, n\,\Gamma_{\rm hot} -\,n_{\rm e}n_{\rm i}\, \Lambda_{\rm hot} +\,n_{\rm H}\,\Gamma_{\rm cold}-\,n^2_{\rm H}\,\Lambda_{\rm cold} 
\end{equation}
\vspace{-0.41cm}
\begin{equation}\label{eos}
P = \left(\gamma -1\right)\rho \left(e -v^2/2\right)
\end{equation}
where $\rho$, $\boldsymbol{v}$, $e$, and $P$ are the gas density, velocity, specific total energy (internal and kinetic), and pressure, respectively.
The adiabatic index is $\gamma=5/3$. The numerical integrator is the recent 3rd order-accurate unsplit PPM solver (\citealt{Lee:2009}).
The cooling ($\Lambda$) and heating ($\Gamma$) rates related to the plasma, neutral and molecular gas are described in \S\ref{s:cold}.
The large 3D box fully encloses the massive galaxy and core of the galaxy group, with a volume $52^3$ kpc$^3$. Boundary conditions are outflow permitted, while inflow prohibited
(inflow from the large-scale halo is unimportant over the short timescale simulated here).
The maximum resolution is $\simeq0.8$\,pc, with radially concentric fixed meshes in cartesian coordinates (cf.~G13, Sec.~2.1). The dynamical range is almost a factor $10^5$. To track more accurately condensation, we doubled the radial extension of each grid level compared with G13 to about 110 cells.
The simulations are extremely computationally expensive, in particular due to the tiny timestep constraint, with a single full run of the order of a few million CPU-hours. We prioritized accuracy and resolution in time and space, at the expense of speed and thus a wide parameter space.

\vspace{-0.41cm}
\subsection{Environment: SMBH, galaxy, and group}
The total gravitational acceleration of the system $\boldsymbol{g}_{\rm tot}$ 
is given by the combination of the central SMBH, the stars within the massive galaxy, and the dark matter halo of the galaxy group ($\boldsymbol{g}_\bullet$, $\boldsymbol{g}_\ast$, and $\boldsymbol{g}_{\rm dm}$, respectively).
The black hole mass is $M_\bullet=3\times10^9$ $\msun$, hence 
the Schwarzschild and Bondi radius are $R_{\rm S}\equiv 2GM_\bullet/c^2 \simeq 3\times10^{-4}$ pc and $r_{\rm B}\equiv GM_\bullet/c_{\rm s,\,B}^2 \simeq 85$ pc, respectively (sound speed $c_{\rm s,\,B}\simeq 390$ km s$^{-1}$ at $r\approx1$ kpc). We integrate the system for 40\,-\,80 Myr, i.e., over 200 Bondi times ($t_{\rm B}\equiv r_{\rm B}/c_{\rm s,\,B}\simeq 210$\, kyr).
By using the pseudo-relativistic \citet{Paczynski:1980} black hole potential, the sonic point is located at a finite distance, near the pc scale, where we sink the inflowing gas mass
with 3 cells radius (being a causally disconnected region; cf.~G13, Sec.~2.2).
Below the subpc region, GRMHD effects are expected to drive the accreted gas toward the final event horizon, in particular via magnetorotational instability (\citealt{Balbus:1998}), the role of which is beyond the scope of the present work.

The massive galaxy is a typical elliptical galaxy analog of NGC 5044 with $M_{\ast}\simeq 3.4\times10^{11}\,\msun$
and $R_{\rm e}\simeq10$\,kpc (\citealt{Buote:2004,David:2009}). As most massive galaxies (\citealt{Sun:2009a}), it centrally dominates the dark matter halo of a massive galaxy group with NFW (\citealt{Navarro:1996}) potential having $M_{\rm vir} \simeq 3.6\times10^{13}$ $\msun$ and concentration 9.5. 
The initial gas density profile is retrieved from hydrostatic equilibrium (neglecting the SMBH), 
using the stellar and dark matter gravitational potentials and the NGC 5044 temperature profile from {\it Chandra} and XMM data (\citealt{Gaspari:2011b}). 

We test a gaseous atmosphere either with negligible or substantial rotation parametrized with $e_{\rm rot}=0$\,-\,0.3, since ETG observations detect either chaotic gas motions (\citealt{Caon:2000}) or inner disks (\citealt{Young:2011}), with the former state being more typical.
In the rotating galaxy, the rotational velocity of the gaseous halo, $v_{\rm rot}\equiv v_\phi$, is 1/3 of the (stellar and dark matter) circular velocity at the equator,
which is roughly constant over the large galactic scale, $v_{\rm rot}\approx\,$100 km\,s$^{-1}$ for $r\simeq1$-13 kpc. 
The centrifugal force effectively lowers the gravitational acceleration along the cylindrical radius $R \equiv (x^2 + y^2)^{1/2}$, thereby slightly flattening the initial equilibrium density profile (G15, Sec.~2.1).

\vspace{-0.41cm}
\subsection{Turbulence} \label{s:turb}
A key component of CCA is turbulence, which enables to drive strongly nonlinear interactions between different physics.
Recent X-ray observations indicate that hot halos are not quiescent, but they are continuously perturbed by chaotic motions with at least subsonic velocities, $\sigma_v > 100$ km\,s$^{-1}$. 
The relative density perturbations in the plasma are linearly related to the turbulent Mach number, $\delta \rho/\rho\approx {\rm Ma_{1D}}$ (\citealt{Gaspari:2013_coma,Gaspari:2014_coma2}).
By using such relation, the hot halos in 33 clusters show an average 3D Mach number ${\rm Ma_{\rm 3D}\approx0.3}$ (\citealt{Hofmann:2016}). Line broadening and shift in X-ray spectra also indicate subsonic motions in the hot plasma
 (e.g., \citealt{Sanders:2013}).
\citet{Hitomi:2016} probed line-of-sight velocity dispersion of $164\pm10$\,km\,s$^{-1}$ in Perseus core.
Mergers, galaxy motions, SNe, and in particular recurrent AGN feedback continuously inject turbulence from the large 100 kpc scale down to the inner kpc core. This has been corroborated by many computational works in the past decade (e.g.~\citealt{Lau:2009,Vazza:2009,Valdarnini:2011,Gaspari:2012b,Miniati:2015,Biffi:2016} and refs.~within).

To model a natural time-correlated and zero-mean acceleration field ($a_{\rm turb}$), turbulence is integrated via a spectral forcing scheme based on Ornstein-Uhlenbeck random process (G13, Sec.~2.4). Such advanced stirring reproduces experimental high-order structure functions (\citealt{Fisher:2008}).
Our reference average 3D Mach number is ${\rm Ma_{\rm 3D}\approx0.4}$, with turbulent velocity dispersion
$\sigma_v\simeq165$ km s$^{-1}$, similar to {\it Hitomi} turbulence measurement. The turbulent kinetic energy is injected only at large scales, $L\gta 4$ kpc, typically associated with AGN feedback (\citealt{Gaspari:2012b,Vazza:2012}; the eddy turnover time is $t_{\rm eddy}\sim25\,$ Myr). After injection, turbulence naturally cascades via self-similar Kolmogorov-like eddies. 
In the turbulence models (\S\ref{s:cca}), 
we start with a fully developed cascade ($t>t_{\rm eddy}$) to follow the long-term evolution of multiphase gas condensation.
In the model with $e_{\rm rot}=0.3$, the turbulent Taylor number is ${\rm Ta_t}\equiv v_{\rm rot}/\sigma_v\approx0.7$.

\vspace{-0.41cm}
\subsection{Plasma cooling and heating} \label{s:hot}
\begin{figure}
    \centering
    \subfigure{\includegraphics[scale=0.39]{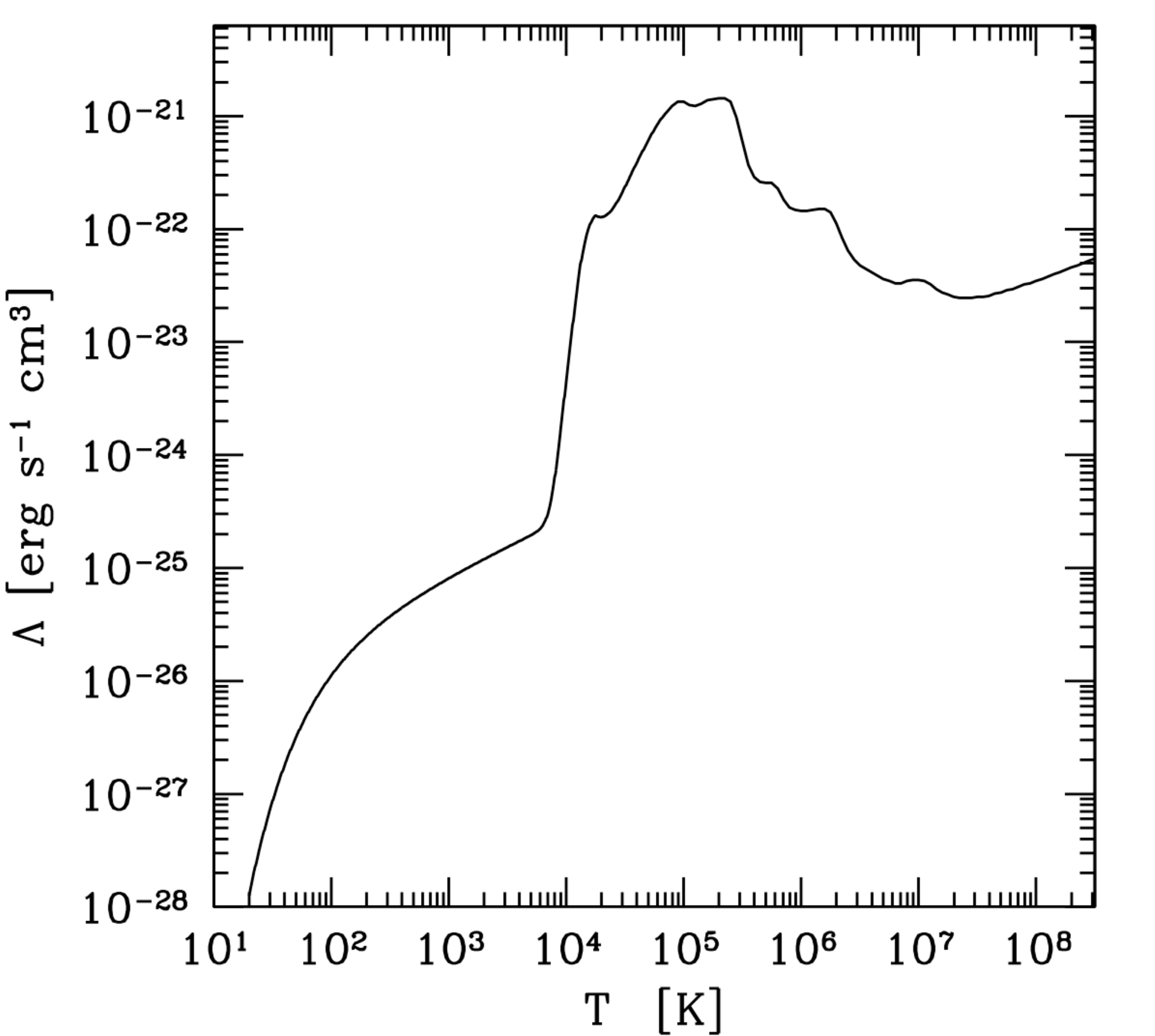}}
    \caption{Bolometric cooling function for solar abundance (emissivity normalized by $n^2_{\rm H}$) covering the plasma ($T\sim10^4$\,-\,$10^8\,{\rm K}$; $\Lambda_{\rm hot}$), neutral and molecular gas regime ($T\sim10$\,-\,$10^4\,{\rm K}$; $\Lambda_{\rm cold}$).}
      \label{f:lambda}  
\end{figure}

The diffuse plasma cools via X-ray and UV radiation from several keV (${\rm 1\,keV}=1.16\times10^7$\,K) down to $\sim\,$$10^4$ K.
The plasma radiative emissivity or cooling rate is $\mathcal{L_{\rm hot}}\equiv n_{\rm e}n_{\rm i}\,\Lambda_{\rm hot}$ (${\rm erg\,s^{-1}\,cm^{-3}}$), where $n_{\rm e}\equiv\rho/(\mu_{\rm e}m_{\rm p})$ and $n_{\rm i}\equiv\rho/(\mu_{\rm i}m_{\rm p})$ are the electron and ion number density, respectively.
The mean particle weight of electrons and ions in the fully ionized plasma are $\mu_{\rm i}\simeq1.32$ and $\mu_{\rm e} \simeq 1.16$, respectively, hence total particle weight $\mu \simeq 0.62$.
The plasma cooling function $\Lambda_{\rm hot}$ is modeled following \citet{Sutherland:1993} cooling curve with solar abundance and in collisional ionization equilibrium (balance between recombination, collisional ionization, and charge transfer rate). The plasma cooling curve incorporates calculations for H, He, C, N, O, Ne, Na, Mg, Al, Si, S, Cl, Ar, Ca, Fe, and Ni and all stages of ionization. 
As shown in Fig.~\ref{f:lambda}, continuum Bremsstrahlung dominates at $T > 10^7$ K ($\Lambda_{\rm hot}\propto T^{1/2}$),
while at lower temperatures line emission boosts the emissivity ($\Lambda_{\rm hot}\propto T^{-1}$), 
creating the characteristic bumps associated with Fe ($T\sim10^6\,{\rm K}$), C, O, and He ($T\sim10^5\,{\rm K}$),
and H ($T\approx10^{4.2}\,{\rm K}$).
Below $T<10^{4.2}\,{\rm K}$, the cooling rate of the plasma drops abruptly due to the rapid H recombination.
While the adopted cooling curve is customary in the literature, we note other types of cooling may be at work, such as the mixing layers proposed by \citet{Fabian:2001_CF}. On the other hand, regardless of the radiative channel (being X-ray or UV), as long as cold clouds form they will follow the same physical principles of CCA shown here (rain, collisions, and boosted accretion).

The initial ratio of the plasma cooling time $t_{\rm cool}\equiv(3/2)\,n k_{\rm B}T/(n_{\rm e} n_{\rm i} \Lambda_{\rm hot})$ and free-fall time $t_{\rm ff}\equiv(2r/g_{\rm tot})^{1/2}$ has a minimum $\approx4.5$ at $r\approx250$ pc.
Below $t_{\rm cool}(r)/t_{\rm ff}(r)\lta10$ thermal instability may grow nonlinear (\citealt{McCourt:2012, Sharma:2012,Gaspari:2012a}; for classic TI studies see \citealt{Field:1965, Krolik:1983, Nulsen:1986, Balbus:1989,Pizzolato:2005}), albeit such radial threshold appears crude in real perturbed halos displaying scatter up to 1 dex (see end of \S\ref{s:cca_TI} for a more efficient ratio and discussion on linear TI).

Over the past decade, {\it Chandra} and XMM observations have discovered that radiative cooling cannot be left unopposed over the long term, as this would induce very high soft X-ray emission and star formation rates
(\citealt{McNamara:2012} for a review).
Mechanical AGN feedback, in conjunction with stellar heating, mergers, and cosmic rays (e.g., \citealt{Brighenti:2003,McNamara:2007,Gaspari:2011a,Gaspari:2011b}), 
appears to maintain the hot, X-ray emitting halo of galaxies, groups, and clusters in global thermal quasi-equilibrium for several Gyr preserving the gentle cool-core structure 
(e.g., \citealt{Vikhlinin:2006, Diehl:2008b, Rasmussen:2009, Sun:2009a}).
With the objective to achieve maximal resolution at the inner radius, it is numerically unfeasible to inject a fast outflow or relativistic jet as this would dramatically decrease the timestep. Moreover,
as shown in \citeauthor{Gaspari:2012a} (2012b; Fig.~9), after several outburst events the core achieves thermal quasi steady state with deposited heating balancing cooling in an isotropic manner.
Therefore, for each timestep, we set the local volumetric plasma heating rate $n\Gamma_{\rm hot}$ equal to the average plasma radiative emissivity in finite radial shells (cf.~G13, Sec.~2.5). 
The hot halo is globally stable but locally unstable.
Albeit simple in spirit, \citet{Gaspari:2015_xspec} analysis showed that such model can suppress the soft X-ray spectrum below 1 keV (Fig.~2 in that paper) consistently with observational data (e.g., \citealt{Peterson:2003,Kaastra:2004}) and mimicking the more complex AGN jet evolution (Fig.~1 in that paper). We remark that if the AGN heating is not quasi isotropically distributed at some stage, a cooling catastrophe will develop.

We are fully aware of the complexity of AGN feedback. 
The AGN outburst injection stage is an important component of feedback, as tackled in-depth in our previous works (\citealt{Gaspari:2011a,Gaspari:2011b, Gaspari:2012a, Gaspari:2012b, Brighenti:2015}), inducing a strong initial uplift which can further facilitate multiphase condensation either via compression and increasing the inflow time (\citealt{McNamara:2016}). The present model does not include such phase, as we want to focus first on the accretion inflow phase in a controlled physical setup and then expand with more complex physics in our ongoing series of investigations on CCA.
It is important to note that, even without a heating term, CCA can develop with the main physical mechanisms remaining intact, the main difference being that the normalization of the accretion rate and total condensed mass becomes higher (cf.~G15, Sec.~6). 

\vspace{-0.41cm}
\subsection{Neutral/molecular gas cooling and heating} \label{s:cold}
Multiphase condensation unfolds from the hot plasma down to the neutral and molecular regime ($\sim\,$10 K).
To model neutral and molecular gas cooling, 
we follow a prescription similar to reference ISM studies
(e.g., \citealt{Inoue:2008,Kim:2008,Kim:2013,Kim:2015}).
Below $T<10^{4.2}\,{\rm K}$\,\footnote{The cooling rate is smoothly, linearly interpolated between the plasma and neutral gas transition.}, 
the volumetric cooling rate is 
$n_{\rm H}^2\,\Lambda_{\rm cold}$
with $n_{\rm H}\simeq\rho/(1.4\,m_{\rm p})$ and cooling function (${\rm erg\,s^{-1}\,cm^3}$; Fig.~\ref{f:lambda})
\begin{multline}\label{e:Koyama}
\Lambda_{\rm cold} = 2\times10^{-19} \exp[-1.184\times10^5/(T+10^3)]\;+\\
+2.8\times10^{-27}\,\sqrt{T}\,\exp(-92/T) \quad\quad\quad\quad\quad\quad\quad
\end{multline}
Eq.~\ref{e:Koyama} approximates the combination of several chemical processes, 
which would make the simulation unfeasible to run at the present moment. 
The related processes mainly include
atomic line cooling from hydrogen Ly$\alpha$, C$^+$, OI, Fe$^+$, and Si$^+$, rovibrational line cooling from H$_2$ and CO, atomic and molecular collisions with grains (\citealt{Koyama:2000}).

Compared with the classic cooling function by \citet{Dalgarno:1972}, 
the cooling curve in Eq.~\ref{e:Koyama} models a gas with ionization fraction $f_{\rm ion}\sim10^{-2}$
as adopted in other key studies (e.g., \citealt{Joung:2006}). 
The post-AGB stars, AGN, cosmic rays, and plasma mixing processes are expected to preserve $f_{\rm ion}$ to non-negligible values; observationally, optical telescopes detect the presence of ionized, H$\alpha$ filaments in several ETGs (\S\ref{s:intro}).
We tested runs with $f_{\rm ion}< 10^{-2}$ (e.g., original \citealt{Koyama:2002} curve is comparable to that of Dalgarno with $f_{\rm ion}\sim10^{-4}$) and the cold phase results to be negligible in the simulated ETG.
Higher values produce more cold gas strengthening the impact of cold accretion. 
We did not attempt to fine-tune the parameter, not granting any further insight.
We are fully aware of the complexity of molecular gas formation. 
The current work is a first step toward a more realistic modeling of 3-phase CCA in massive halos.
Our future works will test additional physics as dust evolution and chemical networks.

The heating rate $\Gamma_{\rm cold}$ -- which is dominated by the photoelectric mechanism -- is proportional to the star formation rate (\citealt{Kim:2011}) and dust abundance in the gas (\citealt{Wolfire:2003}). Compared with spiral galaxies, the star formation rate of ETGs is $\sim$\,1 dex lower,
$\dot M_\ast\approx0.05$\,-\,0.1$\,\msun\,{\rm yr^{-1}}$ (\citealt{Werner:2014}). The ETG dust-to-gas ratio is also 1 dex lower than Milky Way-like galaxies, $M_{\rm dust}/M_{\rm HI+H_2}\sim10^{-3}$ (\citealt{Smith:2012,Hirashita:2015}). For the foregoing reasons, we use $\Gamma_{\rm cold}=10^{-2}\,\Gamma_0$, where $\Gamma_0=2\times10^{-26}$ erg\,s$^{-1}$ is the reference value adopted in disc galaxies (e.g., \citealt{Kim:2015}).
We tested lower $\Gamma_{\rm cold}$ values and do not find major differences in the CCA evolution.
For a condensation dominated by the warm phase, refer to the G15 simulations; together with the present work, these runs bracket the range of cold/warm gas formation. 

The net cooling source term is integrated with a second-order (Runge-Kutta) explicit scheme.
The cooling time often becomes shorter than the hydrodynamic (Courant) timestep $\Delta t$, in particular in very dense, collapsing regions. It is crucial to resolve a small fraction of the cooling time, in order to properly resolve condensation and to avoid overcooling. If the cell $t_{\rm cool} < \Delta t$, the cooling source step is sub-cycled. To avoid spurious cooling, 
we advance each sub-cycle $i$ for $0.1\,t_{{\rm cool},\, i}$, until the full $\Delta t$ is completed. Each sub-cycle explicitly integrates the cooling equation and re-computes the cooling time, thereby self-adapting to the local slope of the cooling curve at each sub-cycle. This is computationally expensive, but we find it prevents significant numerical noise, e.g., compared with implicit methods.

\vspace{-0.41cm}
\subsection{Phases of the gaseous atmosphere} \label{s:phases}
The focus of the investigation is to study the multiphase condensation throughout the entire temperature spectrum, $\sim\,$10\,-\,$10^8$ K.
We select the different phases based on temperature thresholds which are associated with key turning points in the cooling curve and tied to characteristic observational bands. Aiming to simplicity, 
we classify the multiphase structure into 3 major phases: hot gas/plasma, warm/neutral gas, and cold/molecular gas.

At $T>10^{4.2}\;{\rm K}$, the gas is almost fully ionized, thus we call it {\it plasma}.
Within the hot plasma regime, we can differentiate between hard X-ray ($>0.5\,{\rm keV}$) and soft X-ray (0.1\,-\,0.5\,keV).
The plasma can also populate a lower temperature regime, the UV phase ($<0.1$ keV), which is mainly dominated by gas at $\sim10^5$\,K.
According to the \citet{Field:1965} criterium, the plasma is thermally unstable ($d\ln\Lambda/d\ln T< 2$), in particular below the soft X-ray regime, where line cooling becomes dominant ($\Lambda_{\rm hot}\propto T^{-1}$).

Between $200\,{\rm K}<T<10^{4.2}\,{\rm K}$, the gas is dominated in mass by the {\it neutral} phase, mainly neutral hydrogen HI.
Neglecting turbulence, the (first) stable phase settles at $T\approx5\times10^3$ K (e.g., \citealt{Kim:2008}),
which is referred to as {\it warm} neutral medium in ISM studies and mainly observed in optical band (and 21-cm). Between $200\,{\rm K}<T<5\times10^{3}\,{\rm K}$, the warm neutral medium is thermally unstable ($d\ln\Lambda/d\ln T<2$), although turbulence continuously populate such regime. Such warm gas is mainly observed in IR.

Below $T<200\,{\rm K}$, the gas becomes again thermally stable and we call this regime {\it cold} phase, where
H$_2$ is expected to dominate over HI.
The minimum simulated temperature is $20\,{\rm K}$, below which the gas is expected to be fully molecular (\citealt{Girichidis:2016}) and nearly isothermal (\citealt{Koyama:2000}). 
With no explicit chemical network and dust evolution, the molecular masses shall be considered approximate.
Molecular gas is typically traced by CO gas ($J=1$-0, 2-1 lines), although tiny in mass ($\sim\,$3 dex lower), and observed in the microwave and radio bands. 

\vspace{-0.41cm}
\section[]{Multiphase disc}  \label{s:cool_e03}

\begin{figure}
     \subfigure{\includegraphics[scale=0.51]{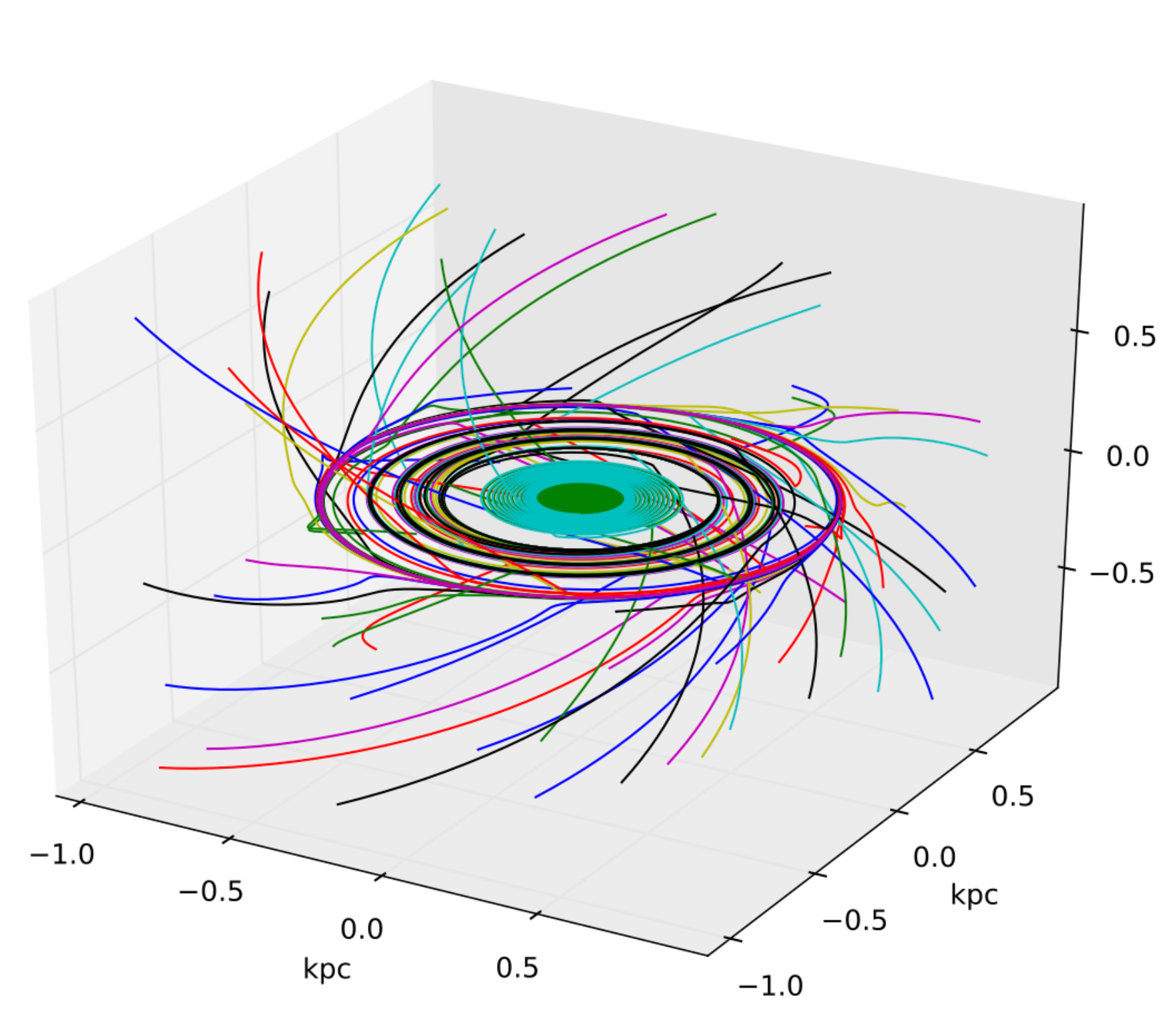}}
     \caption{Accretion with cooling and rotation: a sample of streamlines starting from the outer kpc region ($t=40\,$Myr). 
     Some streamlines share randomly the same color.
     The plot shows the characteristic helical path of the condensing gas converging in an equatorial circularized disc.} 
         \label{f:cool_e03_stream}
\end{figure} 

We now discuss the results of the high-resolution 3D simulation in the regime of pure cooling, with initial rotation $e_{\rm rot}=0.3$, and adopting the full cooling cascade described in \S\ref{s:init} (hot, warm, and cold gas).
This regime applies to quiescent evolutionary periods, in which no major source of heating or turbulence (an AGN outburst or merger) significantly alters the dynamics of the massive galaxy.
The pure cooling runs serve also as a reference for the subsequent fully nonlinear, chaotic models.
We start by analyzing the black hole accretion rate (BHAR) and dynamics of the flow, to proceed on discussing the multiphase condensation cascade and the comparison with observational data.

\begin{figure}
     \centering
     \subfigure{\includegraphics[scale=0.47]{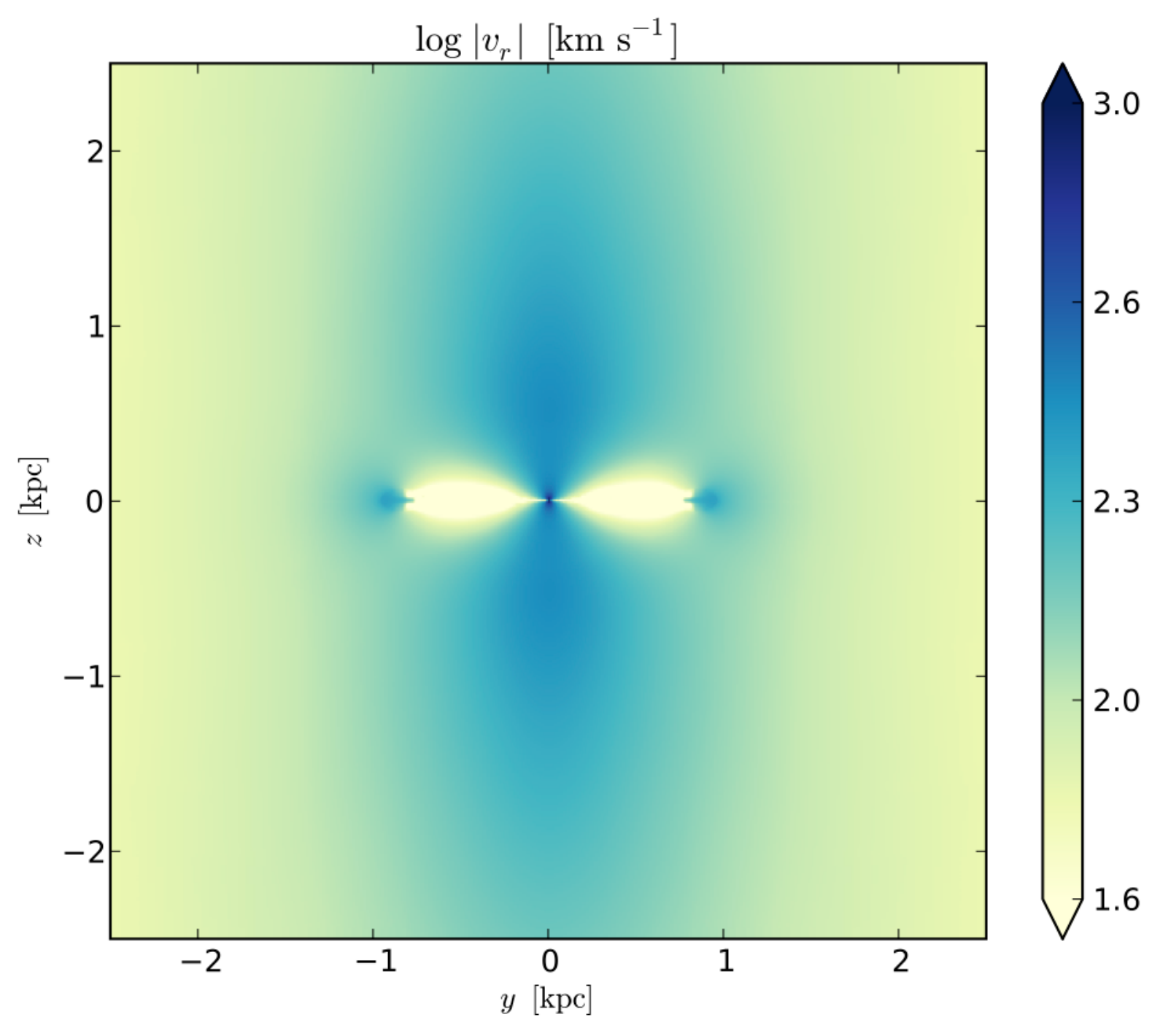}}
     \caption{Accretion with cooling and rotation: absolute value of the 
     radial velocity in the mid-plane cross-section through the $x-$axis 
     (5x5 kpc$^2$; $t=40\,$Myr).}
         \label{f:cool_e03_vrad}
\end{figure}

\begin{figure}
      \begin{center}
      \subfigure{\includegraphics[scale=0.31]{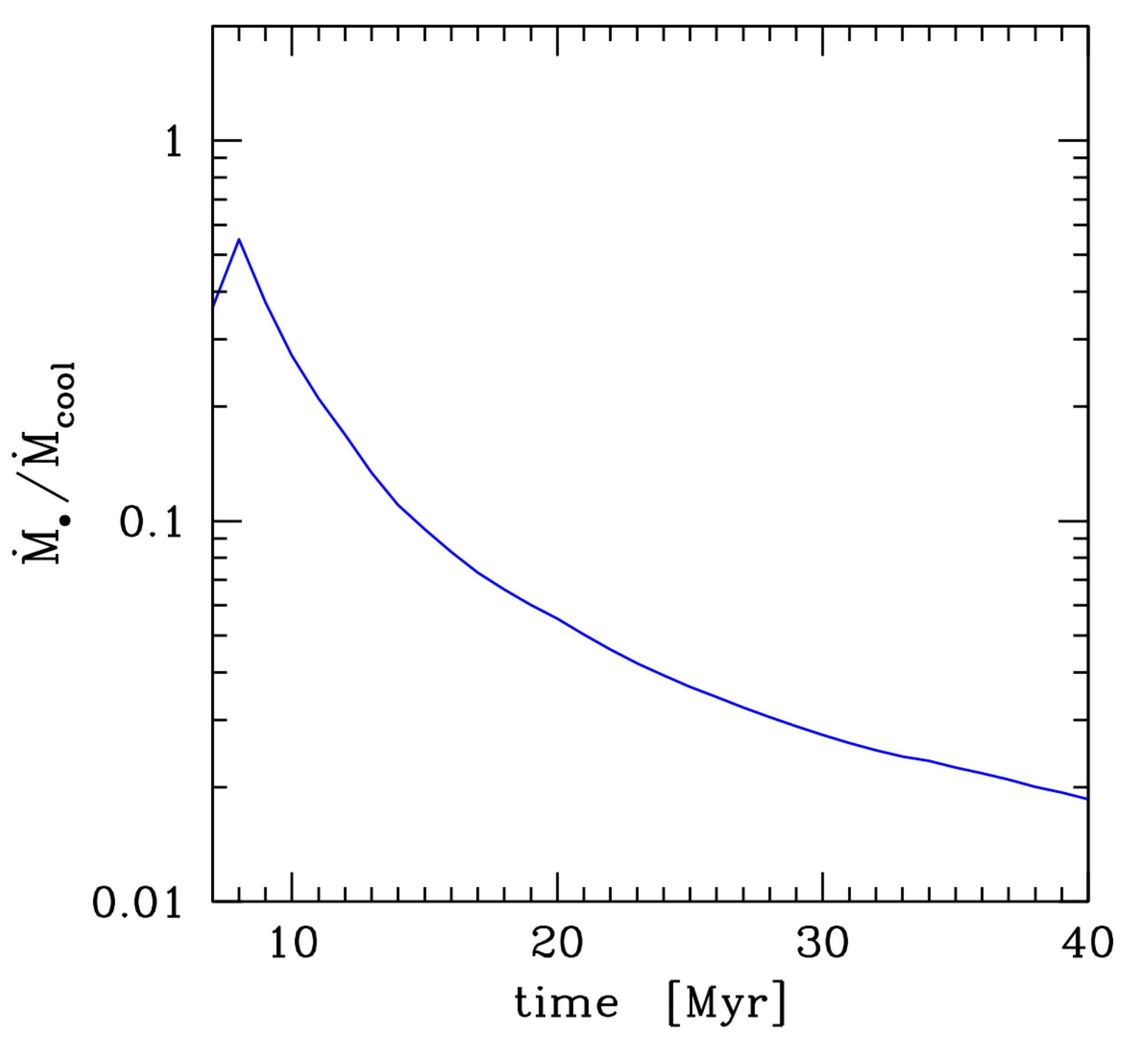}}
      \subfigure{\includegraphics[scale=0.31]{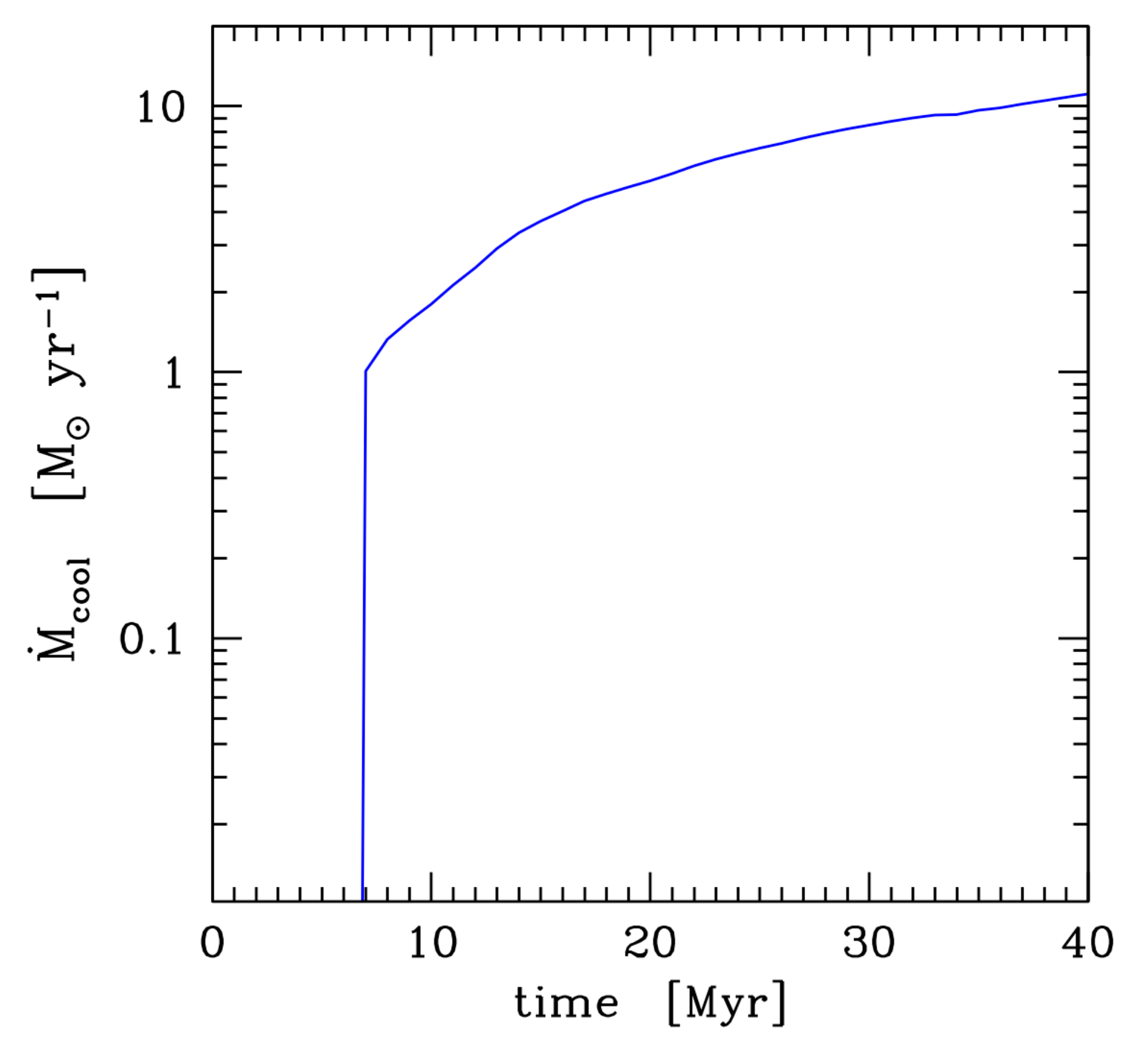}}
      \end{center}
      \caption{Accretion with cooling and rotation: evolution of the average accretion rate as a fraction of the cooling rate (top). Average cooling rate in $\msun$\,yr$^{-1}$ (bottom; 1 Myr step) -- directly computed from the run summing the mass variations per timestep over the UV, warm, and cold gas cascade, while accounting for the sinked mass.
      The gas spirals in and forms a rotating multiphase disc. Only the low angular momentum gas contained in a very narrow polar funnel can eventually accrete. The total accretion rate is suppressed by 2 dex compared with the pure cooling flow.
      }
      \label{f:cool_e03_mdot}
\end{figure}

\vspace{-0.41cm}
\subsection[]{Dynamics \& accretion} \label{s:cool_e03_dyn}
\noindent
The  hot plasma atmosphere emits photons mostly via Bremsstrahlung emission and loses pressure support. 
The loss of internal energy is more severe in the inner, denser region ($\mathcal{L}\propto n_{\rm e} \,n_{\rm i}\, T^{1/2}$).
Unlike in a classic pure cooling flow, the current rotating flow has a centrifugal force which initially balances 1/3 of the gravitational force along $R$. 
As the plasma cools, it starts to collapse following a non-radial pattern.
The dynamics can be decomposed into two components.
Along $R$, the infalling gas spins up increasing the rotational velocity
in order to conserve the initial specific angular momentum $l_z = R\,v_\phi>0$. 
Along $z$, the gas gradually goes into free fall.
This drives a motion following a conical helix, as depicted by the sample of streamlines\footnote{Curves instantaneously tangent to the flow velocity vector, describing the direction a massless fluid element would follow.}
in Fig.~\ref{f:cool_e03_stream}.
The 3D helix can be understood in terms of net torque 
$\boldsymbol{\tau} = \boldsymbol{r} \times \boldsymbol{a}_{\rm net}$.
In the rotating frame, the net acceleration is the sum of the gravitational acceleration $\boldsymbol{g}_{\rm tot}$, 
the pressure acceleration $-{\boldsymbol \nabla} P/\rho$, and the centrifugal acceleration $v_\phi^2/R$ (in the $x$-$y$ plane only).
As condensation develops, such net acceleration becomes non-zero with a non-radial component, which is associated with a torque injecting $l_x$ and $l_y$, hence the slowly shrinking helix.
We inspect such torques and find that the related timescale $t_{\tau}=l/\tau$ is relatively slow in most of the volume; 
at final time, $t_{\tau}\approx10,\,40$ Myr at $r\simeq$\,1,\,2 kpc, respectively.
Runaway condensation develops during the infall and the UV, neutral, and molecular phases  (discussed in \S\ref{s:cool_e03_multi}) emerge with significant rotational velocity.
The envelope of the multiphase disc is the region with maximal loss of pressure support and thus of major torques with $t_{\tau}\lta1$\,Myr, stretching the final path of the helix. 
The generated $l_x$ and $l_y$ are symmetric compared to the equatorial plane and cancel out as the infalling multiphase gas converges on such plane forming the disc with negligible internal torques.  

The equatorial disc has on average $v_{\rm rot}\approx v_{\rm circ}$. As shown in Fig.~\ref{f:cool_e03_vrad}, such multiphase disc has negligible inflow/outflow radial velocity along its overall structure (numerical viscosity is minor). The dense disc later blocks the inflow of gas at large $R$, thus accumulating more mass and expanding through time; at $t=40\,$Myr the disc has reached $R\simeq1\,$kpc. 
Not all the warm/cold gas attains rotational velocity equal to the circular velocity (Fig.~\ref{f:cool_e03_vphi}).
For polar angles $\theta\lta10^\circ$ the gas condensing out of the extended hot halo has low angular momentum and can accrete through the narrow funnel or after settling onto the equator. In a classic spherically-symmetric cooling flow the accretion rate is equal to its characteristic rate, the cooling rate, $\dot M_\bullet \simeq \dot M_{\rm cool}$ (\S\ref{s:cool_e00}), which is typically two decades larger than the Bondi rate $\dot M_{\rm B}$ -- tied instead to hot flows. 
For our galaxy, the asymptotic cooling rate is $\dot M_{\rm cool}\simeq10\ {\rm \msun\,yr{^{-1}}}$ (Fig.~\ref{f:cool_e03_mdot}).
In the disc model the effective accretion rate is reduced by the narrow funnel emerging out of the anisotropic cooling flow. Modeling the accreting funnel as a double cone\footnote{Solid angle of a single cone is $\int^{2\pi}_0\int^\theta_0\sin\theta'd\theta'd\phi=2\pi(1-\cos\theta)$.} we obtain 
\begin{equation}\label{e:disc}
\dot M_\bullet=\frac{\Delta \Omega}{4\pi}\,\dot M_{\rm cool}=(1-\cos\theta)\,\dot M_{\rm cool}\approx0.018\,\dot M_{\rm cool},
\end{equation}
where $\theta\simeq11^\circ$ is the apex half-opening angle, in line with the simulated accretion rate at late times (Fig.~\ref{f:cool_e03_mdot}) when the multiphase structure has mostly found circularization.
The average Bondi accretion rate for the non-rotating hot flow is $\dot M_{\rm B}\approx0.08\ \msun$\,yr$^{-1}$ (G13, Sec.~3), which is about 1\% of the cooling rate and thus comparable (incidentally) to the disc accretion rate. By using instead the rotating hot flow as reference (G15, Sec.~3), the current disc BHAR is 1 dex higher. 
Compared with the 2-phase cooling model in G15, the BHAR has increased by 30\%, as the cooling gas becomes more compressive via neutral and molecular cooling.  

To summarize, the halo and accretion dynamics is shaped by the rotational barrier forcing highly suppressed $\dot M_\bullet\ll\dot M_{\rm cool}$. The final accretion rate normalized to the Eddington rate for our SMBH, $\dot M_{\rm Edd}\simeq 66\, \msun\,{\rm yr^{-1}}$, is $\dot M_\bullet/\dot M_{\rm Edd} \approx 3\times10^{-4}$, which marks the lower envelope of massive galaxies with active mechanical AGN feedback (\citealt{Russell:2013}).
Assuming mechanical efficiency $\varepsilon=10^{-3}$ (\citealt{Gaspari:2012b}), the released AGN jet/outflow power would be 
$P_{\rm out}=\varepsilon\, \dot M_\bullet c^2\simeq 10^{42}\,$erg\,s$^{-1}$, which is a fraction of the bolometric X-ray luminosity of NGC 5044 group, $L_{\rm x}\approx2\times10^{43}\,$\,erg\,s$^{-1}$. The CCA rain will instead drive accretion peaks up to the cooling rate (\S\ref{s:cca_acc}), and thus substantially boosted AGN power, 
promoting the natural counterpart to the low-level, smooth disc accretion stage. 

\begin{figure}
      \subfigure{\includegraphics[scale=0.4]{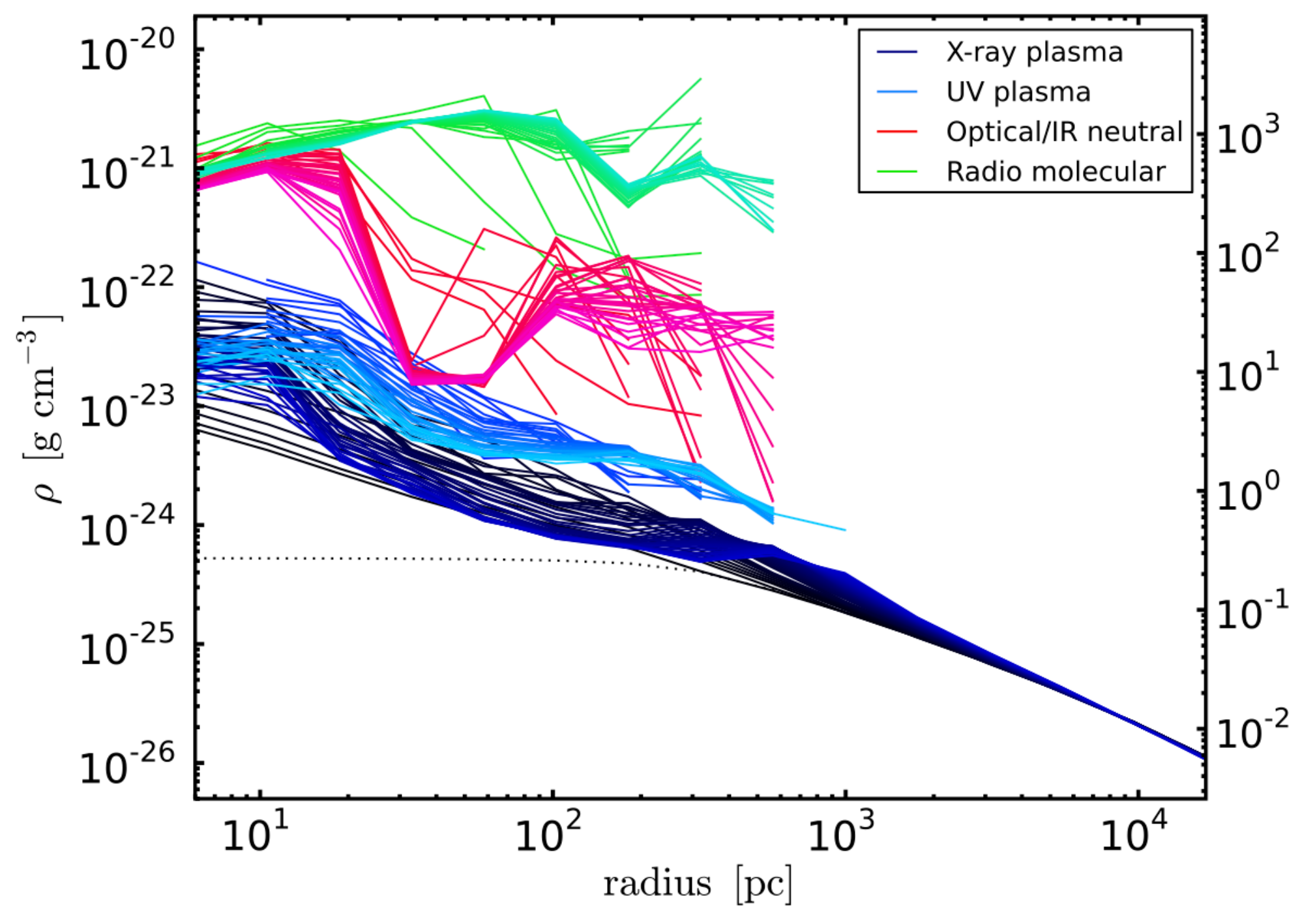}}
      \subfigure{\includegraphics[scale=0.403]{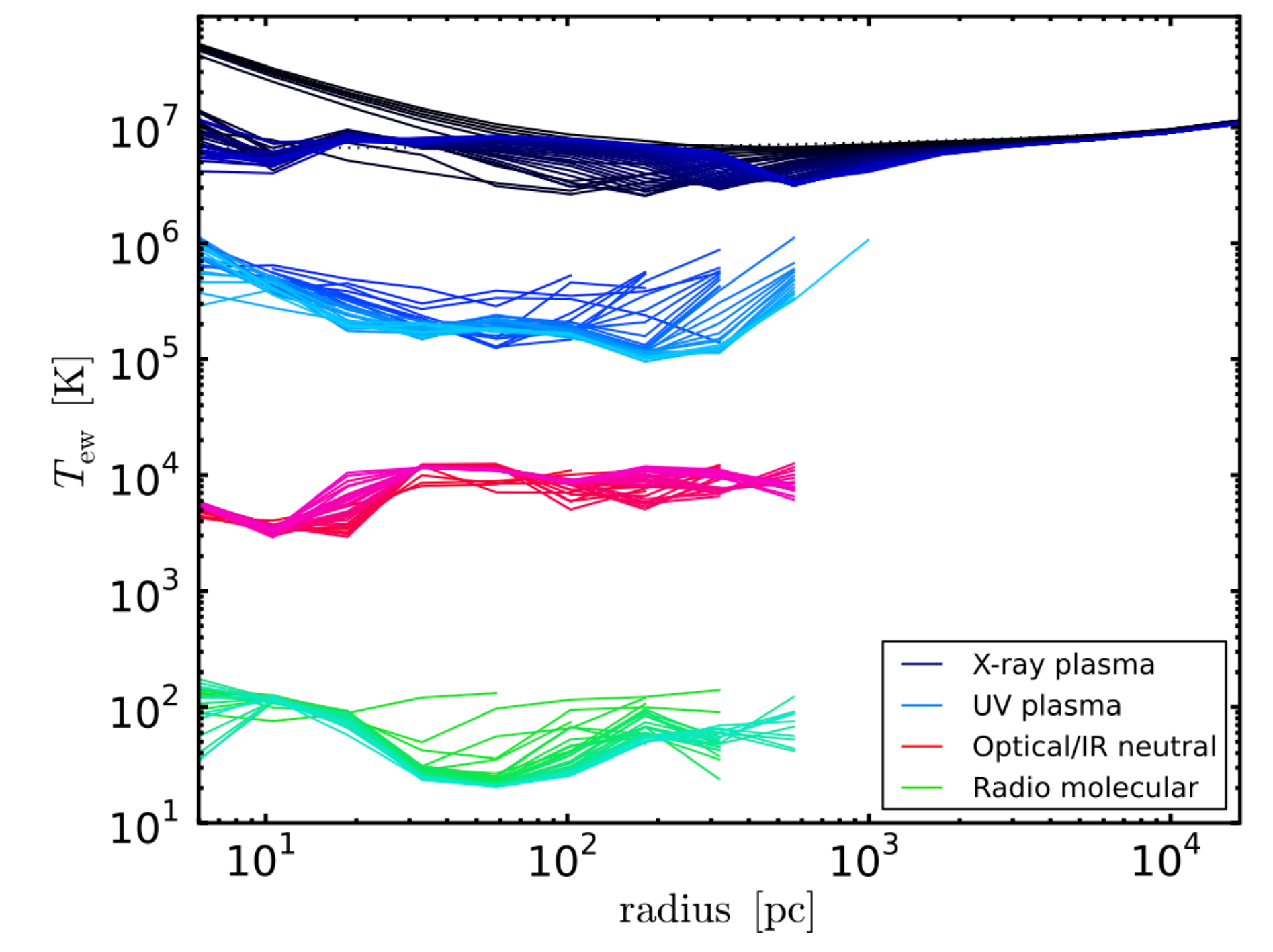}}
      \caption{Accretion with cooling and rotation: 3D emission-weighted radial profiles of density and temperature. 
      The multiphase flow is mainly comprised of plasma (blue), neutral gas (red), and molecular gas (green). The plasma is further separated into the X-ray ($T\ge0.1\,$keV) and UV component ($10^{4.2}\,{\rm K}<T<0.1\,$keV). Number of bins is 15 to emulate observational post-processing. The 40 Myr evolution is color-coded from dark to lighter color with 1 Myr step (the dotted line is the initial profile). The mass-weighted profiles follow similar trends. The top right y-axis denotes the electron number density $n_{\rm e, x}=\rho_{\rm x}/(\mu_{\rm e}\,m_{\rm p})$, valid only for the hot plasma. Notice the warm/cold phase is not volume filling as the plasma halo; the related emission-weighted density profiles thus represent values internal to the phase.}
       \label{f:cool_e03_prof}      
\end{figure}      
      
\begin{figure}
      \centering
      \subfigure{\includegraphics[scale=0.42]{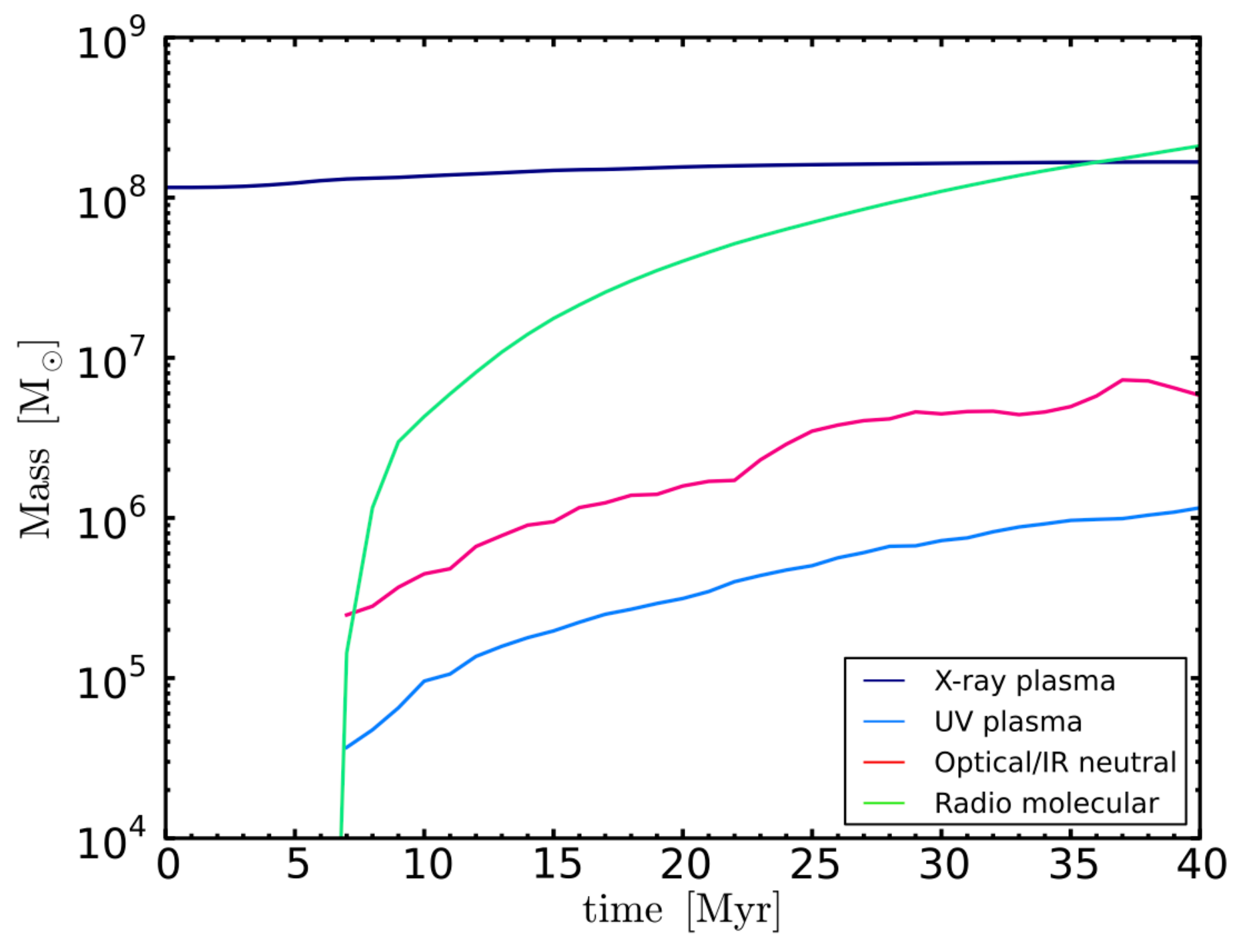}}
      \caption{Accretion with cooling and rotation: evolution of mass for each of the major multiphase components (\S\ref{s:phases}) within $r\lesssim2\,{\rm kpc}$: plasma (blue), warm neutral gas (red), and cold gas (green). }
       \label{f:cool_e03_masses}      
\end{figure} 

\vspace{-0.41cm}
\subsection[]{Multiphase condensation} \label{s:cool_e03_multi}
\begin{figure}
      \centering
      \subfigure{\includegraphics[scale=0.405]{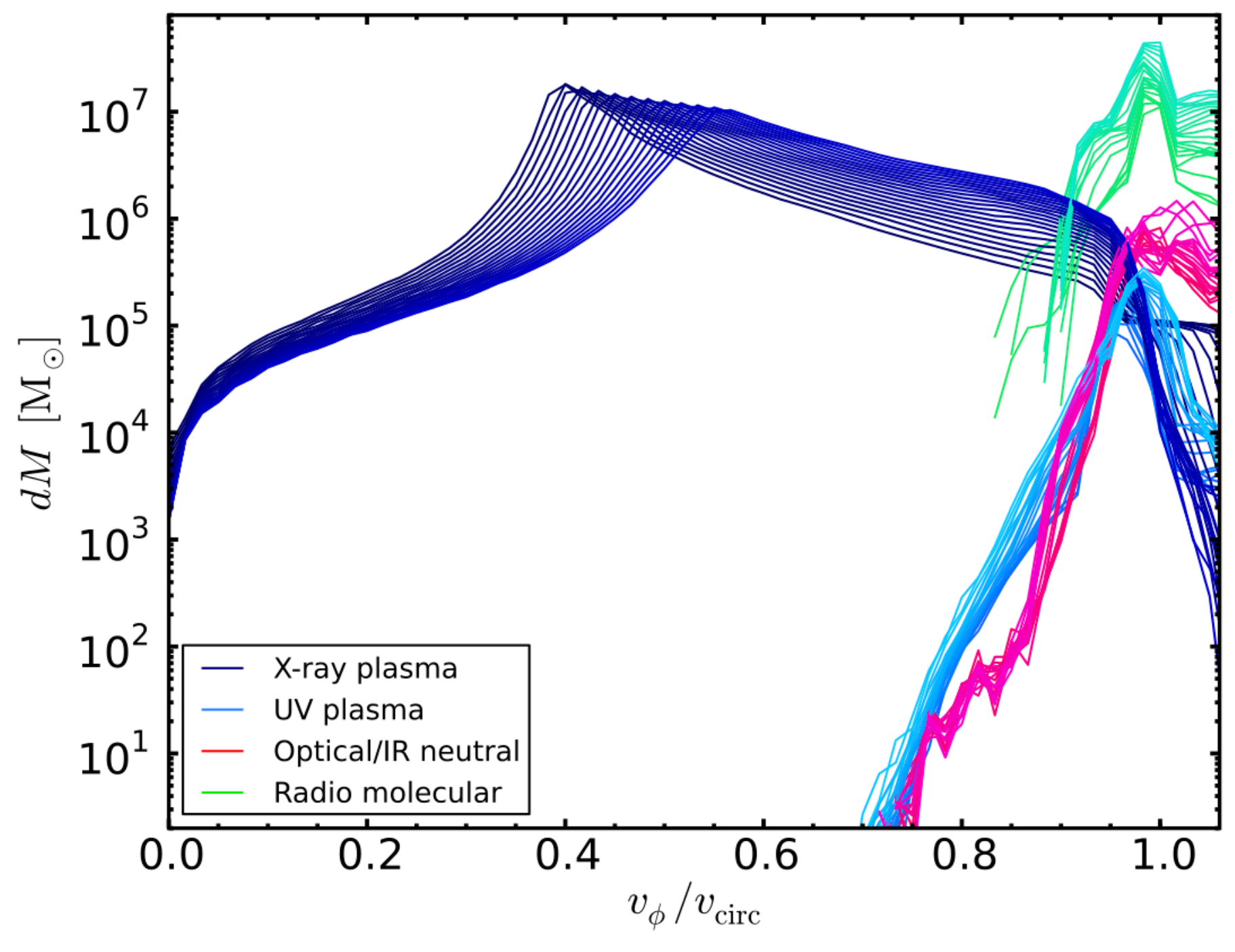}}
      \caption{Accretion with cooling and rotation: multiphase mass distribution per bin of rotational velocity divided 
      by circular velocity $v_{\rm circ}=({g_{\rm tot}\,r})^{1/2}$.
      The bin spacing is 0.02 and the extraction radius is 2 kpc. 
      Lines start from $t=\,$20 Myr, after most of the warm/cold halo has circularized and condensation has developed.
         }
      \label{f:cool_e03_vphi}    
\end{figure}

\noindent
Besides BH accretion, a primary goal of our investigation is to assess the condensation and evolution of the different multiphase components (\S\ref{s:phases}). 
We present several diagnostics which help us to dissect the multiphase structure.
In Fig.~\ref{f:cool_e03_prof}, we analyze the 3D radial profiles of gas density and temperature weighted by the emissivity in each of the main observable bands: X-ray, UV, optical/IR, and radio band.
Fig.~\ref{f:cool_e03_masses} tracks the temporal evolution of the masses within $r<2\,$kpc. 
Fig.~\ref{f:cool_e03_vphi} shows the multiphase mass distribution per bin of rotational velocity normalized to the circular value. 

Three main phases clearly emerge from the full cooling cascade: the plasma (blue), the neutral gas (red), and the molecular gas (green). The UV plasma (light blue) is differentiated from the X-ray plasma (dark blue) to be as close as possible to observations.
The gaseous atmosphere starts to condense from the hot X-ray emitting halo with $T_{\rm x}\approx0.6$\,-\,1\,keV. After reaching a quasi equilibrium from the initial numerical condition (dotted), the X-ray density is  
$\rho_{\rm x}\approx4\times10^{-25}$\,g\,cm$^{-3}$ (electron number density $n_{\rm e, x}\approx0.2$\,cm$^{-3}$) at 1\,kpc radius.
After 7 Myr, the cooling plasma reaches the UV regime ($10^{4.2}\,{\rm K}<T<0.1\,$ keV) and line cooling boosts the cooling curve by 2 dex, accelerating the condensation cascade (Fig.~\ref{f:lambda}). 
The typical density and temperature of the UV plasma are 
$\rho_{\rm uv}\approx10^{-24}$\,-\,$2\times10^{-23}$\,g\,cm$^{-3}$
and $T_{\rm uv}\approx2\times10^5\,$K.
The slopes of the X-ray and UV plasma density profiles are analogous $\propto r^{-1}$, shallower than the galactic gradient ($\propto r^{-1.3}$) and the BH free-fall scaling ($\propto r^{-1.5}$; see \S\ref{s:cool_e00_multi}) due to the continuous gas dropout.
The temperature slopes track each other too, nearly flat at $r>200$\,pc, with a mild increase toward the center. The X-ray and UV plasma may be thus considered one phase (hot gas) with different normalizations.
As the system evolves, the hot gas density steadily drops by $\sim$\,1 dex, as the hot halo becomes more diffuse.
The X-ray mass within 2 kpc is however constant $M_{\rm x}\sim\,$$10^8\,\msun$ (Fig.~\ref{f:cool_e03_masses}), since condensation is balanced by the inflow of plasma from the outer shells. For reference, the X-ray mass within 5 kpc and 20 kpc is on average $10^9\,\msun$ and $10^{10}\,\msun$, respectively.
The UV mass is always subdominant compared with all other phases, rising from $10^5\,\msun$ to $10^6\,\msun$ after 33 Myr, albeit the emissivity can be substantial due to the peak in $\Lambda$.

\begin{figure}
     \centering
     \subfigure{\includegraphics[scale=0.5]{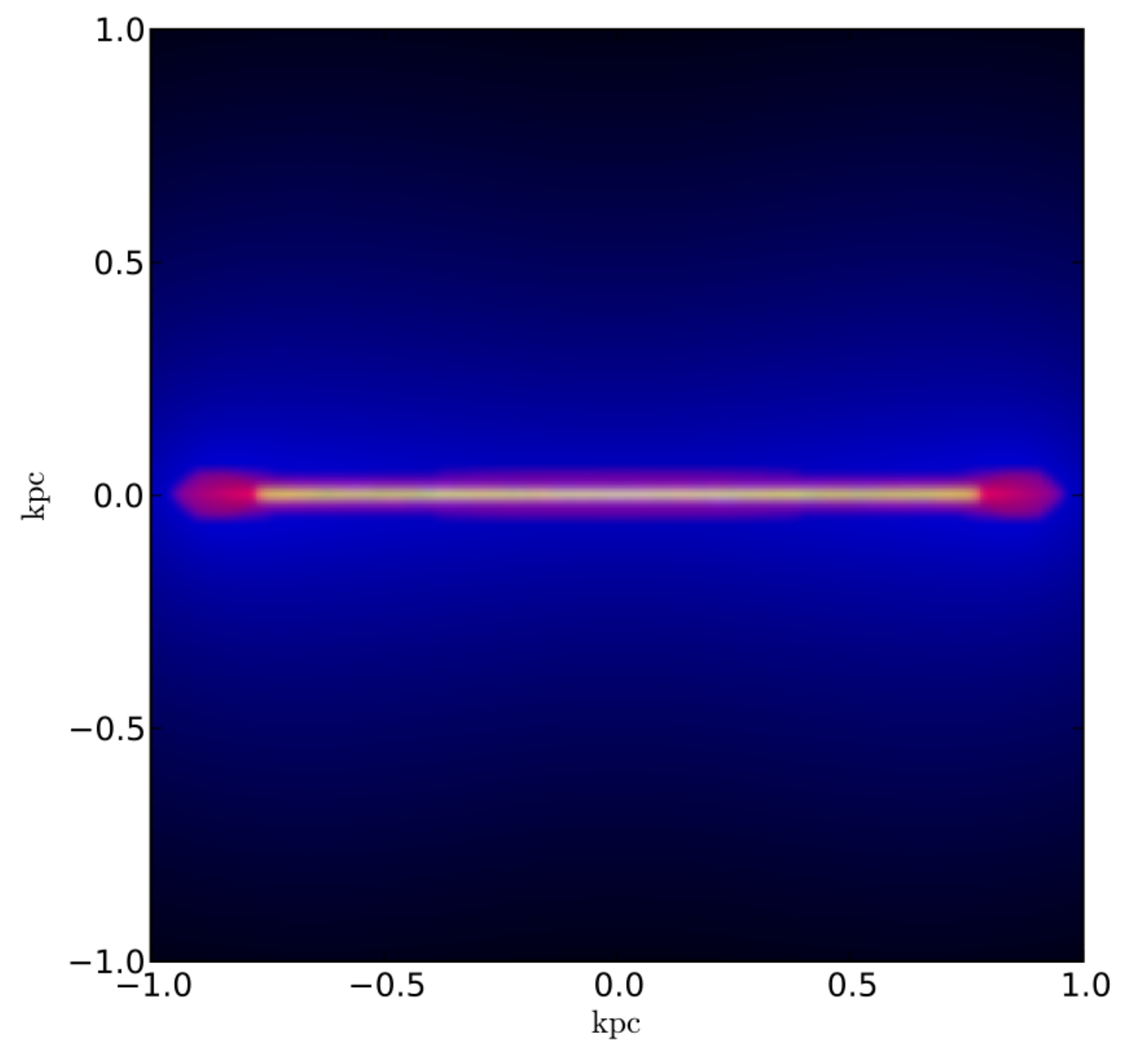}} 
     \caption{
     Accretion with cooling and rotation: composite RGB image of the surface density related to the plasma (blue), warm gas (red), and cold gas (green), showing the edge-on projected view of the multiphase disc ($t=40$\,Myr).
        }
         \label{f:cool_e03_RGB}
\end{figure}

In a few kyr after the UV phase emerges, the centrally condensing gas settles to the first plateau at $T\simeq5\times10^3$\,-\,$10^4$\,K where the emissivity drops abruptly due to hydrogen recombination (Fig.~\ref{f:lambda}). Most of the gas is expected to be neutral ($f_{\rm ion}\sim0.01$) and is associated with the outer layer of the rotating disc (Fig.~\ref{f:cool_e03_RGB})
experiencing major torques (\S\ref{s:cool_e03_dyn}).
The optical/IR gas (red lines in Fig.~\ref{f:cool_e03_prof}) does not display a smooth density profile.
In the inner 20 pc, the strong influence of the SMBH gravitational potential induces emission-weighted neutral gas densities up to 
$10^{-21}$\,g\,cm$^{-3}$, which decline by 2 dex at 30 pc. The second increase delimits the equatorial edge of the disc, where the warm gas accumulates and the flow is blocked by the inner circularized gas. 
This effect and the increasing $P$ scale height at larger radii create a mild convective flare, as seen in the projected RGB image (red lobes in Fig.~\ref{f:cool_e03_RGB}) and radial velocity map (Fig.~\ref{f:cool_e03_vrad}). 
The temperature radial profile is relatively smooth, with gas at 5000-8000 K dominating the warm phase: below this regime the cooling function becomes shallow allowing the warm gas to quickly condense to the molecular phase.
The total mass of the warm gas (Fig.~\ref{f:cool_e03_masses}) reaches $6\times10^6\,\msun$ after 40 Myr, $\sim$\,1 dex higher than the UV mass and 1.5 dex lower than the molecular mass. Less than 1/3 of the warm mass is due to gas with $200<T<5000$\,K.

The warm gas which has settled on a disc structure continues to cool down to the radio/molecular regime (green lines), as there is no major balancing heating (photoelectric heating just slightly delays cooling). The radial profile gradients of the molecular gas are fairly smooth, with characteristic temperature and internal density 30\,-\,50\,K and a few $10^{-21}$\,g\,cm$^{-3}$ ($10^3$\,cm$^{-3}$ in molecular number density), respectively. 
Below the UV regime, cooling departs from the isobaric mode as the gas cooling time is much shorter than the plasma sound crossing time.
Furthermore, the strong radiative cooling tends to quickly dissipate any shock induced variation in the cold phase.
At $r<20$ pc, part of the gas with low angular momentum flows in and is re-heated via compressional heating above $100$\,K, repopulating the warm phase. Most (90\%) of the inner accreted gas is thus warm gas. The total molecular mass is $4\times10^7\,\msun$ after 20 Myr and reaches $2\times10^8\,\msun$ at 40 Myr, which is comparable to the X-ray plasma mass within $r<2\,$kpc. The average molecular surface density of the disc is $\approx100\,\msun$\,pc$^{-2}$, up to 1.5 dex higher than that of the warm phase.

The UV plasma, neutral and molecular phases show all the same radial extension, forming a {\it multiphase disc}. 
Fig.~\ref{f:cool_e03_vphi} shows that the 3 components have on average reached circularization, $v_{\rm rot}\sim v_{\rm circ}$. The mass distributions have an extended left tail with $v_{\rm rot}/v_{\rm circ}<1$, which is related to the accreting gas (\S\ref{s:cool_e03_dyn}). A fraction of gas mass exhibits slightly $v_{\rm rot}/v_{\rm circ}\gta1$ motions, in particular the warm phase, due to the expanding outer disc edge which blocks infalling fluid elements. 
The warm phase is more subject to fluctuations as it is located in the intermediate transition layer experiencing the largest $P$ gradient and thus torques due to condensation.

\vspace{-0.41cm}
\subsection[]{Observations} \label{s:cool_e03_obs}
The composite RGB image of projected density (Fig.~\ref{f:cool_e03_RGB}) captures the multiphase structure:
the molecular gas defines the inner pancake (green), the neutral phase is the intermediate layer (red), and the UV/soft X-ray plasma is the envelope. The multiphase and stratified disc is a relevant phase of the condensation mechanism, as the multiphase rain will be for the turbulent CCA (\S\ref{s:cca}). The two modes (coherent versus chaotic) cycle through the evolution as a function of the dynamical state, albeit the incidence of coherent rotation is expected to be much smaller in massive galaxies and clusters due to the presence of frequent perturbers.
The size of the disc is a probe of how long the quiescent cooling-dominated phase has lasted: every 40 Myr the disc radius and mass may grow by $\sim$\,1\,kpc and $10^8\,\msun$, respectively.
AGN feedback will at some stage restore heating and turbulence, preventing further growth or destroying the disc (\citealt{Gaspari:2012a,Barai:2016}).
A limitation of the pure cooling model is indeed that overcooling starts to become relevant after 30\,-\,40\,Myr increasing the nuclear plasma density, and thus X-ray brightness, toward $n_{\rm e,x}\sim10$\,cm$^{-3}$. Nevertheless, integrating the related X-ray emissivity profile ($\propto \rho_{\rm x}^2$) within 150 pc, we retrieve a nuclear X-ray luminosity $1.1\times10^{40}$\,erg\,s$^{-1}$ which is still consistent with spectroscopic data of AGN point sources at the center of massive galaxies  (Fig. 13 in \citealt{Russell:2013}).
Current X-ray imaging typically resolves the kpc or larger scale; at such radius the simulated plasma density mildly varies by $\lta2\times$ ($\lta4\times$ in brightness) around $n_{\rm e,x}\approx0.1$\,cm$^{-3}$.

The presence of a disc or not is key to assess if the galactic atmosphere is rotation or turbulence dominated; in dimensionless terms, if turbulent Taylor number ${\rm Ta_t}\equiv v_{\rm rot}/\sigma_v>1$ or $<1$. \citet{Young:2011} find double-horned line profiles in a third of the ETGs detected in CO, suggesting that the molecular gas is in a disc with a flat rotation curve, although the data quality may not rule out absorption in some cases.
The typical size, mass, and surface density of the detected molecular discs are 1-2\,kpc, 1-5$\times10^8\,\msun$, and 50\,-130\,$\msun\,{\rm pc^{-2}}$, respectively. Very similar molecular disc fraction and masses are found in the IRAM sample of 52 massive elliptical galaxies by \citet{Ocana:2010}.
These observed values are in good agreement with the predictions of our disc simulation. 
For the warm gas, there is evidence of rotating disc in ETGs as NGC 5077 (\citealt{Caon:2000}), NGC 6868, and NGC 7049 (\citealt{Werner:2014}; see also \citealt{Brighenti:1997}), reaching radii up to 3 kpc, albeit the dynamics of the warm gas is typically irregular over the full sample.
More massive galaxies in clusters can show discs with molecular masses up to a few $10^9\,\msun$ and $R\lta 5$ kpc as in Hydra-A (\citealt{Hamer:2014}), again overcoming the warm phase mass by 1-2 dex, albeit it may be an exceptional case.
\citet{Hamer:2016} find several clusters harbor H$\alpha$/warm gas with rotating structures correlated with undisturbed condensation mode.
On the other hand, evidence for discs is lacking in the ALMA observations of A1664 (\citealt{Russell:2014}) and PKS 0745-191 (\citealt{Russell:2016}), corroborating the trend that irregular dynamics appears more frequent in massive systems (see also \citealt{Heckman:1989}).

In this internal multiphase condensation model, spiral galaxies may be viewed as an extreme version of the ${\rm Ta_t \gg 1}$ regime, in which rotation overcomes turbulence, thereby inducing a very extended and massive neutral and molecular disc (surface densities are similar to that of ETG discs; cf.~\citealt{Krumholz:2009}).
The associated suppression of the BHAR and the divergence from the cooling rate (\S\ref{s:cool_e03_dyn}) may help to explain why spiral galaxies have much smaller SMBH masses compared with elliptical galaxies (\citealt{Kormendy:2013}) and why central SMBHs appear to be disconnected from the properties of the spiral host, as suggested by the megamaser discs observed with HST (\citealt{Lasker:2016}).

\vspace{-0.41cm}
\section[]{Multiphase cloud}  \label{s:cool_e00}
In the previous Section, we probed a massive ETG with a rotating hot halo consistent with the upper envelope of the observed range. We now probe the regime of pure cooling with negligible rotation, i.e., a very quiet and passive ETG.

\begin{figure}
     \subfigure{\includegraphics[scale=0.5]{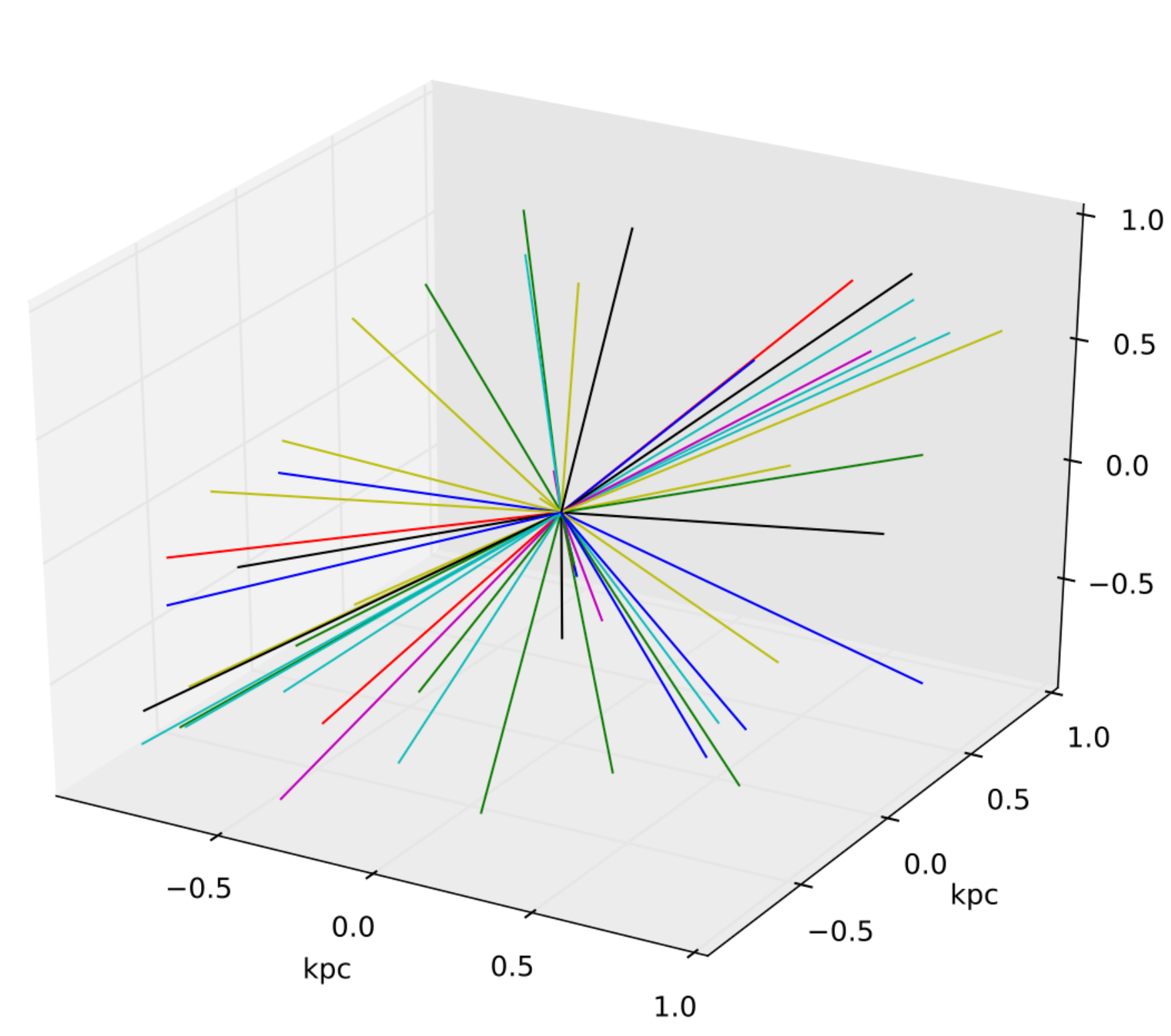}}
     \caption{Accretion with cooling and no rotation: a sample of streamlines integrated from the outer kpc region ($t=40\,$Myr), which shows the perfectly spherically symmetric accretion.
     }
         \label{f:cool_e00_stream}
\end{figure} 

\begin{figure}
     \centering
     \subfigure{\includegraphics[scale=0.47]{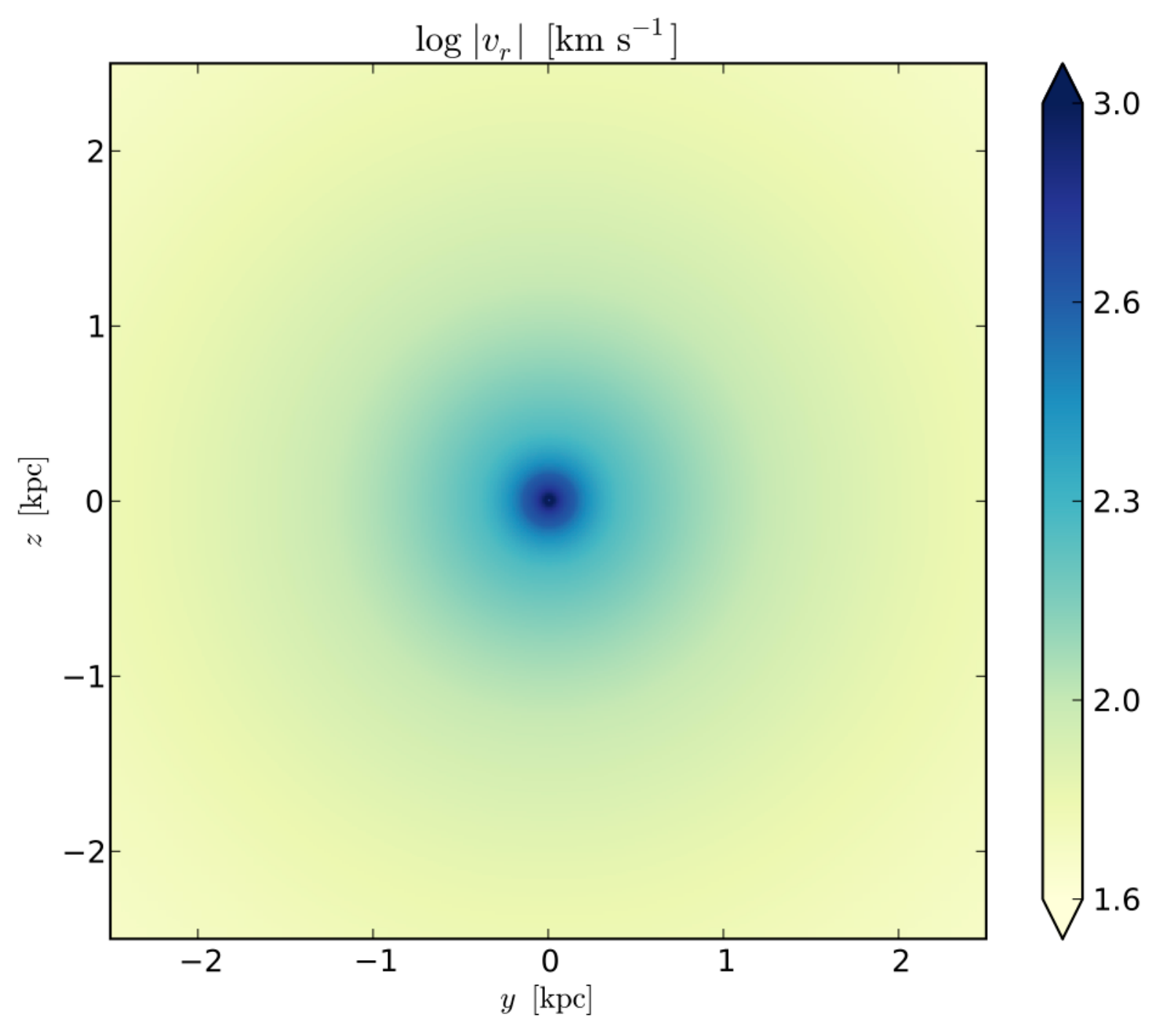}}
     \caption{Accretion with cooling and no rotation: absolute value of the 
     radial velocity in the mid-plane cross-section through the $x-$axis (5x5 kpc$^2$; $t=40\,$Myr). The condensing gas is spherically inflowing, approaching the free-fall scaling.}
         \label{f:cool_e00_vrad}
\end{figure}

\vspace{-0.41cm}
\subsection[]{Dynamics \& accretion} \label{s:cool_e00_dyn}
\begin{figure}
      \begin{center}
      \subfigure{\includegraphics[scale=0.31]{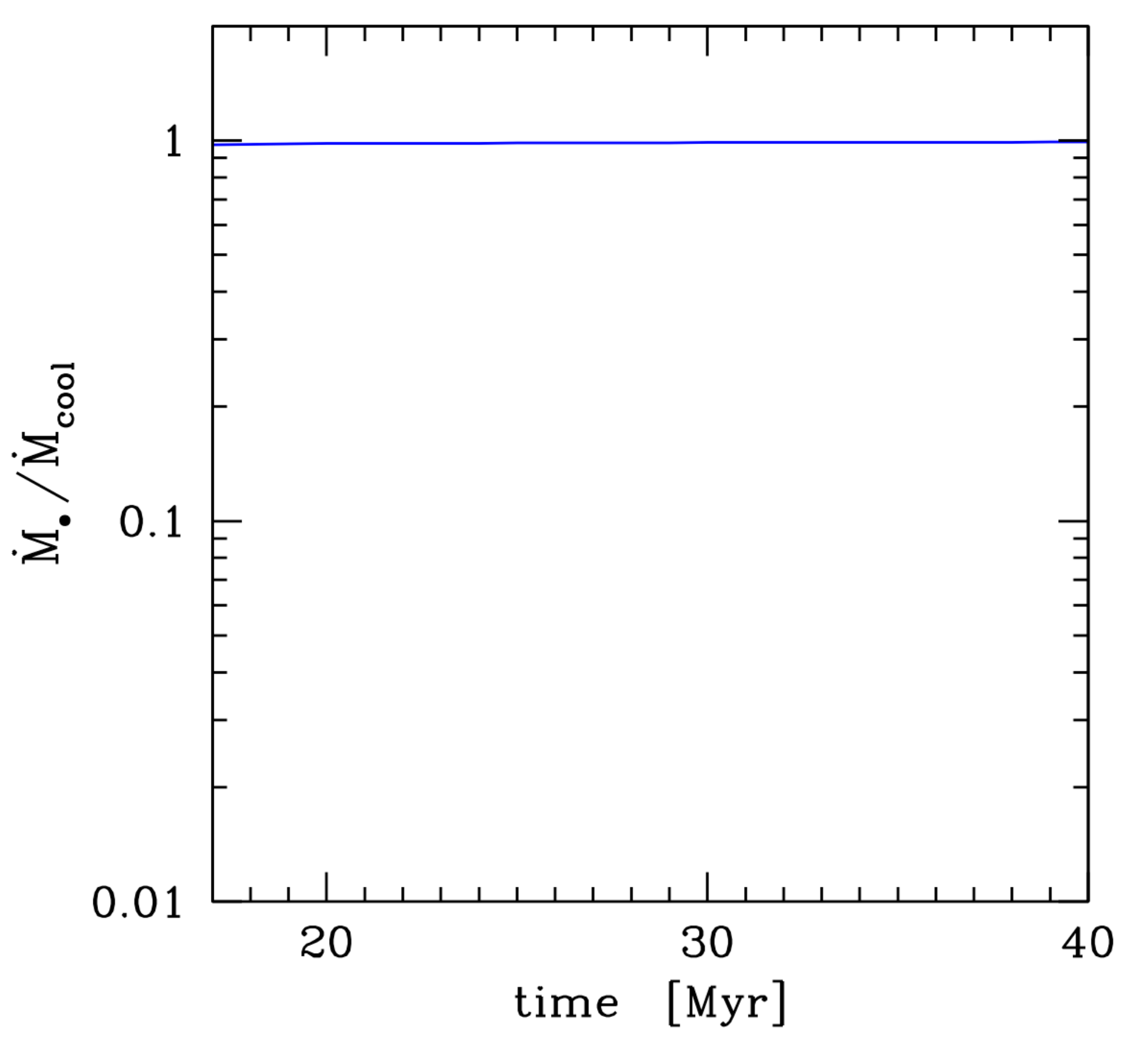}}
      \end{center}
      \caption{Accretion with cooling and no rotation: evolution of the average accretion rate normalized to the cooling rate (cf.~Fig.\ref{f:cool_e03_mdot}). Without centrifugal barrier, the accretion flow becomes a pure cooling flow with accretion rate equal to the maximal dynamical value, the cooling rate. 
      }
      \label{f:cool_e00_mdot}
\end{figure}

The hot plasma loses pressure via radiative emission and, lacking the centrifugal barrier driven by rotation, it accretes as a spherically symmetric flow. The condensing gas radially converges toward the central point-like SMBH (Fig.~\ref{f:cool_e00_stream}).
As seen in the radial velocity map (Fig.~\ref{f:cool_e00_vrad}), the flow approaches the free-fall velocity 
$v_{\rm ff}=[2GM(<$$\,r)/r]^{1/2}$,
e.g., at 20 pc the inflow velocity has reached $-v_r\simeq1100$\,km\,s$^{-1}\,\simeq v_{\rm ff}$.
The gas is thus able to accrete at the maximal dynamical rate $\dot M_\bullet = \dot M_{\rm cool}$ (Fig.~\ref{f:cool_e00_mdot}), where $\dot M_{\rm cool}$ is the actual simulated cooling rate emerging from the UV to cooler phases condensation.
The relative difference between the accretion and cooling rate is $\lta1$\%, meaning that the flow is equivalent to a pure cooling flow. Compared with the multiphase disc evolution and related 99\% suppressed accretion (\S\ref{s:cool_e03_dyn}), the multiphase sphere model represents the exact opposite.
Choosing the spherical Bondi rate as reference, the final accretion rate is $\dot M_\bullet \simeq 200\,\dot M_{\rm B}$. We remark that the Bondi rate formula is only suited for hot adiabatic flows, which is a rare state for the cores of galaxies, groups, and clusters.

From a numerical perspective, the run probes the good stability and symmetry preservation of the code. The noise of the grids is negligible and does not introduce spurious fluctuations in the accretion and condensation flow, as shown in Fig.~\ref{f:cool_e00_stream}-\ref{f:cool_e00_vrad}. We inspected the torque maps (e.g., due to pressure gradients) and found to be null to machine precision. 

\vspace{-0.41cm}
\subsection[]{Multiphase condensation} \label{s:cool_e00_multi}
The multiphase condensation follows the cascade described in \S\ref{s:cool_e03_multi}, with the major difference that it is halted at the warm phase regime.
Because of the strong convergence of streamlines at progressively smaller radii (Fig.~\ref{f:cool_e00_stream}), the compressional heating driven by the gravitational force plays an important role in the condensation dynamics (Eq.~\ref{e:comp}; more below). First, it is able to delay the cooling time of the warm phase.
Second, it prevents the emergence of the cold/molecular phase. This model is consistent with the imposed 2-phase G13 model.

\begin{figure}
      \subfigure{\includegraphics[scale=0.4]{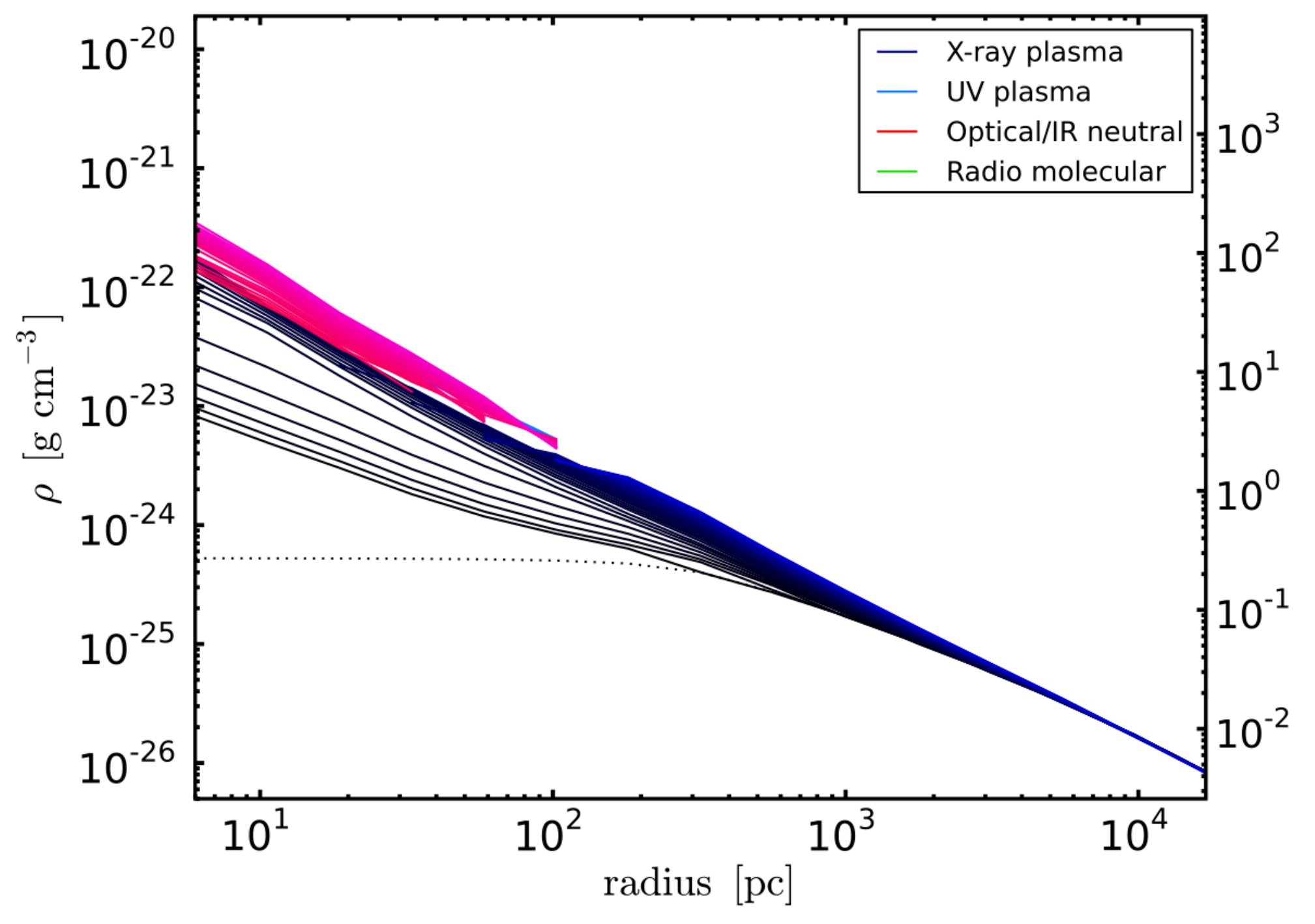}}
      \subfigure{\includegraphics[scale=0.403]{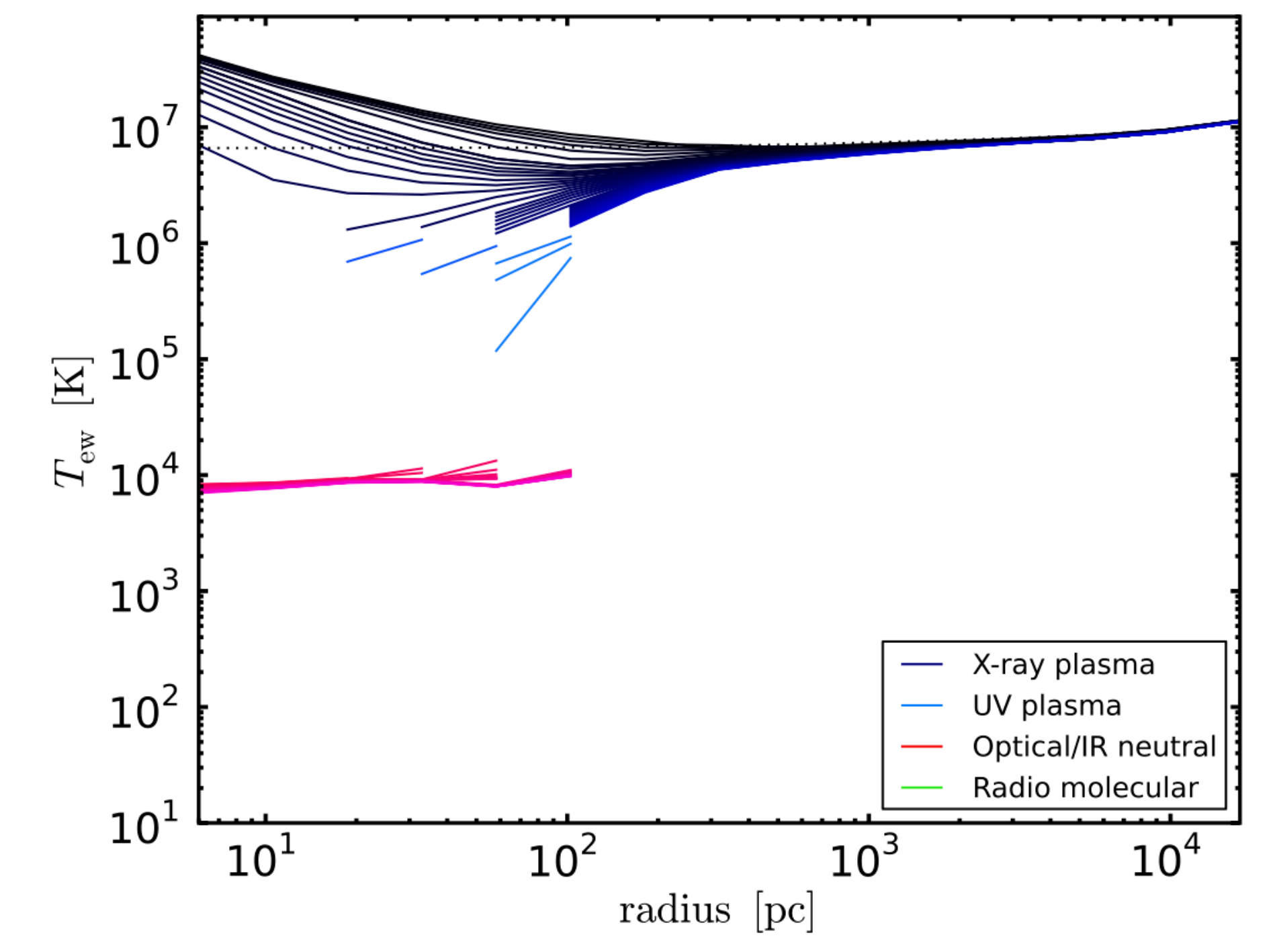}}
      \caption{Accretion with cooling and no rotation: 3D emission-weighted radial profiles of density and temperature, probing the multiphase structure of the accretion flow (cf.~Fig.~\ref{f:cool_e03_prof}). The top right y-axis denotes the electron number density $n_{\rm e, x}$ for the hot plasma.
      At late times the warm phase has dropped out of the hot phase and fully dominates within 100 pc radius. Beyond such radius the hot phase always dominates the mass content (see Fig.~\ref{f:cool_e00_masses}). The molecular phase is inhibited by compressional heating in such spherical cooling flow. 
      }
       \label{f:cool_e00_prof}      
\end{figure}

\begin{figure}
      \centering
      \subfigure{\includegraphics[scale=0.42]{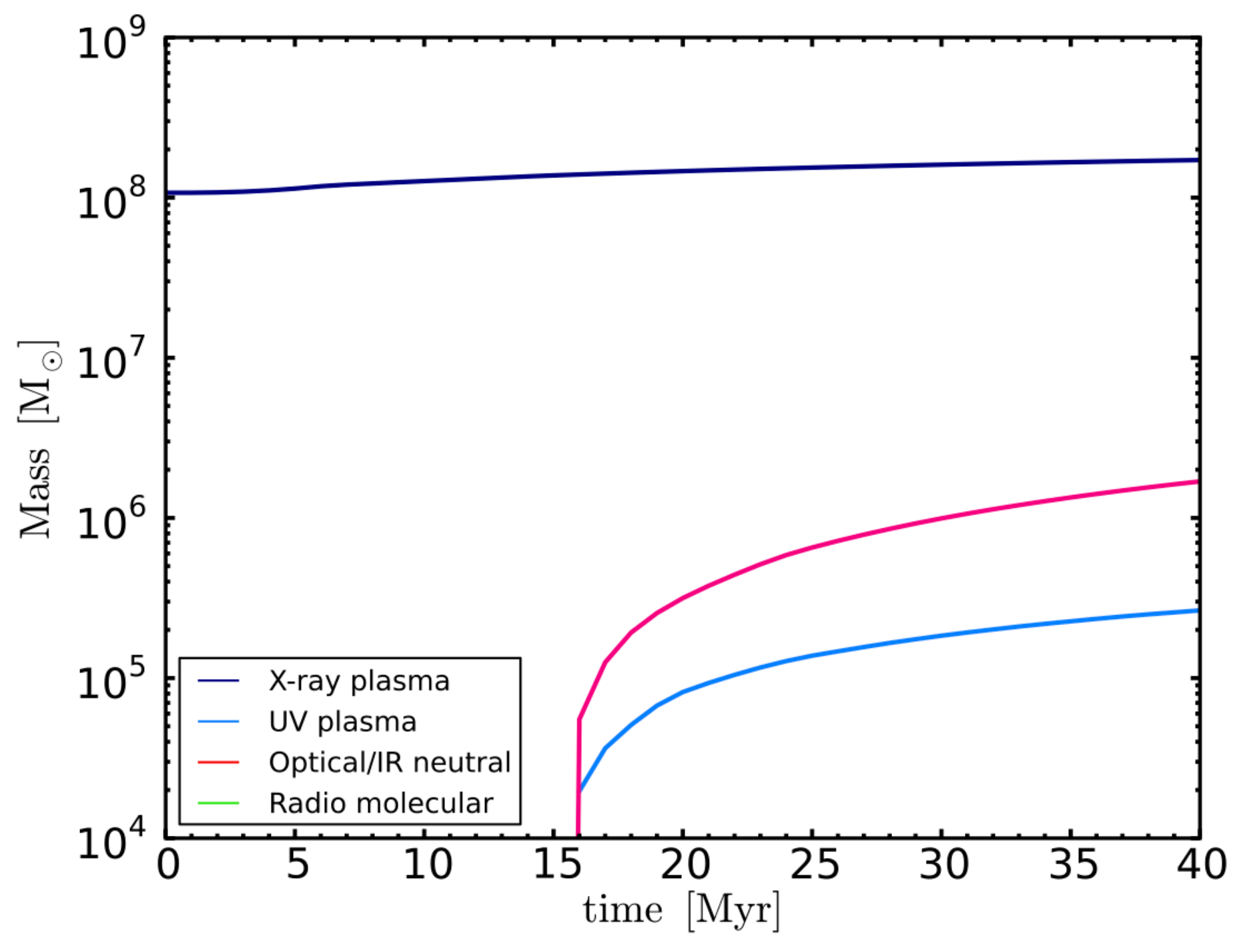}}
      \caption{Accretion with cooling and no rotation: evolution of mass for each of the multiphase gas residing within $r\lesssim2\,$kpc (\S\ref{s:phases}). Notice that most of the condensed mass has been sinked by the SMBH.}
       \label{f:cool_e00_masses}      
\end{figure}

The hot plasma cools from $T_{\rm x}\simeq0.7$\,keV and as it collapses the density increases to 10s cm$^{-3}$  (Fig.~\ref{f:cool_e00_prof}) within 20\,pc radius.
The nuclear plasma density is almost 1 dex higher than that of the multiphase disc.
The system is entering a significant cooling flow regime, which cannot be the characteristic long-term state of massive galaxies according to X-ray observations.
The X-ray plasma density profile evolves toward $\propto r^{-3/2}$, which is the expected profile for a free-falling flow within the SMBH influence region -- using the above free-fall velocity in the mass continuity equation $\dot M = 4\pi\,r^2\rho v$ leads to $\rho\propto r^{-2}v^{-1}\propto r^{-3/2}$. The X-ray temperature profile initially tries to approach the free-fall scaling too, $T\propto r^{-1}$ -- via the adiabatic internal energy equation $\partial \ln T/\partial \ln r = (\gamma-1) \,\partial \ln \rho/\partial \ln r = -1$. 
However, the continuous central hot gas accumulation raises the cooling rate, decreases first the intermediate (50-100\,pc) temperature, and then induces the inner collapse, until the major phase transition to the warm gas occurs at $t\gta16$\,Myr.
The inner gas fully drops out of the hot phase and quickly passes through the UV regime (light blue).
The UV phase has no extended radial profile as it is compressed in the thin external layer of the multiphase cloud, which rapidly morphs into the warm phase. This layer roughly marks the sonic point transition, which expands with time.
The final UV mass is small, $M_{\rm uv}\simeq3\times10^5\,\msun$ (Fig.~\ref{f:cool_e00_masses}), which is $3\times$ lower than that of the disc evolution. In the rotating model the plasma fills part of the inner volume, with $\rho(r)$ shallower than the free-fall scaling due to the continuous dropout (Fig.~\ref{f:cool_e03_prof}). 

The inner condensed gas settles to the warm phase regime.
The warm/optical gas has an average temperature $T_{\rm warm}\approx8000$\,K, with minimum 5500 K, 
and inner density $2\times10^{-22}$\,g\,cm$^{-3}$; no major oscillations are observed. 
Given the fast collapse, the condensing gas cools in nearly isochoric -- and not isobaric -- mode, 
even more efficiently than in the disc evolution.
The density profile of the neutral gas ($\sim$\,1\% ionized) continues the trend of the hot phase, reaching the above free-fall scaling.
The total available mass (residual from sinking) at final time is $M_{\rm warm}=1.7\times10^6\,\msun$, which is 2 dex lower compared with the hot plasma within $r<2$\,kpc and $3.5\times$ lower than the warm mass of the multiphase disc run.
The expanding, dropped-out warm cloud has reached $r\simeq110$\,pc at 40 Myr and fully dominates the inner gas content (that is why the profiles appear `truncated' in Fig.~\ref{f:cool_e00_prof}). Beyond such radius the hot plasma always dominate the gas mass content (Fig.~\ref{f:cool_e00_masses}).

Compressional heating can be written as
\begin{equation}\label{e:comp}
\mathcal{H}_{\rm comp} \equiv-\,P\,\left(\boldsymbol{\nabla} \cdot \boldsymbol{v}\right) = -P\,\frac{v_r}{r}\left(\frac{\partial\ln v_r}{\partial\ln r} +2 \right) \approx \frac{3}{2}\,P\,\frac{v_{\rm ff}}{r},
\end{equation}
where the third step comes out assuming spherical symmetry, and the last using the free-fall velocity profile.
Dividing Eq.~\ref{e:comp} by the cooling rate of the warm phase, $\mathcal{L_{\rm warm}}=n^2_{\rm H}\Lambda_{\rm warm}$, and using the simulated radial profiles we obtain
${\mathcal{H}_{\rm comp}}/{\mathcal{L_{\rm warm}}} \simeq 2.4\times10^{-29}\,{T_{\rm warm}}/{\Lambda_{\rm warm}}\simeq 1$, which is independent of radius since $T_{\rm warm}\simeq8000$\,K is fairly constant with $r$ (Fig.~\ref{f:cool_e00_prof}) and the free-fall density scaling cancels out with the $r^{3/2}$ term at denominator.
In the soft X-ray/UV regime, instead, the cooling rate is always greater than $\mathcal{H}_{\rm comp}$ by 0.5-2 dex, since $-v_r \ll v_{\rm ff}$ (X-ray) or $\Lambda_{\rm hot}/T_{\rm hot}\gg\Lambda_{\rm warm}/T_{\rm warm}$ (UV). 
Overall, compressional heating prevents the formation of molecular gas and can only partially delay the plasma condensation.
The rotating halo (\S\ref{s:cool_e03_multi}) and CCA evolution (\S\ref{s:cca_multi}) lack such extreme radial inflow convergence (embodied in the compressive term $\boldsymbol{\nabla} \cdot \boldsymbol{v}$), allowing the 3rd phase to condense out and to generate larger UV/warm gas masses, as the gas settles vertically onto a disc or drifts non-radially in the turbulent field. 

\begin{figure}
     \centering
     \subfigure{\includegraphics[scale=0.5]{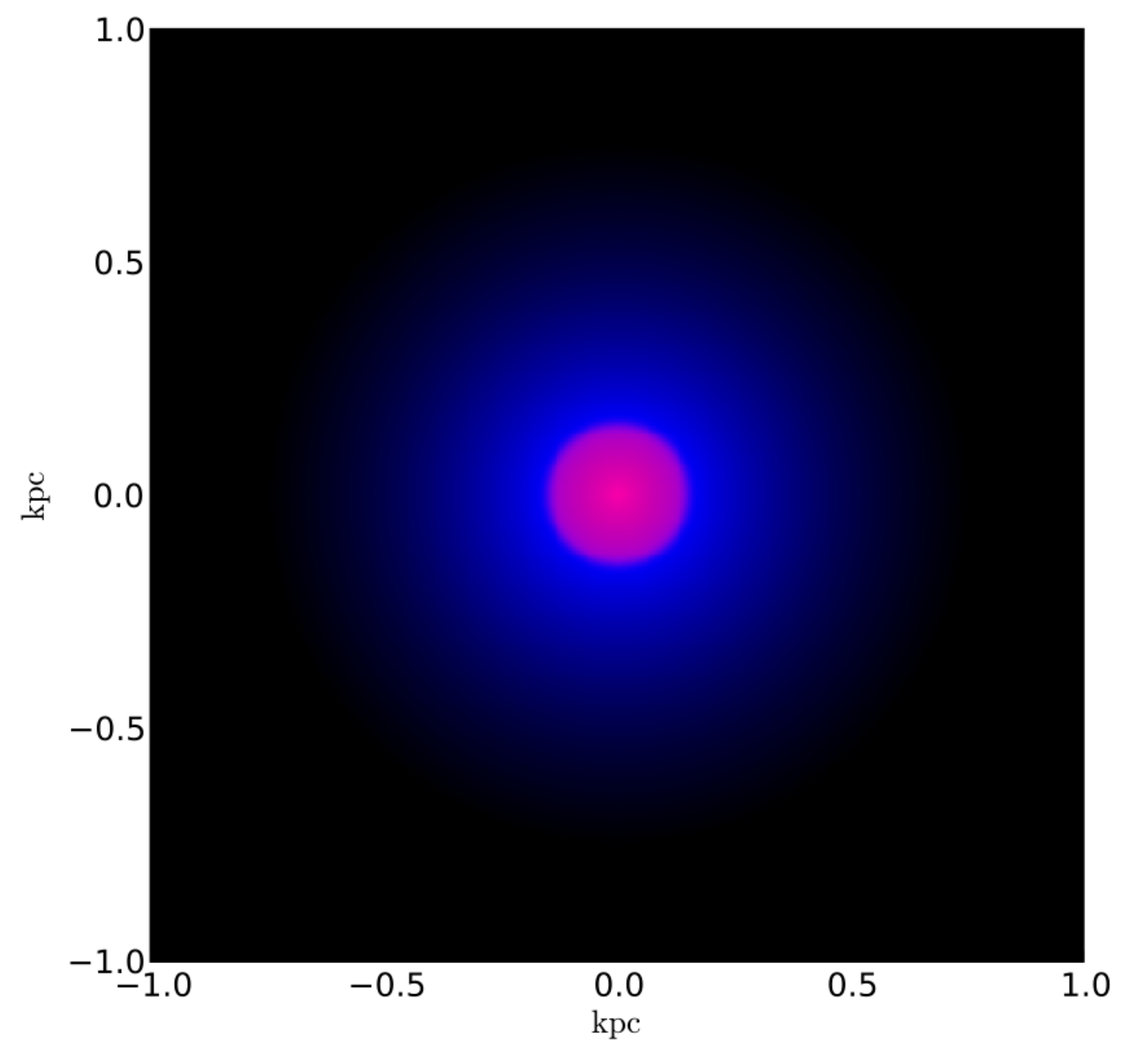}} 
     \caption{Accretion with cooling and no rotation: composite RGB image of the plasma (blue), warm gas (red), and cold gas (green; absent) surface density ($t=40$\,Myr). 
     In a pure cooling flow lacking rotation, the dominant phase is the warm gas (8000 K). Compressional heating prevents the formation of the cold molecular phase. The observed ETG related to this stage is very quiet and (thermo)dynamically passive.
        }
         \label{f:cool_e00_RGB}
\end{figure} 

\vspace{-0.41cm}
\subsection[]{Observations} \label{s:cool_e00_obs}
The RGB image of projected density (Fig.~\ref{f:cool_e00_RGB}) depicts how such atmosphere would be overall perceived, i.e., a central warm gas cloud with a thin UV layer embedded in the outer X-ray envelope. While the multiphase disc grew by 1\,kpc after 40 Myr, the multiphase spherical core has only reached 100\,pc due to the higher compression.
The typical effective radius of massive ETGs is $R_{\rm e}\gta10$\,kpc, thus the central warm cloud is a tiny fraction of the galactic and group volume and would be challenging to detect. Clearly, if left unopposed over the long term, this scenario would lead to a massive pure cooling flow, which is inconsistent with X-ray constraints (\S\ref{s:intro}), such as a substantial central X-ray brightness increment. Nevertheless, X-ray telescopes are usually limited to kpc scales; at such resolved radii the current profiles are dominated by the hot phase with reasonable number densities around 0.1\,cm$^{-1}$(Fig.~\ref{f:cool_e00_prof}).
We remark the objective of this model is primarily to test the physical effect of molecular cooling on a spherical cooling flow, so we limit the comparison to data in this Section only to a few salient aspects.

The ATLAS$^{\rm 3D}$ survey finds that ETGs with low angular momentum have low CO detection rates (\citealt{Young:2011}). If the galaxy is very quiet and passive -- negligible turbulence and AGN outbursts -- the impact of compressional heating increases and is able to prevent the internal formation of molecular gas. As seen in the \S\ref{s:cool_e03}, even in the disc evolution compressional heating helps to partially reheat the very inner cold phase.
While the warm phase (traced by H$\alpha$) often shows irregular velocity profiles and perturbed morphologies (\S\ref{s:intro}), there are ETGs in which the H$\alpha$ distribution is fairly spherically symmetric and compact in the core with smooth X-ray emission and negligible cold gas. A few examples are NGC 3379, NGC 5898, and NGC 4649 (\citealt{Caon:2000,Werner:2012,Werner:2014}), albeit we note stellar light contamination is often complicated to remove for some systems (e.g., NGC 1399).

Overall, the multiphase cloud regime describes the most passive and (thermo)dynamically quiet massive galaxies, as very low molecular gas masses lead to very low star formation rates (a `red and dead' galaxy). The multiphase disc and cloud stage represent two opposite dynamical regimes. 
However, massive galaxies more commonly reside in the intermediate state, in which the hot gas velocity field is neither fully rotating nor fully radially infalling, while experiencing recurring perturbations. At the same time, AGN heating tend to prevent over the long term a massive cooling flow. The remaining central part of the paper tackles such more realistic heated and turbulent regime.

\vspace{-0.41cm}
\section[]{Multiphase CCA  - chaotic cold accretion}  \label{s:cca}
\noindent
We tackle now how the accretion flow and multiphase condensation process evolve in the more common state of a cooling gaseous halo perturbed by turbulent motions and balanced by AGN heating.
X-ray halos of galaxies, groups, and clusters are recurrently perturbed by the action of AGN feedback, galaxy motions, and/or mergers (\S\ref{s:turb}). 
We will focus primarily on the models including the full physics: rotation, subsonic turbulence ($\sigma_v\approx165$\,km\,s$^{-1}$, a level similar to what {\it Hitomi} telescope has recently probed), radiative cooling down to the molecular regime, and AGN heating. All other partial physics models (e.g., no heating or no rotation) produce a similar normalized CCA evolution (\S\ref{s:cca_norot_noheat}). 

\begin{figure}
      \centering
      \subfigure{\includegraphics[scale=0.31]{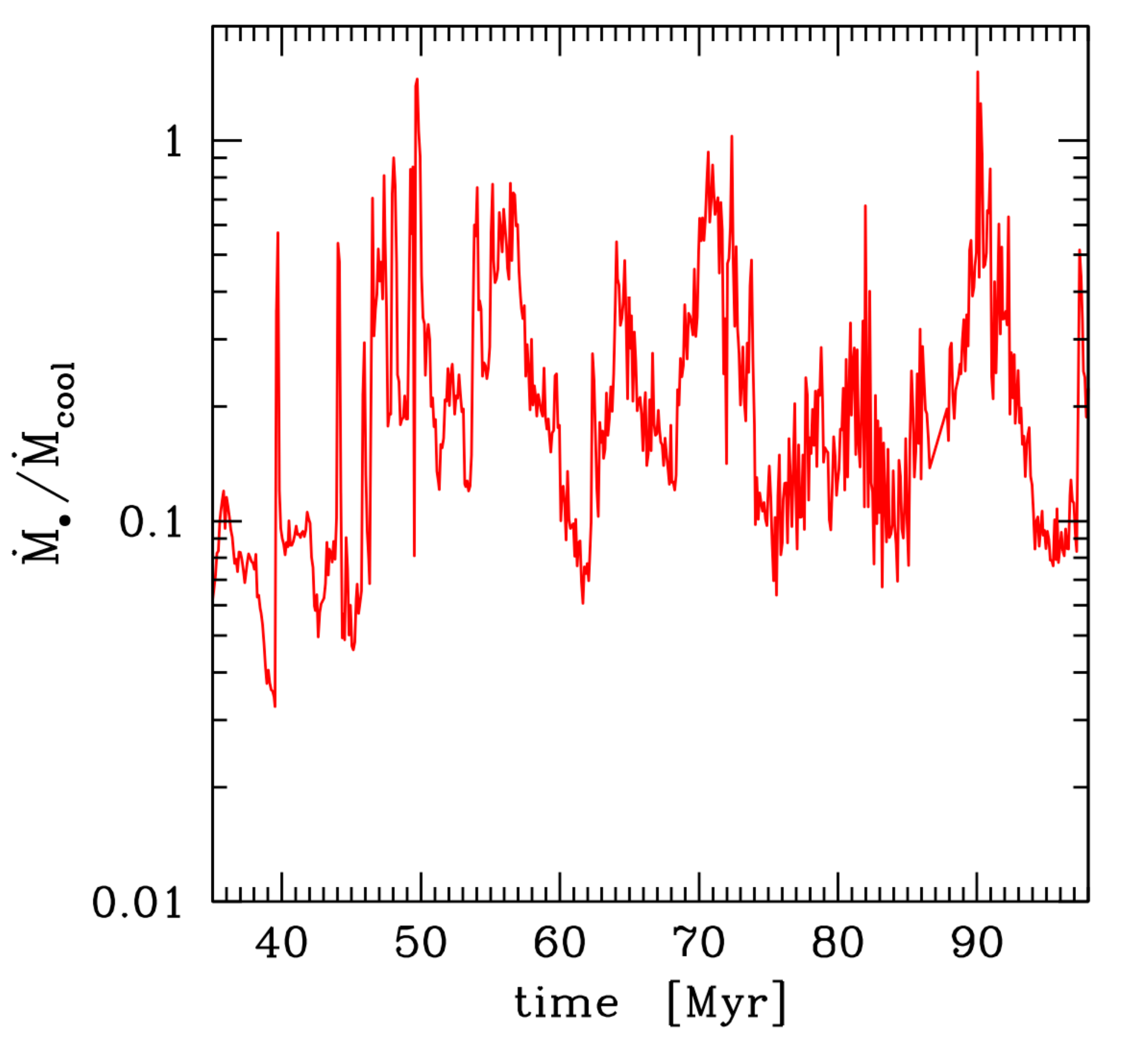}}
      \subfigure{\includegraphics[scale=0.31]{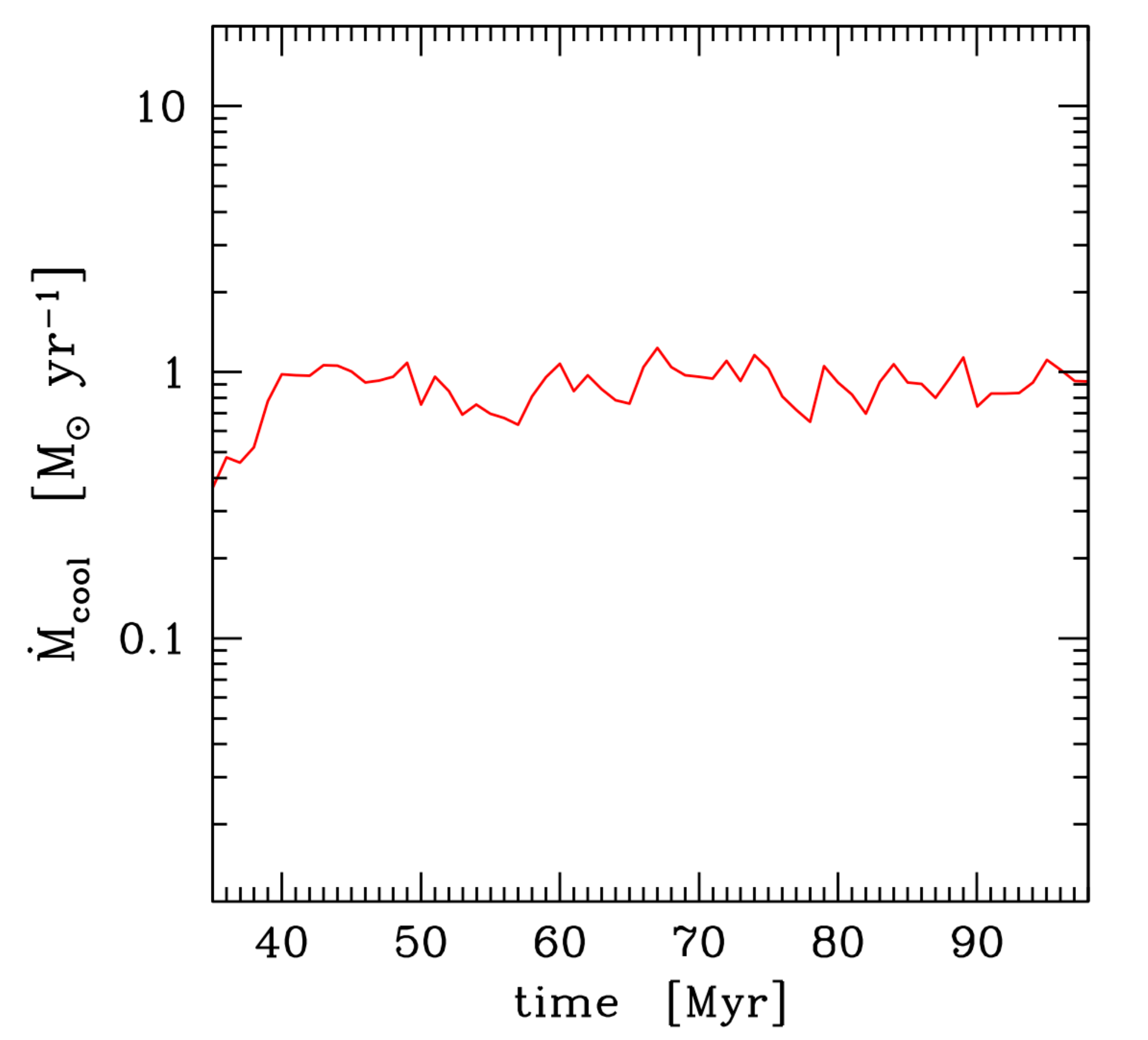}}
      \caption{Accretion with turbulence, cooling, AGN heating, and rotation: (top) evolution of the accretion rate as a fraction of the cooling rate (100\,kyr step); (bottom) average net cooling rate (1\,Myr step) -- 1 dex lower than the pure cooling flow.
       In CCA, the chaotic motions and inelastic collisions between the multiphase filaments recurrently boost the BHAR up to the cooling rate, which is $\approx100\times$ the Bondi rate computed at $r\approx1$\,kpc.
      }
      \label{f:cca_e03_mdot}
\end{figure}  

\begin{figure}
      \centering
      \subfigure{\includegraphics[scale=0.3]{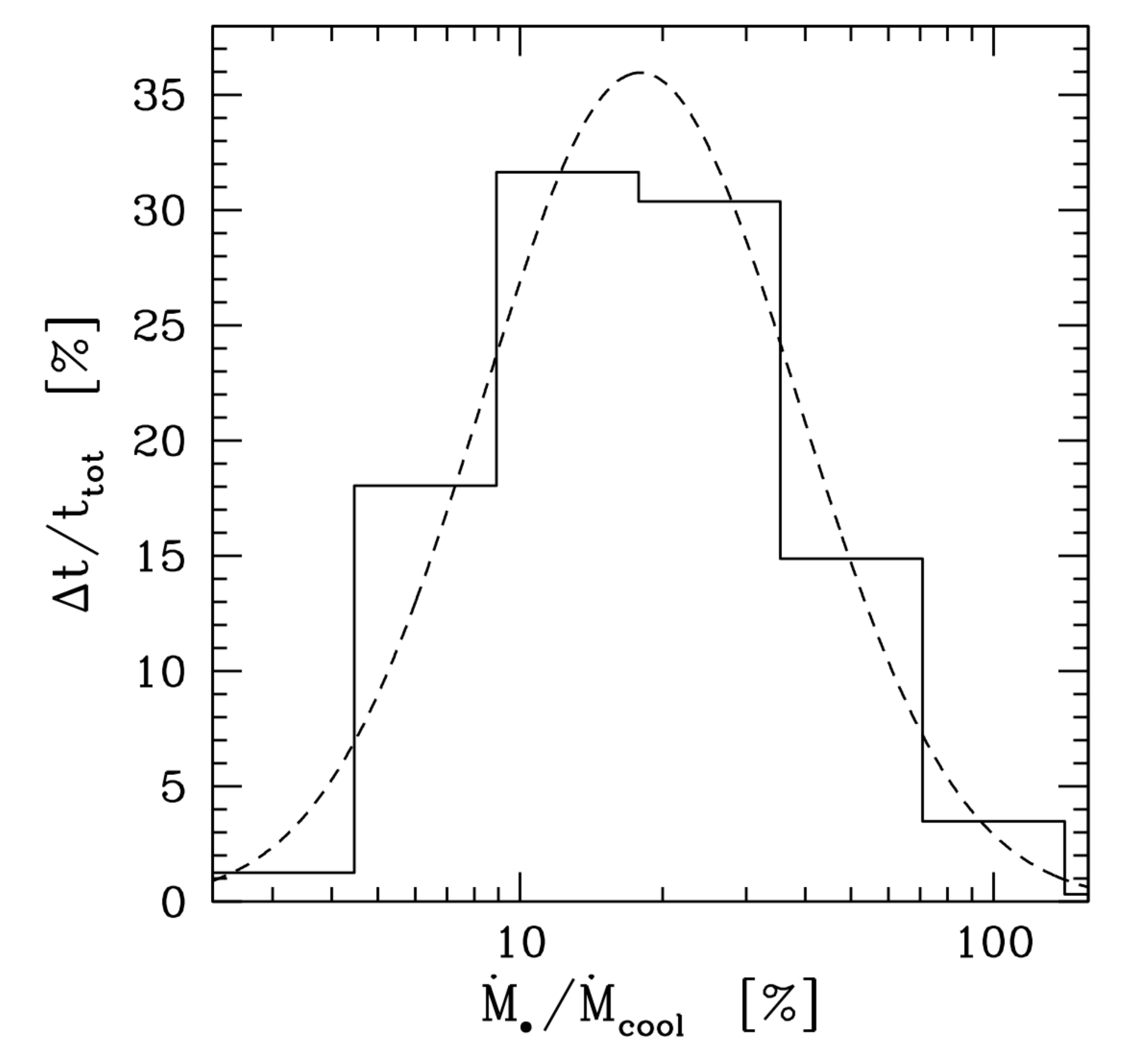}}
      \subfigure{\includegraphics[scale=0.3]{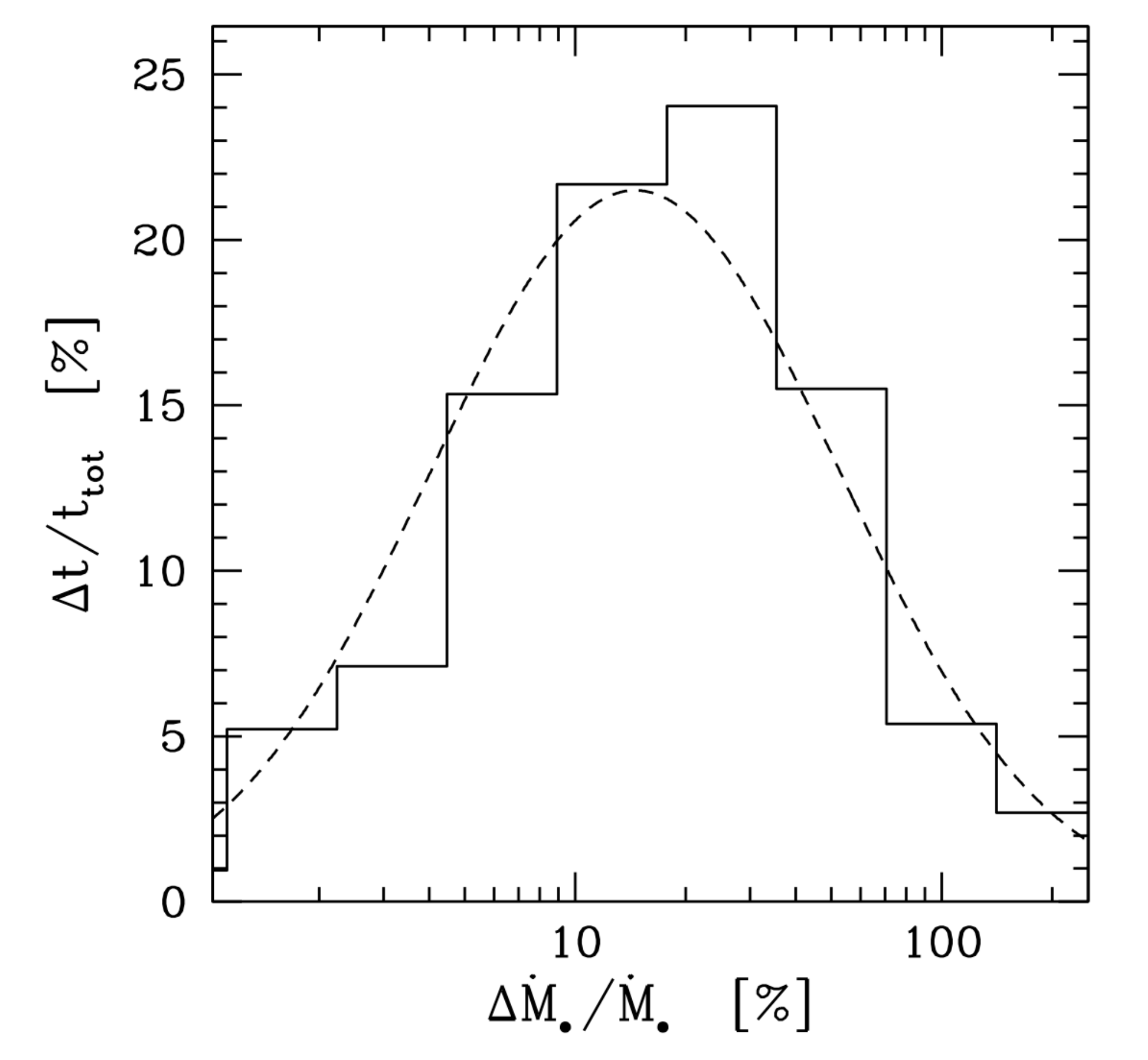}}
      \caption{Accretion with turbulence, cooling, AGN heating, and rotation: (top) temporal distribution of the normalized accretion rate; (bottom) temporal distribution of the relative BHAR rate of change over 100 kyr.
      Both are computed in logarithmic bins ($=$\,0.3) and 
      tend toward a lognormal distribution
      (a gaussian with variance and normalization from the data is overplotted -- not identical to a best-fit model due to the mild non-zero skewness and kurtosis).
      A lognormal distribution is the statistics related to thermodynamic fluctuations imprinted by turbulence.
      }
      \label{f:cca_e03_mdot_PDF}
\end{figure}

\vspace{-0.41cm}
\subsection[]{Accretion rate and distribution}  \label{s:cca_acc}
\noindent
As shown in Fig.~\ref{f:cca_e03_mdot}, the accretion flow evolution is very different from the multiphase disc or spherical cloud stage. In the turbulent atmosphere, the accretion rate chaotically oscillates between 5-100 percent of the cooling rate value. Because of the AGN heating balancing the plasma radiative emissivity, the net cooling rate reaches a quasi-stable configuration at $\dot M_{\rm cool}\simeq\,1\;\msun$\,yr$^{-1}$, which is more than 1\,dex lower than that of the pure cooling flow model (\S\ref{s:cool_e03_dyn}), consistently with observed X-ray spectra (e.g., \citealt{Peterson:2006,Gaspari:2015_xspec} and refs.~therein). 

At variance with G13/G15, we start here from a fully developed turbulence cascade ($t>t_{\rm eddy}$) and we are able to follow the long-term evolution up to $\sim$\,100 Myr, albeit numerically highly expensive due to the tiny timestep. The chaotic nature of the accretion rate persists throughout the evolution, with recurrent boosts up to $\dot M_\bullet \approx \dot M_{\rm cool}\simeq100\,\dot M_{\rm B}$, where the latter is the Bondi rate calculated at $r\simeq1$\,kpc. The accretion rate valleys tend toward the suppressed multiphase disc regime (\S\ref{s:cool_e03_dyn}), while the peaks briefly touch the multiphase cloud limit (\S\ref{s:cool_e00_dyn}). Thereby, CCA dynamically represents the intermediate regime, in analogy to the weather condition of rain. We note the cooling rate value is sometimes surpassed as cold clouds can accumulate in the core and experience a delay in sinking.

Compared with previous G13/G15 2-phase models, the overall CCA mechanism is preserved. On the other hand, the scatter in the accretion rate has increased by 0.5 dex, with deeper valleys and more impulsive peaks. This is tied to the more realistic 3-phase halo, as described in-depth below.
From Fig.~\ref{f:cca_e03_mdot} (top), it is clear that the flow has a complex fractal structure, i.e., based on `peaks within peaks'.
By visual inspection, the largest peaks approaching the cooling rate occur every 15-20 Myr. The $\dot M_\bullet/\dot M_{\rm cool}$ zone around 50 percent is touched every 5-10\,Myr, while the 20 percent range appears to be a common, Myr regime.
More quantitatively, Fig.~\ref{f:cca_e03_mdot_PDF} (top) shows the temporal distribution of the normalized accretion rate in logarithmic bins, as a fraction of total evolution time. 
The mean and standard deviation of the (percent) distribution are $\log(\dot M_\bullet/\dot M_{\rm cool})+2=1.25\pm0.33$, i.e., the average normalized rate is 18 percent. This is $\sim\,$$2\times$ lower than G15 model. 
Compared with the Eddington rate, the average BHAR is $2.7\times10^{-3}\,\dot M_{\rm Edd}$.
Notice that while CCA is on average sub-Eddington at low redshift, at high redshift -- when the BH mass is orders of magnitude smaller -- the BHAR can easily reach Eddington values. At low redshift, we expect only $\lta1$ percent of nuclear sources to become X-ray bright.

The CCA accretion rates follow a lognormal distribution (dashed line) with mild deviations. The skewness and kurtosis is 0.24 and $-0.34$, respectively.
Remarkably, a lognormal distribution with mild non-Gaussianities is also the statistics which describes density perturbations $\delta\rho/\rho$ driven by subsonic turbulence (\citealt{Gaspari:2013_coma,Porter:2015}) and which is observed in the ICM by X-ray (\citealt{Kawahara:2008}) and SZ studies (\citealt{Khatri:2016}). This tells us that turbulence directly puts its imprint on the condensing filaments and clouds (stronger subsonic turbulence leads to larger $\delta\rho/\rho$ variance; \citealt{Gaspari:2014_coma2}) which later become fuel for the accretion rate. 
We note that the system will cycle through cold and hot accretion mode over the long term; the latter mode will populate the left tail of the distribution ($\dot M/\dot M_{\rm cool}< 0.1$; cf.~G15) over several Gyr. The {\it total} BHAR distribution over a large AGN sample may thus tend to a power law with a high-end steepening.

\begin{figure}
      \centering
      \subfigure{\includegraphics[scale=0.23]{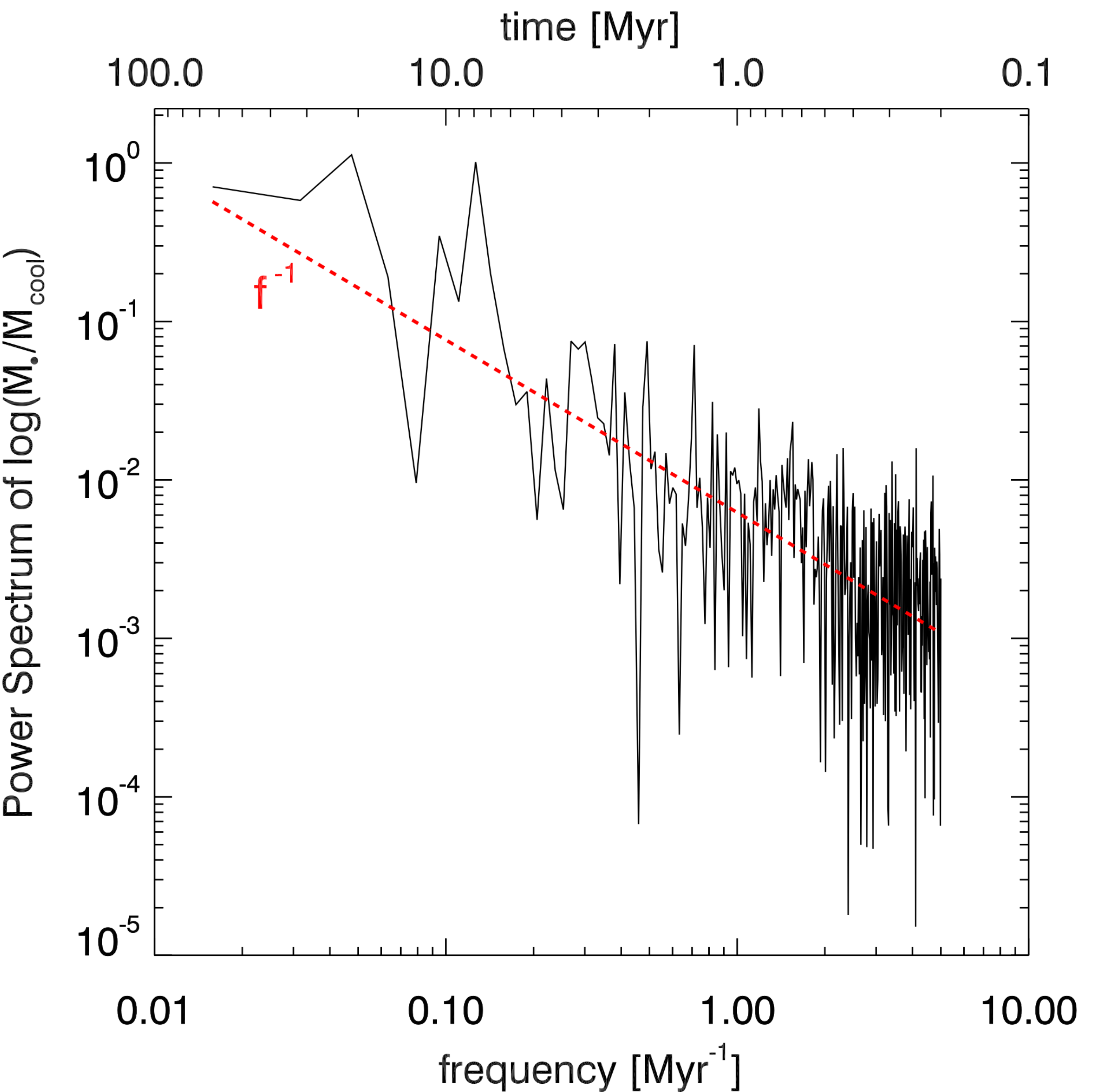}}
      \caption{Accretion with turbulence, cooling, AGN heating, and rotation: power spectral density (dex$^2$/Myr$^{-1}$) of the logarithmic normalized accretion rate as a function of frequency $f =1/t$. The red line is the linear regression fit to the log power spectrum, with slope $\simeq-1.09$. The $1/f$ fluctuations (pink noise) is ubiquitous in nature and is the signature of nonlinear, chaotic fractal processes. CCA is self-similar and able to induce rapid variability on various scales, as seen in most AGN light curves.
      }
      \label{f:cca_e03_PS}
\end{figure}  

About 15-20 percent of time CCA resides in the very large or low accretion state (tails of the distribution). 
It is ill-posed to ask what is the global `duty cycle' of AGN feedback: the actual duty cycle is a function of what threshold is used to define the `active' state. Choosing a high/medium/low state ($\dot M_\bullet/\dot M_{\rm cool}>0.5,\,0.2,\,0.05$), the cumulative duty cycle is 18, 48, 99 percent, respectively.
In other words, AGN feedback is always active to some level showing minor bubbles and turbulent perturbations -- as found by observations (\citealt{Birzan:2012}) -- while major outbursts such as 30-100\,kpc bubbles occur sporadically (\citealt{Hlavacek-Larrondo:2015}) albeit easier to detect in X-ray brightness. In analogy with weather forecast, Fig.~\ref{f:cca_e03_mdot_PDF} may be interpreted as a probability of precipitation (PoP) onto the BH and hosting galaxy: heavy rain (20\%), light rain (60\%), and `sunny' (20\%). As further support of the weather similarity, the lognormal distribution well describes the rainfall rates on earth as measured by tipping bucket instruments (\citealt{Segal:1980}).

In Fig.~\ref{f:cca_e03_mdot_PDF} (bottom), we introduce another important diagnostic tool which helps us to understand the characteristic of the chaotic flow: the temporal distribution of the relative BHAR variations (over the short 100 kyr scale). This can be interpreted as a mass accretion {\it acceleration} (or deceleration). As above, the relative variations tend toward a lognormal distribution with mild non-Gaussianities (skewness $-0.63$ and kurtosis 0.86). The (percent) logspace mean is 1.16 with 1 standard deviation of 0.55 dex. 
The similarity between the distributions of the BHAR and BHAR rate of change corroborates the fractal structure of CCA described by self-similar peaks within peaks. 
Since the logspace standard deviation provides a rough estimate for the relative variation, 
$\sigma_{\log \dot M_\bullet}\sim\delta \dot M_\bullet/\dot M_\bullet$, 
a similar $\sigma_{\log \dot M_\bullet}$ implies that even the variation of the BHAR variation will be of the same order.
Such recurring self-similarity has its roots in the main driver, turbulence, which is a fractal structure composed of rapidly stretched, randomly folded vortex tubes with typical dimension $D\approx2.6$ (\citealt{Procaccia:1984}).

Although common BHAR oscillations are of the order of 10-30 percent (about half of time), the right tail of the BHAR rate of change distribution can reach short-term amplitudes $\gta$\,200 percent ($\gta2$ standard deviations). Considering the cumulative effect over a few Myr, the minimum to maximum BHAR variation can reach a factor 30 (Fig.~\ref{f:cca_e03_mdot}, top).
CCA can induce rapid variability up to a factor of several in the accretion rate and thus 
efficiently powering self-regulated AGN outflows or jets, at the same time rapidly varying
the AGN luminosity and related light curves
(see \S\ref{s:time_domain}).
For instance, the generated AGN outflow power would be $P_{\rm out}=\varepsilon\, \dot M_\bullet c^2\simeq 10^{43}-10^{44}\,$erg\,s$^{-1}$, for the low and high state respectively, which is comparable to the bolometric X-ray luminosity of NGC 5044.
As shown in previous work (e.g., \citealt{Gaspari:2012a,Gaspari:2012b}), CCA is a very efficient self-regulation mechanism preserving hot halos in quasi thermal equilibrium for several Gyr.
Observationally, the strong link between ICM properties (e.g., high luminosity or central density) and the AGN jet power (traced via synchrotron radio emission) has been confirmed in several radio surveys (e.g., \citealt{Ineson:2015} and refs.~therein). 

\vspace{-0.41cm}
 \subsubsection{Frequency power spectrum} \label{s:cca_PS}
A further important question is what is the power in the frequency domain.
Fig.~\ref{f:cca_e03_PS} shows the power spectral density of the BHAR (or simply power spectrum, $P_f$) as a function of frequency ($f=1/t$). The power spectrum is retrieved via the discrete Fourier transform of the time series in $\log \dot M_\bullet/\dot M_{\rm cool}$ ($P_f$ of linear BHAR is similar).
The first key characteristic is that the median power spectrum is self-similar, i.e., it follows a power-law $P_f\propto f^{-1}$. 
Such kind of self-similarity is often referred to as {\it pink} noise. 
Remarkably, $1/f$ power spectra appear to be ubiquitous in nature, including not only physics but also technology, biology, economics, psychology, language, and music (\citealt{Ward:2007} for a review). Notable examples are semiconductor fluctuations, meteorological data series, heart beat rhythms, brain neural activity, and stock market variations. 
Interesting astrophysical examples of 1/f noise are the intensity variability of solar sunspots (\citealt{Polygiannakis:2003}) and quasar variability (more below).
Closer to our study, hydrodynamic and magnetohydrodynamic turbulence follows $1/f$ fluctuations in time (e.g., \citealt{Mininni:2014,Herault:2015}). As turbulence is the main process in CCA, this is directly reflected in the accretion rates tied to the condensing chaotic clouds (\S\ref{s:cca_dyn}).
Although there is no trivial mathematical explanation, pink noise arises from a superposition of multiple relaxation processes and is often the signature of chaotic, nonlinear fractal phenomena (\citealt{Milotti:2002}), explaining the {\it peak within peak} structure that we find in CCA. 

A self-similar negative spectral slope implies that the largest power {\it density} resides in the lowest frequencies. This is analogous to the power spectrum as a function of lengthscale $l$ (or mode $k=1/l$) of the turbulent velocity -- often called energy spectrum -- $P_k\propto k^{-5/3}$, where $k^2=k^2_x+k^2_y+k^2_z$. The $P_k$ maximum occurs at the turbulence injection scale which is typically driven at large lengthscales. The $1/f$ frequency power spectrum conveys that waiting longer time we can expect larger peak variability, exceeding even 1 dex.
Single, smaller frequencies have progressively smaller amplitude; however, the {\it sum} power (or variance) in 1 dex logarithmic bin remains roughly constant ($f\,P_f$).
This strengthens the self-similarity of CCA, i.e., the {\it mean} relative BHAR variations ($\delta\dot M_\bullet/\dot M_\bullet\sim\sigma_{\log \dot M_\bullet}$) will be of order 10-20 percent over each 1 dex range, regardless of scale.

A final point to highlight in the unfiltered power spectrum (Fig.~\ref{f:cca_e03_PS}) is related to the characteristic frequencies at which CCA is more likely to cycle. Over large timescales, typical cycles are 20, 8, and 4 Myr. Below 1 Myr, the BHAR does not display a preferred frequency. Small BHAR variations are very frequent because of the common interactions between the warm/cold filaments and clouds. To produce large BHAR variations, an increasing number of collisions is required to reduce angular momentum. Such process takes longer time to develop, up to 20 Myr for the largest peaks\footnote{A longer, 1~Gyr evolution may produce larger BHAR peaks with low frequency; such outlier events up to a few dex may be the cause of the strongest quasar or blazar events observed.}.

\vspace{-0.41cm}
\subsubsection{Time-domain observations: AGN, quasars, HMXB} \label{s:time_domain}
The distribution and power spectrum analysis shows that CCA can produce rapid, strong, and recurrent variability in the accretion rate (and thus luminosity or kinetic power), which are similar to fractal structures found in Earth weather (e.g., \citealt{Blender:2011}) and numerous natural systems. From an astrophysical point of view, we propose that CCA can solve the problem of the rapid variability of AGN and other compact objects (e.g., high-mass X-ray binaries).

Besides boosting the kinetic power and thus efficiently creating a tight self-regulated AGN feedback loop (\citealt{Gaspari:2016_IAU} for a review), rapid variability implies variations in luminosity. 
After several decades of investigations from X-ray to radio band, it is widely accepted that most of AGN exhibits some degree of rapid variability (\citealt{Ulrich:1997,Peterson:2001_AGN_var} for a few reviews), including those of BCGs (\citealt{Russell:2013}).
Such variations in the observed light curves can span a century, years, hours, or even minutes. Typical variations are of the order of 10-30\%, up to extreme cases of several $100$\% (the latter called `changing-look quasars'; \citealt{LaMassa:2015}). 
Variability can be constrained directly in the continuum or from the lines, e.g., through reverberation mapping. Interestingly, the latter is tied to the so-called  broad-line-region (BLR) clouds, which are a natural outcome of CCA.
It is remarkable that, as in the CCA mechanism, AGN variability has chaotic behavior and amplitude which is inversely correlated with timescale. As predicted by CCA, the observed distribution of AGN luminosity variability is lognormal (\citealt{Gaskell:2003}). 

One of the most famous quasars, 3C273, has been monitored for over 80 years. Figure 1 in \citet{Press:1978} shows 3C273 light curve with chaotic variations of 5 to 60 per cent over years and decades, respectively, following a pink noise spectrum.
In the same paper, Press highlights the ubiquity of pink (or `flicker') noise. 
Other exemplary AGN are NGC 4151 (30\% up to 300\% variations over a month), NGC 5548 (20\% to 80\% variations over a year), Ark 564 (15\% up to 100\% variability over a month), and Mrk 590 (13\% to 46\% variations over 25 years).
Regarding BCGs, the AGN flux in A2052, Hydra A, and M84 has been observed to vary by 50\%\,-\,200\% over a decade.
Observations in the time domain are unfortunately limited by a time window which can usually span years or days.
Nevertheless, since relative $\delta \dot M_\bullet/\dot M_\bullet$ remains roughly the same over each temporal decade, the same relative variability structure of $\sim\,$20\% with peaks up to $\sim\,$100\% is repeated over and over. The true is valid in the opposite direction, i.e., the AGN can dim by over 1 dex, as the cold clouds are consumed, thus explaining the recently discovered changing-look quasars (\citealt{Runnoe:2016}).
In conclusion, the rapid AGN variability can be simply understood in terms of CCA variability, without invoking complex obscuration or microlensing models, or unrealistic modifications of smooth accretion models as Bondi or $\alpha$-discs.

Similarly to AGN, the rapid variability in high-mass X-ray binaries (HMXBs) is widely debated and can not be reproduced with smooth Bondi-Hoyle models. HMXBs can be considered a scaled-down version of AGN, reaching luminosities up to $L_{\rm x}\sim10^{38}$ erg s$^{-1}$.
In a HMXB, the OB star does not typically fill its Roche lobe and accretion can not occur via a disc, as in low-mass binaries,
but it is expected to occur in a quasi-spherical way from the stellar wind (\citealt{Shakura:2014}). 
Similar to galactic hot halos, the hot wind is expected to be clumpy via radiative cooling.
The large, fast, and self-similar variability driven by CCA may help to explain the HMXB fluctuations.
The classical HMXB example is Vela X-1, which shows light curves varying by 30\% up to several 100\%, on timescales of a few hours and days, respectively (\citealt{Martinez:2014}), which is a typical variability magnitude induced by the CCA mechanism.
Extreme cases of variability are 
Supergiant Fast X-ray Transients, varying by several dex, possibly because of cold gas depletion, similarly to the changing-look quasars.
We investigate in depth such scenario in a separate work.

\vspace{-0.41cm}
\subsubsection{Flow with no rotation or no heating}\label{s:cca_norot_noheat}
We ran two additional simulations: the no-rotation and no-heating models with the new 3-phase cooling module.
We confirm the G13/G15 results, so we briefly describe them. The accretion rate normalized to the cooling rate (not shown) is similar to that in Fig.~\ref{f:cca_e03_mdot}. In the no-heating run, the cooling rate is 1 dex higher and the warm/cold filaments are comparably more massive.
We remark subsonic turbulence is able to naturally induce nonlinear perturbations in the plasma, which do not require a heated atmosphere to grow (cf.~G13-G15).
Thereby, AGN heating is mainly an agent to preserve the global atmosphere from pressure collapse. 
In a non-heated halo the cold phase is more prone to sediment into a partial rotating structure with slightly lower average $\dot M_\bullet/\dot M_{\rm cool}$.

In the other run with full physics but no rotation, the CCA evolution remains unaltered, as long as  ${\rm Ta_t} \equiv v_{\rm rot}/\sigma_v \lta 1$ (cf.~G15). This is because turbulence continuously generate chaotic vorticity which overcomes coherent rotation, thereby chaotic collisions between filaments and clouds (with opposing angular momentum) are frequent.
As turbulence weakens and ${\rm Ta_t} > 1$, the rotating disc becomes the dominant structure and the accretion rate is progressively reduced; the dynamics follows that described in \S\ref{s:cool_e03}. 
Intermediate cases, with a major central disc and a partial CCA rain, can appear in nature, in particular in S0 galaxies, as seen in PKS B1718-649 (\citealt{Maccagni:2016}).

\vspace{-0.41cm}
\subsection[]{CCA dynamics}  \label{s:cca_dyn} 

\begin{figure}
     \hspace{-0.56cm}
     \subfigure{\includegraphics[scale=0.51]{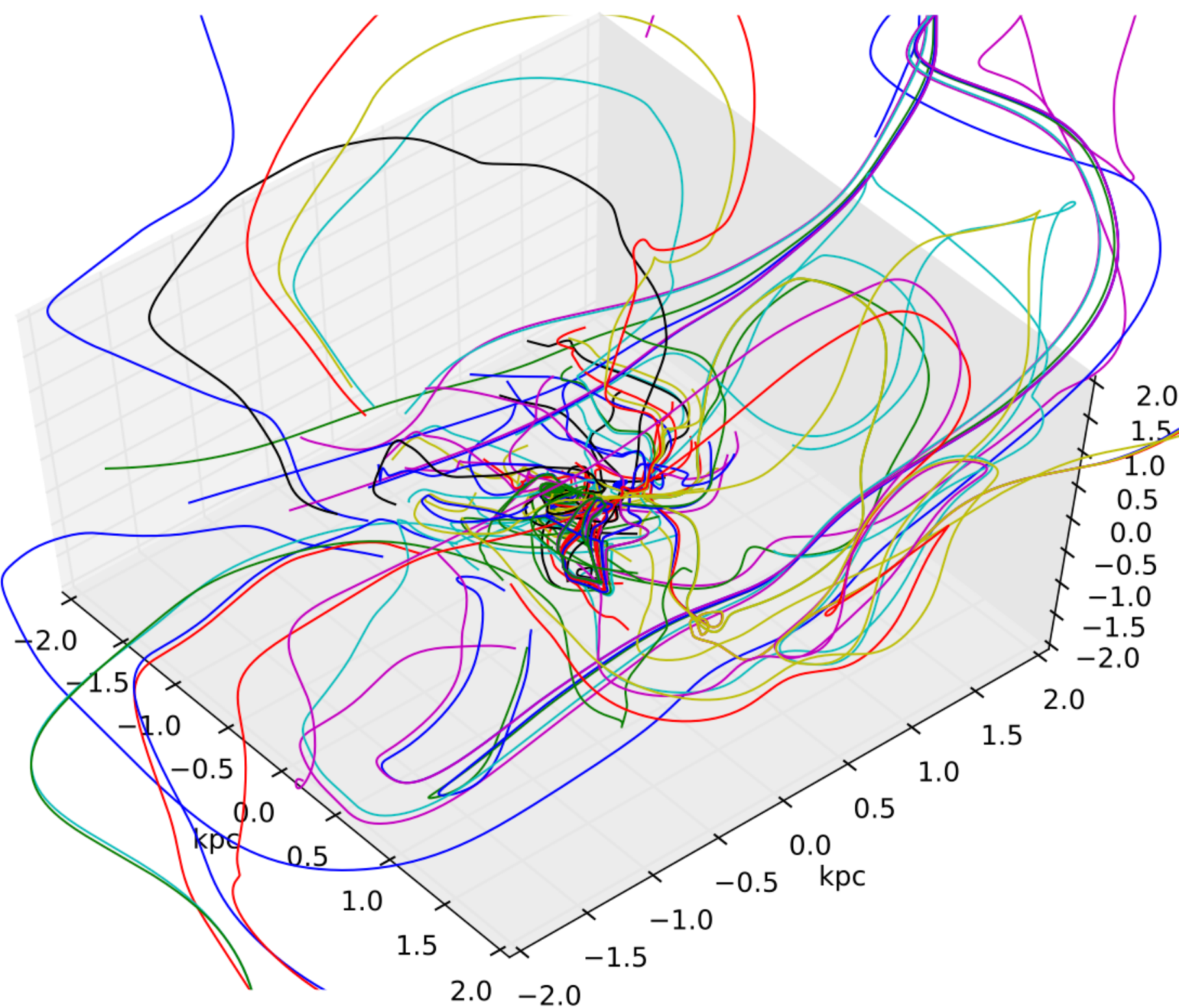}}
     \caption{Accretion with turbulence, cooling, AGN heating, and rotation:~a sample of streamlines integrated from the outer kpc region (at final time).
     The plot shows the characteristic chaotic motions defining CCA, which lead to recurrent multiphase gas elements interactions increasing toward smaller radii. } 
         \label{f:cca_stream}
\end{figure} 

\begin{figure}
     \centering
     \subfigure{\includegraphics[scale=0.47]{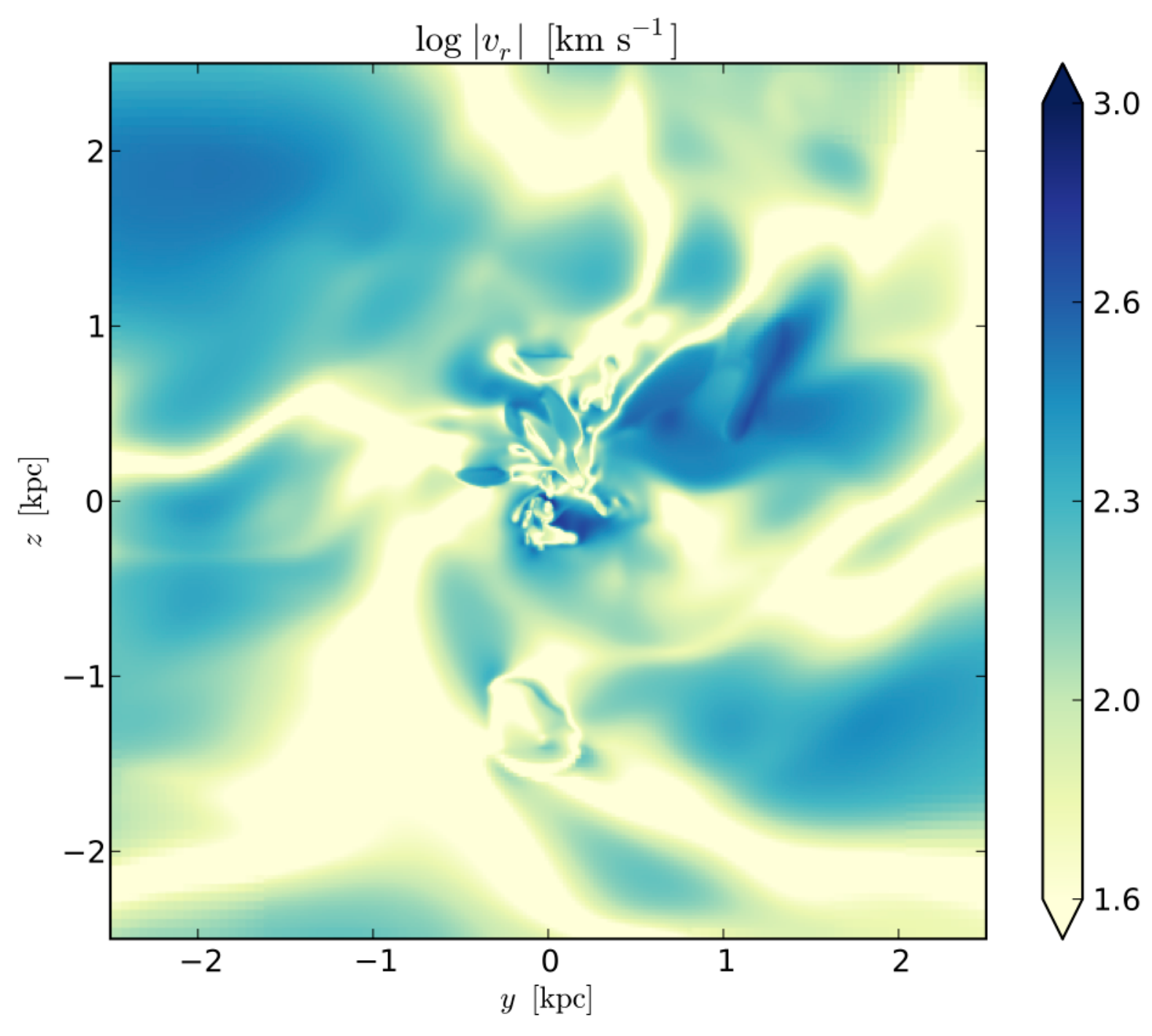}}
     \caption{Accretion with turbulence, cooling, AGN heating, and rotation: absolute value of the 
     radial velocity in the mid-plane cross-section through the $x-$axis 
     (5x5 kpc$^2$; final time).
     The turbulent eddies are evident. 
     The cold and warm phases retain the imprint of the hot halo turbulence and kinematics.
     In the inner 500 pc, the warm and cold filaments recurrently collide in inelastic and nonlinear way, leading to angular momentum cancellation, thus boosting the subsequent accretion rate. }
         \label{f:cca_vrad}
\end{figure} 

\noindent
We analyze now the CCA dynamics, in terms of the velocity field, 
the role of collisions between the multiphase clouds, and the dominant mechanism of angular momentum transport.
Fig.~\ref{f:cca_stream} shows a sample of the characteristic streamlines defining CCA. It is clear that the dynamics is very different from that of the multiphase disc (\S\ref{s:cool_e03_dyn}) or the monolithic cloud (\S\ref{s:cool_e00_dyn}). The flow does not display a coherent rotation nor a coherent radial inflow, rather it is comprised of chaotic streamlines.
The turbulence injected at $L\gta4\,$kpc generates large eddies, which cascade into smaller self-similar vortices 
following $\sigma_{v,\rm int}\approx\sigma_v\,(l'/L)^{1/3}$.
Fig.~\ref{f:cca_stream} depicts part of such kpc-scale chaotic eddies and turbulence inertial cascade.
Because of the strong inner gravitational attraction and related large escape velocity, 
the central streamlines experience a higher degree of entanglement.
This implies that the finite multiphase elements (which are not massless nor point-like) experience increasing collisions while raining toward the central SMBH. Such high-interaction region is evident within $r<500$\,pc, which is the common raining locus for the warm and cold filaments and clouds (\S\ref{s:cca_dyn}).
The recurrent inelastic interactions continuously mix and partially cancel angular momentum, leading to the {\it peak within peak}, time-varying accretion described in \S\ref{s:cca_acc}.
In the inner 100 pc a clumpy toroidal structure can occasionally form -- residual of the initial rotation -- decreasing the accretion rate. Such `torus' is not permanent during the CCA rain, as it is continually bombarded, warped, and dismantled by the infalling clouds and filaments, reigniting the accretion rate. 
AGN obscuration thus varies rapidly, in anti-correlated manner with the accretion rate or nuclear X-ray luminosity. This is indeed the case in the XMM COSMOS survey of 1310 AGN (Fig.~6 in \citealt{Merloni:2014}).
CCA can self-consistently explain the presence of the so-called fast BLR and slow NLR clouds (broad/narrow-line region) in the sub-pc and 10\,-\,100 pc scale, respectively. The clumpy nature of CCA thus corroborates the new advancements in the standard AGN Unification model, which require a non-homogeneous medium to explain the plethora of obscuring observational properties of AGN (\citealt{Bianchi:2012,Netzer:2015}).

Fig.~\ref{f:cca_vrad} shows the cross-section of the radial velocity field within $r<2.5$\, kpc.
Typical inflow velocities are 100\,-\,300 km\,s$^{-1}$.
The variations within 500 pc are related to the condensed clouds which continuously collide in inelastic way\footnote{Two colliding elements coalesce in one entity without preserving kinetic energy
and often decreasing the velocity due to the mass increase; in 1D, $v_{\rm fin}=(m_1 v_1 + m_2 v_2)/(m_1+m_2).$}.
The interactions in the 3 different phases are dynamically linked, as the cold clouds are embedded in the warm filaments which are embedded in the hot eddies. 
Fig.~\ref{f:cca_vrad} shows the condensing cold and warm phase retain the imprint of the turbulent kinematics of the progenitor hot halo. 
The dynamics qualitatively works as follows.
Turbulence broadens the angular momentum distribution of the plasma (Fig.~\ref{f:cca_e03_vphi}, more below).
Condensation occurs in the plasma density peaks driven by turbulence, carrying on the hot gas kinematics.
Several clouds are born with low angular momentum (near the median of the distribution). 
Those born from the tail will have either large positive or negative angular momentum and require multiple collisions to be accreted. This is why in the $\rm {Ta_t}>1$ case, turbulent diffusion is so weak that the clouds only condense with coherent rotation (the angular momentum distribution is only positive), inducing a lower average BHAR as a function of ${\rm Ta_t}$ (cf.~G15).

\begin{figure}
      \centering
      \subfigure{\includegraphics[scale=0.29]{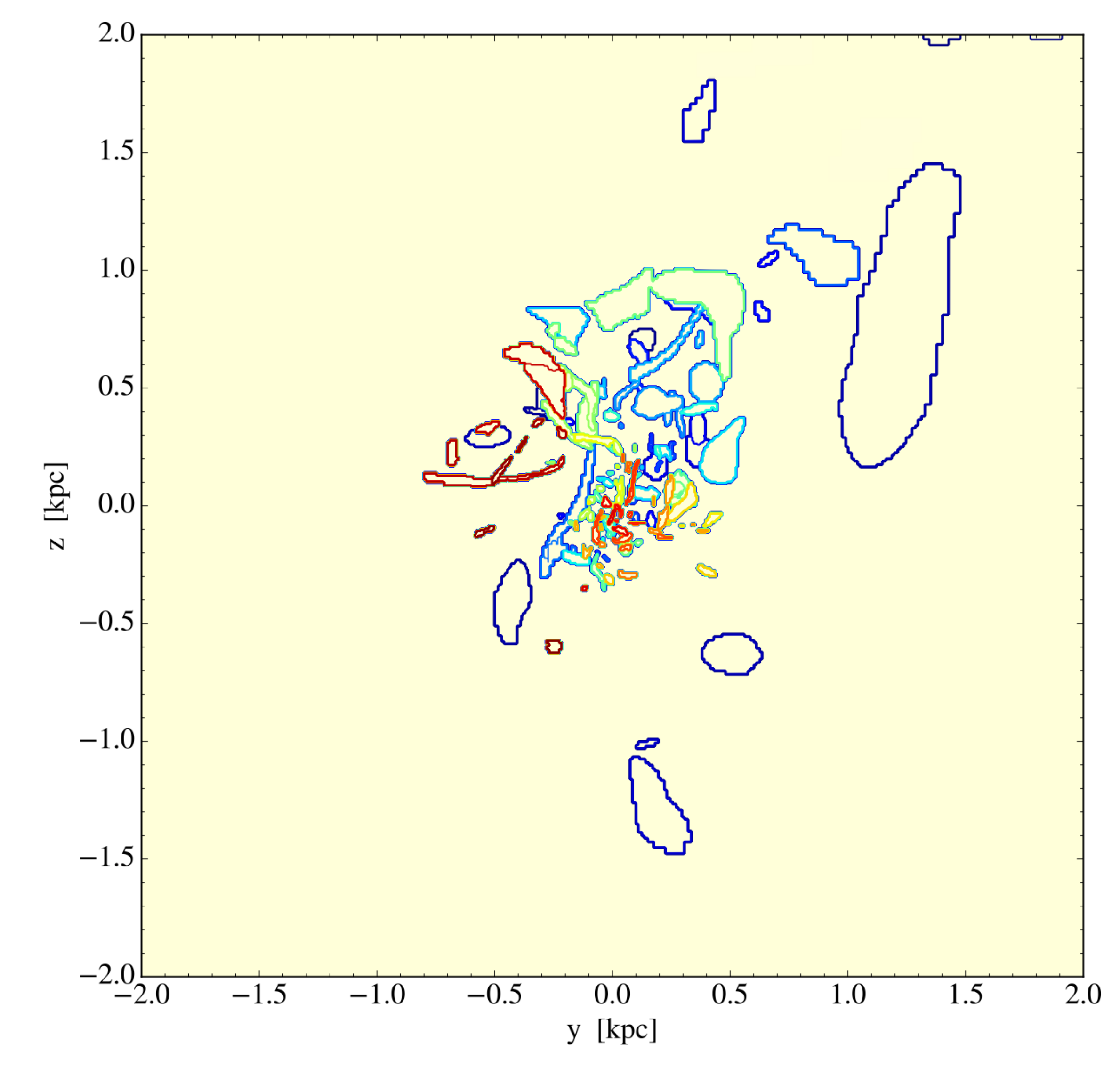}}
      \caption{Accretion with turbulence, cooling, AGN heating, and rotation: projected spatial distribution of the leaf clouds selected via the clump finder algorithm within 2 kpc region at final time.
      Clump color-coding is randomly chosen. The network of condensed substructures is characteristic of the developed CCA rain, which tends to significantly fill the core and thus recurrently obscure the AGN, as invoked by recent AGN Unification models. 
      }
      \label{f:cca_clumps_map}
\end{figure}  

\vspace{-0.41cm}
\subsubsection{Collisional viscosity and torques}\label{s:cca_visc}
To get insights into the detailed CCA collisional mechanism, we study the cloudlet properties by using a clump finder algorithm as commonly employed in cosmological studies (cf.~\citealt{Smith:2009}). The clump finder identifies topologically disconnected substructures by recursively shrinking density contouring with 0.3 dex step. The final result is a family tree comprised of parent and children clumps.
Fig.~\ref{f:cca_clumps_map} shows the network of clouds selected via the clump finder algorithm at the bottom of the hierarchy.
The projected map captures the fully developed CCA rain with cloudlets of varying size, which are continuously shaped by the recurrent collisions.
The leaf clumps typically combine in extended filaments/giant clouds and are further enclosed in larger parent structures (up to several kpc; not shown), creating a complex multiphase web of elements which strongly interact within 100s pc distance from the SMBH
(the NLR in AGN observations).
Notice that the CCA rain can be fairly anisotropic, i.e., one region may show enhanced cloud precipitation (e.g., here the NE sector), while another may be poor in cold gas. This variance is induced by the local turbulent motions. 
It is thus best to use sector analysis in observational data, instead of averaging over large azimuthal annuli.
The anisotropic, clumpy distribution may help to explain the large zoology of AGN classifications which arise primarily because of the line-of-sight inclination angle (\citealt{Netzer:2015} for a review).

\begin{figure}
      \centering
      \subfigure{\includegraphics[scale=0.34]{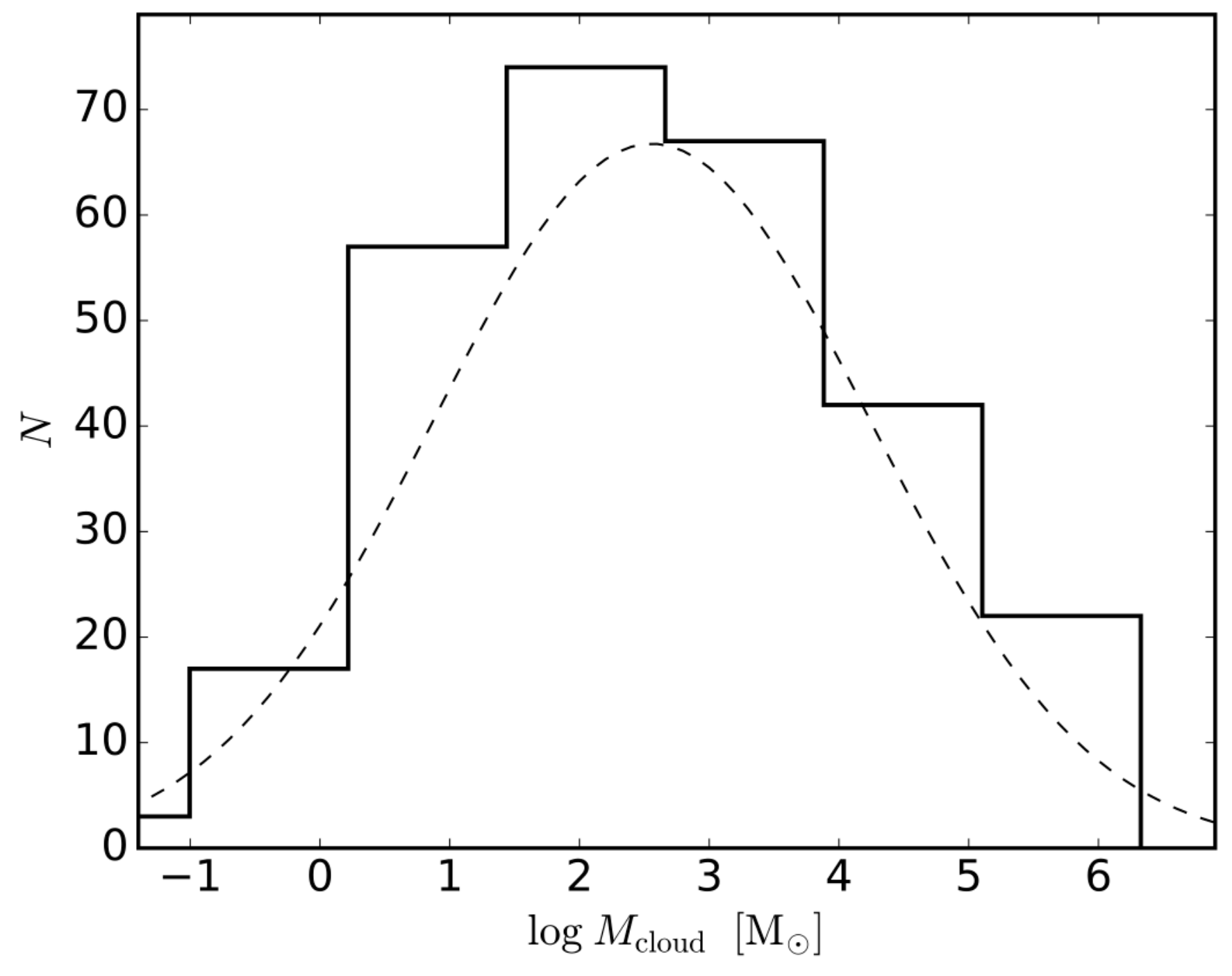}}
      \subfigure{\includegraphics[scale=0.34]{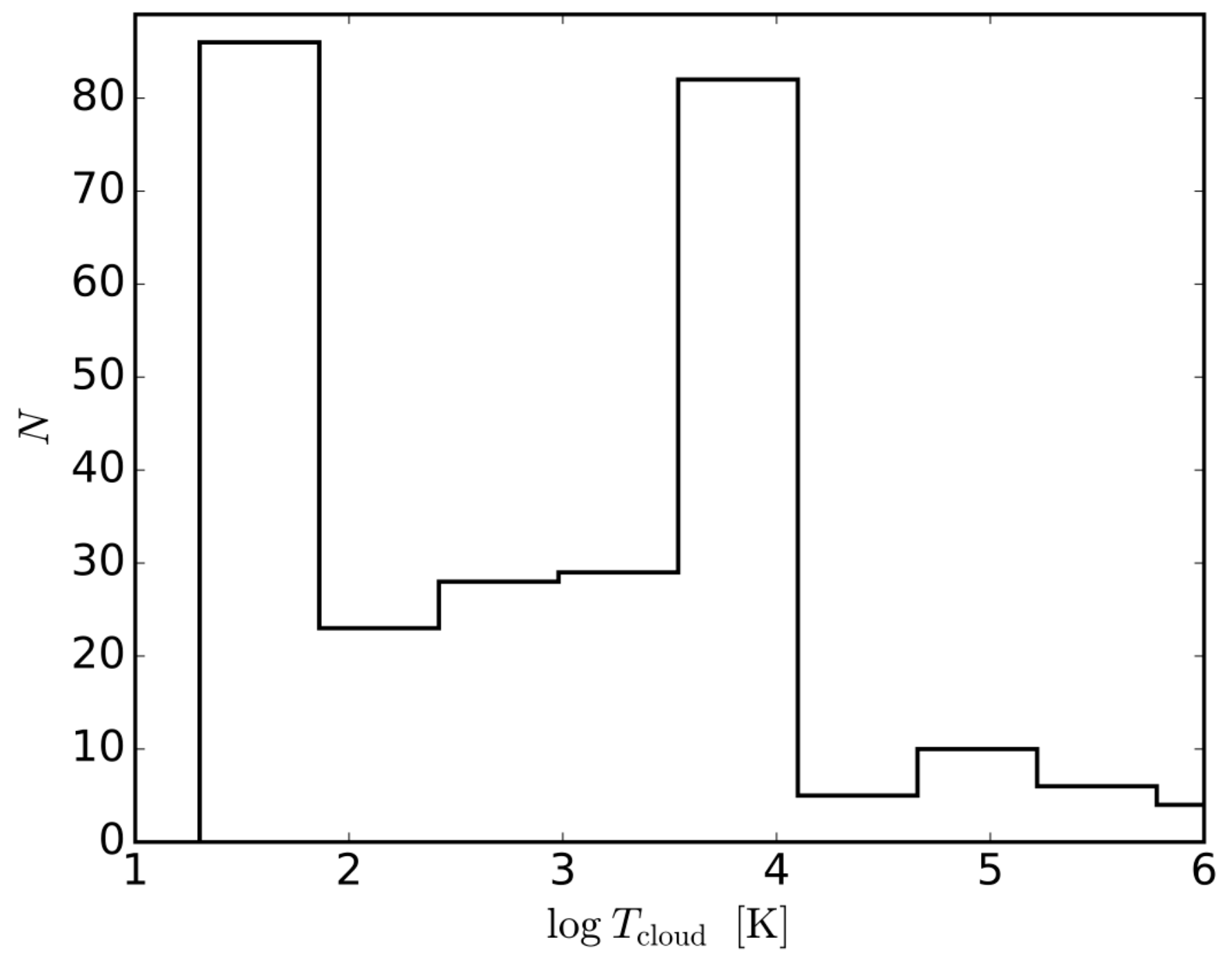}}
      \subfigure{\includegraphics[scale=0.34]{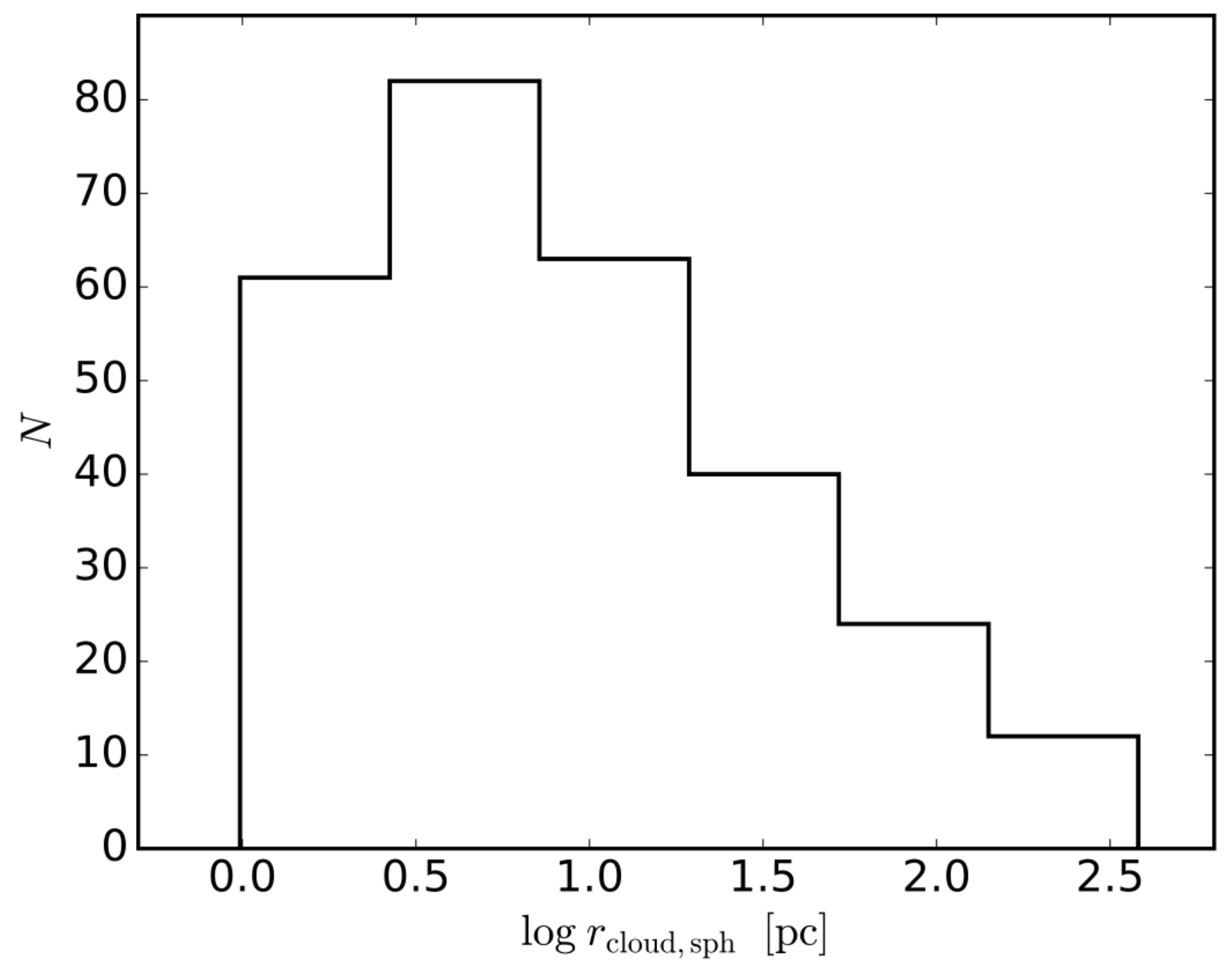}}
      \subfigure{\includegraphics[scale=0.34]{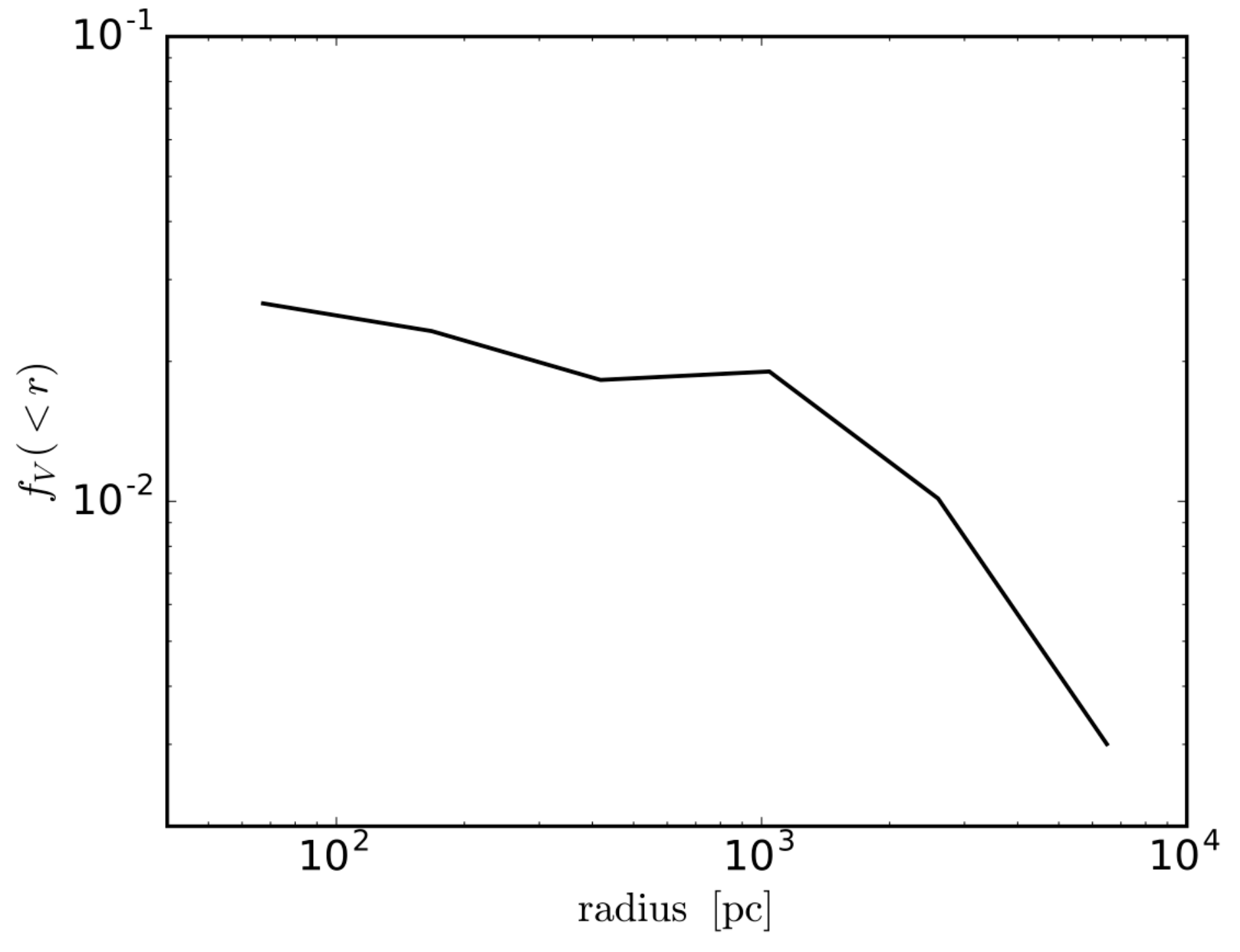}}
      \caption{Accretion with turbulence, cooling, AGN heating, and rotation: (top to bottom) distribution of the leaf clump mass, temperature, effective spherical radius, and volume filling within enclosed radius (final time).
      The mass distribution is quasi lognormal with the high-end tails dominated by molecular gas (a guideline gaussian with variance from the data is overplotted). The clumps have bimodal $T$ distribution, with effective sizes up to 100 pc and volume filling up to a few percent. 
      The concentration is highest within $r<500$\,pc, where it stays fairly unaltered, leading to efficient collisions.
      }
      \label{f:cca_clumps_PDF}
\end{figure}

Figure \ref{f:cca_clumps_PDF} shows the distribution of the leaf clump properties during the developed CCA rain, including mass, temperature, effective spherical radius, and the volume filling as a function of radial distance (top to bottom panels, respectively). 
The mass of the clumps spans a large range, from 10s $\msun$ to the $10^6\,\msun$ of the cold molecular clouds. 
The retrieved cloud mass distribution is quasi lognormal (cf.~\S\ref{s:time_domain}) corroborating CCA as a multiplicative process, i.e., the clouds build up via continuous stratification. 
For instance, the recent ALMA observation by \citet{Tremblay:2016} finds 3 giant molecular cloud with mass $10^5\,\msun$ accreting in the direction of the central SMBH in A2597 and embedded in a multiphase condensation rain. The mean and standard deviation of our clump mass distribution (characteristic of a central galaxy of a massive group) is $\log M_{\rm c}/\msun = 2.6\pm1.7$. 
This implies that Cycle 0/1 ALMA is currently able to to detect only giant molecular associations (e.g., \citealt{David:2014}).
Besides the limited beam and exposure, the small cold clouds are typically observed as giant molecular structures due to line-of-sight projection (\S\ref{s:cca_comp}).
The tiny, 550\,K G2 cloud accreting onto the Milky Way SMBH (\citealt{Gillessen:2013}) may be an exceptional example of resolved CCA droplet given Sgr A$^*$ proximity. 

The clump temperature distribution (Fig.~\ref{f:cca_clumps_PDF}, second panel) shows bimodality with larger number of molecular clouds at 50-100\,K then warm filaments near $10^4$\,K. 
The accreted gas clumps are typically 60\% cold gas and 40\% warm gas, with the accreted hot mass remaining below 1\%.  
The clouds between 100\,-\,1000\,K are highly unstable; the same can be said for the $10^5\,$K regime in which only a few clouds can persist before rapid condensation.  
It is useful to define an effective spherical radius $r_{\rm c}=[V_{\rm c}/(4\pi/3)]^{1/3}$, where $V_{\rm c}$ is the cloud volume.
The spherical radius of single clumps ranges between 5\,-\,100 pc (third panel), with mean and large dispersion $\log r_{\rm c}/{\rm pc}=0.9\pm0.6$.
Filaments and the clump network can be much more extended in one direction, reaching up to kpc length.
The bottom panel quantifies the volume filling of the non-overlapping leaf clumps enclosed within $r$. The volume filling reaches 3 percent value in the core and steadily declines beyond 1 kpc below 0.5 percent, as the CCA rain vanishes due to the more diffused hot phase. Similar large-scale decline from a few percent is observed in the volume filling of the molecular clouds detected by ALMA in A2597 BCG (\citealt{Tremblay:2016}).

In analogy with the kinetic theory of gas, the CCA cloud mean free path can be written as
\begin{equation}\label{e:mfp}
\lambda_{\rm c}\equiv\frac{1}{n_{\rm c}\,\pi(2\,r_{\rm c})^2}=\frac{1}{3}\frac{r_{\rm c}}{f_V}\simeq88^{+263}_{-66}\,{\rm pc}, 
\end{equation}
where $n_{\rm c}=f_V/[(4/3)\pi r_{\rm c}^3]$ is the cloud number density; the last step uses the retrieved clump properties in the core region (Fig.~\ref{f:cca_clumps_PDF}).
Beyond 1\,kpc, the decline in the volume filling increases the above mean
free path by an order of magnitude, rendering CCA collisions ineffective (and thus the next diffusion model shall not be applied here).
The mean free path is important to estimate the effective collisional viscosity $\nu_{\rm c}$.
The velocity dispersion of the ensemble multiphase gas is directly inherited from the turbulent motions driven in the plasma (Fig.~\ref{f:cca_stream}) and it is found to be fairly stable at
$\sigma_v\simeq165\,$km\,s$^{-1}$.
We note the {\it internal} velocity dispersion of single clouds is smaller than such large-scale dispersion -- following the turbulence cascade -- and typically ranges $\sigma_{v,\rm int}\simeq\,8$\,-\,30 km\,s$^{-1}$ for cold and warm clouds, which are thus supported by dynamical, instead of thermal pressure. 
Overall, the effective viscosity in the inner core, where the mixing length approach is applicable, results to be 
\begin{equation}\label{e:visc}
\nu_{\rm c}\equiv\sigma_v\,\lambda_c\simeq4.5^{+13.4}_{-3.4}\times10^{27}\,\rm{cm^2\,s^{-1}}
\end{equation}
Such property is key to understand the accretion rate.
The accretion process can be modeled as a quasi-spherical viscous accretion 
which drives an inflow velocity $v_r=-\,\nu_{\rm c}/r$ and accretion rate $\dot M_\bullet \simeq \pi\,\rho_{\rm en}r\,\nu_{\rm c}$ (the derivation is given in Appendix \ref{app:visc}).
The mass density\footnote{Total clump mass per enclosed spherical volume -- to not be confused with the emission-weighted, internal density of the single gas phases.} 
of the ensemble clumps decreases as $\propto r^{-1}$ with average $\rho_{\rm en}\simeq6\times10^{-23}$\,g\,cm$^{-3}$ at 5\,pc.
By combining the above retrieved values, the analytic accretion rate translates into
\begin{equation}\label{e:mdot_visc}
\dot M_\bullet = 2.9\times10^{-3}\,\nu_{\rm c} \simeq 0.21^{+0.61}_{-0.15}\ \msun\,{\rm yr}^{-1},
\end{equation}
which well reproduces the numerical results (\S\ref{s:cca_acc}). Notice that the large variance in the cloud radius 
-- due to the turbulent condensation -- induces the large variations in the accretion rate. In other words, the recurrent major variability discussed in \S\ref{s:cca_acc}
can be interpreted as a continuously contracting and expanding mean free path during the CCA rainfall.
Such simple 1D model can be easily incorporated in analytic and semi-analytic studies 
(in similar spirit as alpha-disc prescription is employed), as well as subgrid models for large-scale cosmological simulations which aim to include AGN feedback tied to a realistic multiphase accretion.

We are now in a position to investigate the dominant source causing the transport of angular momentum.
Let us analyze the angular momentum conservation equation. Taking the cross product of $\boldsymbol{r}$ and the compressible Navier-Stokes momentum conservation equation (Euler equation with the stress tensor as additional source term; Batchelor 1967), we can write in convective form 
\[
\frac{\partial{\boldsymbol{l}}}{{\partial t}}  =
\boldsymbol{r}\times\frac{\partial\boldsymbol{v}}{\partial t} =
-\,\boldsymbol{r}\times\left[(\boldsymbol{v}\cdot\boldsymbol{\nabla}){\boldsymbol{v}}\right]\,
+\boldsymbol{r}\times \left(\boldsymbol{g}_{\rm tot} - \frac{\boldsymbol{\nabla}P}{\rho}+\boldsymbol{a}_{\rm drag}\right)+
\]
\vspace{-0.41cm}
\begin{equation}\label{e:tor}
\quad\quad\quad\quad\quad\quad\;\;\,+\, \boldsymbol{r}\times \nu_{\rm coll} \left[\boldsymbol{\nabla}^2\boldsymbol{v} +\frac{1}{3}\,\boldsymbol{\nabla}(\boldsymbol{\nabla}\cdot\boldsymbol{v})\right]
,
\end{equation}
that is the specific angular momentum $\boldsymbol{l}$ in the control volume evolves because of advection (first term), the specific torques $\boldsymbol{\tau} \equiv \boldsymbol{r}\times \boldsymbol{a}\,$ tied to pressure, gravity, and the plasma drag (second term), and the cloud effective collisional viscosity (third term).
The gravitational torque is zero as our potential remains spherically symmetric\footnote{In a long-term cosmological evolution, gravitational torques can further facilitate accretion (\citealt{Hopkins:2010}).}, while the pressure force contributes to torques as it is rendered asymmetric via condensation and the chaotic rain. The stirring acceleration (not included above) is tiny, typically $\lta10^{-4}$ of the pressure gradient acceleration.
The second line in Eq.~\ref{e:tor} adds the cloud collisions through a stress term with the effective (shear and bulk\footnote{As per Stokes hypothesis, bulk viscosity coefficient is assumed to be $-(2/3)\nu$ to preserve the mean normal stress, i.e., pressure.}) viscosity tied to the above kinematic coefficient $\nu_{\rm c}$.
We find the compressive $\boldsymbol{\nabla}\cdot\boldsymbol{v}$ term to be sub-dominant, $<10$\% of the total viscous acceleration in most of the volume, except near convergent streamlines where it increases up to a fraction of 60\%. The viscous torque is thus mainly driven by the laplacian term $\propto\boldsymbol{\nabla}^2\boldsymbol{v}$, analogous to an incompressible flow affected by shear viscosity. Interestingly, in an incompressible flow, the laplacian term can be substituted with $\nu\,\boldsymbol{\nabla}\times(\boldsymbol{\nabla}\times\boldsymbol{v})$, thereby the larger the local vorticity 
the larger the viscous stresses. 
The dense clouds interact with the diffuse hot atmosphere via ram-pressure drag. Such drag force can be written as ram pressure times the cloud cross section, $F_{\rm drag}=\rho_{\rm hot}\,v^2_{\rm rel}\,(\pi\,r^2_{\rm c})$, where we use as relative velocity the chaotic $\sigma_v$.
The related acceleration is thus given by $a_{\rm drag}=F_{\rm drag}/M_{\rm c}\simeq(\rho_{\rm h}/\rho_{\rm c})\,(\sigma_v^2/\,r_{\rm c})$, with timescale 
$t_{\rm \tau}=(\rho_{\rm c}/\rho_{\rm h})\,(r_{\rm c}/\sigma_v)$.

\begin{figure} 
      \centering
      \subfigure{\includegraphics[scale=0.4]{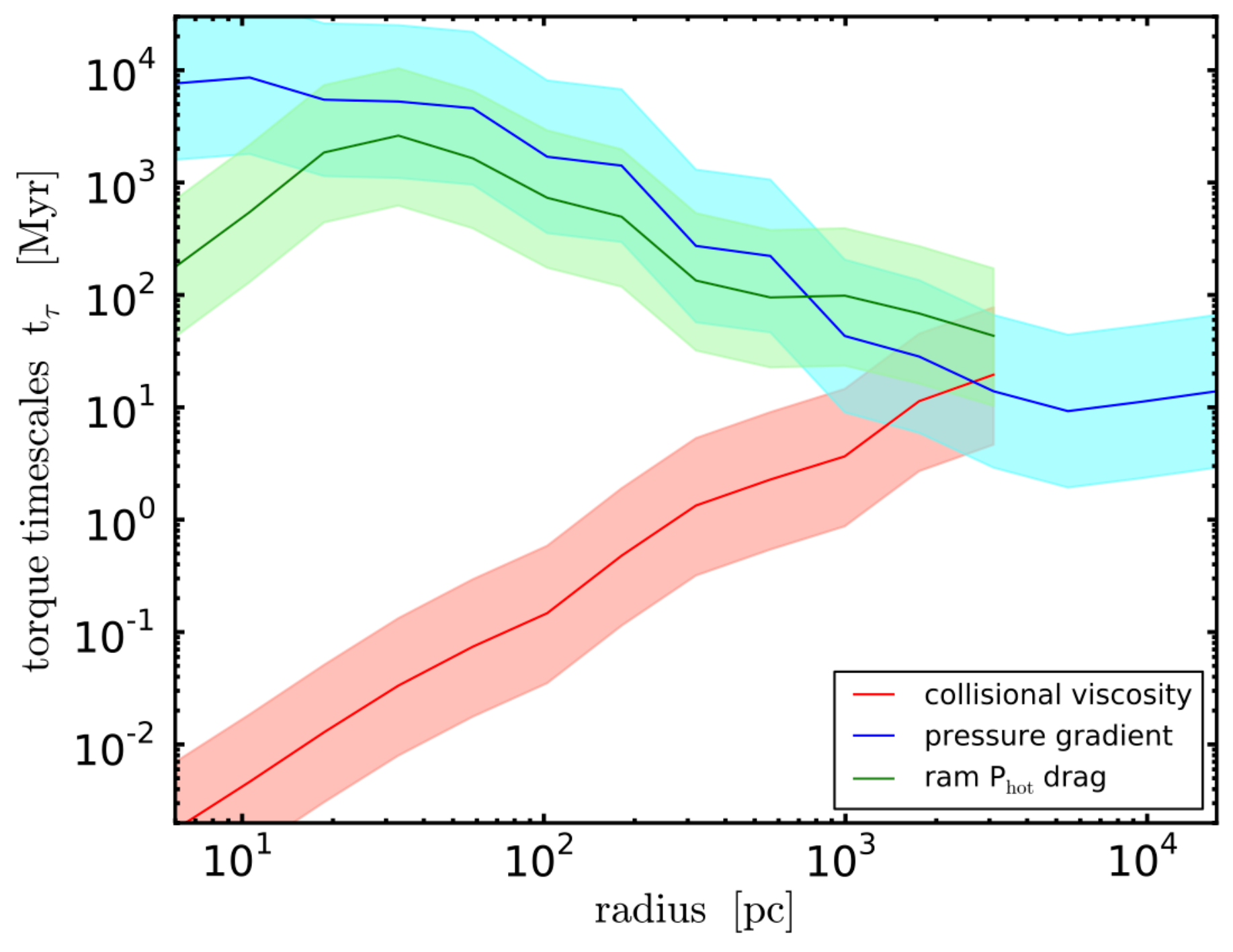}}
      \caption{Accretion with turbulence, cooling, AGN heating, and rotation: timescale of the involved torque mechanisms as a function of radius, $t_{\rm \tau}\equiv l/\tau$, including the cloud collisional viscosity (red), the pressure gradient torque (blue), and the plasma ram-pressure drag (green) during the developed CCA rain. The red and green bands include the dispersion in the effective cloud radius. The blue band represents the logarithmic standard deviation of the pressure torque within the box.
      The cloud collisions dominate the transport of angular momentum in the core, while pressure torques become relevant beyond several kpc where clouds can not easily condense. The ram pressure drag is sub-dominant.
      }
      \label{f:cca_torques}
\end{figure} 

Figure \ref{f:cca_torques} shows the timescales $t_{\rm \tau}\equiv l/\tau$ related to the three torque mechanisms and computed from the final simulation cube by using the terms indicated in Eq.~\ref{e:tor} and the previously retrieved clump properties. Notice that in quasi steady state
the advection term (and thus accretion) is equal to the dominant torque. The cloud collisions (red; second line in Eq.~\ref{e:tor}) dominate the transport of angular momentum within 1 kpc radius, with the viscous torque term increasing toward smaller radii. This clarifies why the SMBH can quickly swallow clouds in $<\,$100\,kyr (\S\ref{s:cca_acc}).
Pressure torques (blue) become relevant beyond 3 kpc, where the plasma has not fully condensed in the chaotic weather. The typical timescale is 10 Myr, a slow, yet significant fraction of total simulation time ($\sim$\,80\,Myr). Pressure torques also affect rapidly the transition layers of clouds (as seen for the disc evolution; \S\ref{s:cool_e03_dyn}), promoting further chaotic interactions.
The ram pressure drag of the diffuse medium on the clouds (green) is always sub-dominant and slow, except near the transition region at 2\,-\,3 kpc where the cloud volume filling is minimal. The large contrast ratio between the internal clump density and the hot plasma ($10^{2}$\,-\,$10^{3}$) substantially hinders the ram-pressure drag.

\begin{figure}
      \subfigure{\includegraphics[scale=0.4]{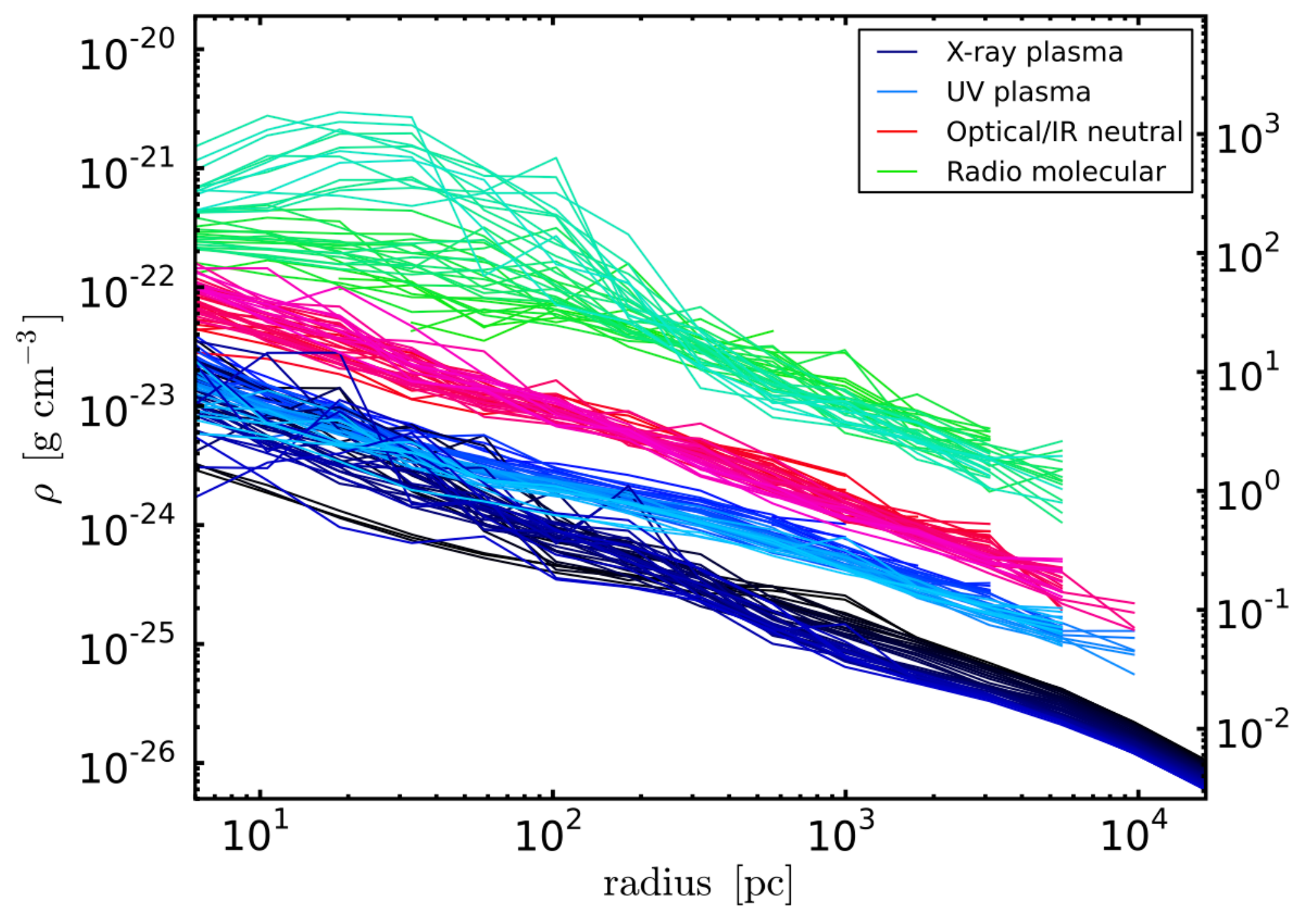}}
      \subfigure{\includegraphics[scale=0.403]{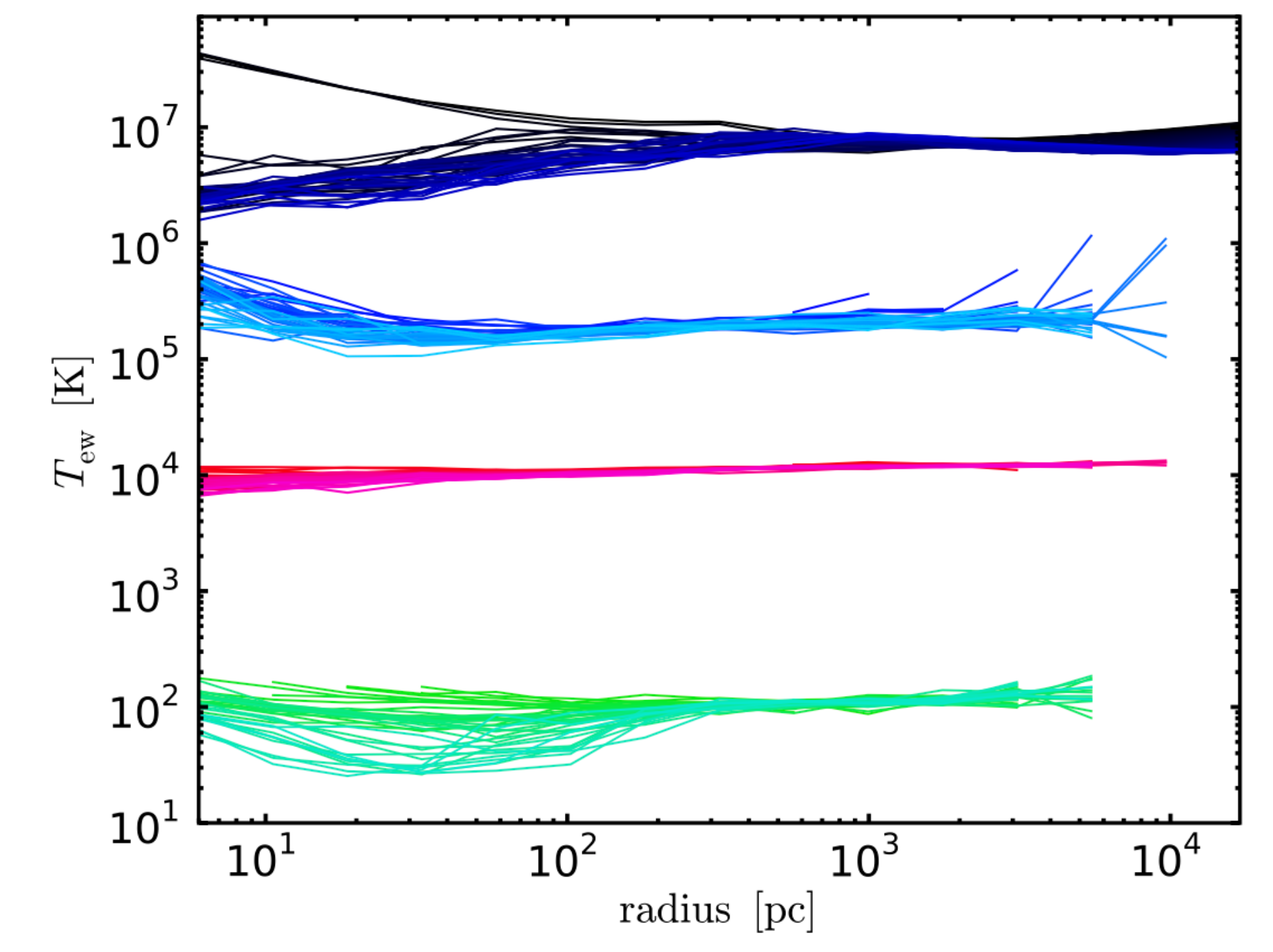}}
      \caption{Accretion with turbulence, cooling, AGN heating, and rotation: 3D emission-weighted radial profiles of density and temperature, probing the multiphase structure of the CCA flow. The flow is mainly comprised of 3 phases: plasma (blue), neutral gas (red), and molecular gas (green). The plasma is further separated into the X-ray ($T\ge0.1\,$keV) and UV component ($10^{4.2}\,{\rm K}<T<0.1\,$keV).  The top right y-axis denotes the electron number density $n_{\rm e, x}$ for the hot plasma. Only the plasma phase is volume filling ($f_V\simeq1$); the cold/warm gas density profiles represent the internal densities of such phase.
      The 80 Myr evolution is color-coded from dark to lighter color with 2 Myr step.}
       \label{f:cca_prof}      
\end{figure}

\vspace{-0.41cm}
\subsection[]{Multiphase condensation}  \label{s:cca_multi}
We explore now the multiphase structure of the CCA flow, focusing in particular on the observational predictions in various bands.
As elaborated in-depth in the disc scenario (\S\ref{s:cool_e03_multi}), the diffuse X-ray medium is the crucible out of which all other phases condense out in a top-down cascade through 1\,keV $\rightarrow$ $10^4$\,K $\rightarrow$ 20\,K. The multiphase cascade establishes a tight thermodynamical link between all phases and thus observational bands.
At variance with the disc and monolithic cloud evolution, however, in a heated and turbulent atmosphere the cold and warm phase condense in localized high $\delta\rho/\rho$ regions, forming clouds and extended filaments, which are mainly comprised of molecular and neutral gas. For a CCA evolution dominated by warm gas, refer to G13-15 work where we assumed that the cold gas cooling is prevented by strong photoionization heating.

Fig.~\ref{f:cca_prof} shows the emission-weighted density (top) and temperature (bottom) radial profiles of the multiphase gas in the X-ray, UV, optical and radio band during the 80 Myr evolution. 
Overall, all phases have density profile with logarithmic slope oscillating around a value of -1, and fairly flat temperature profile.
The X-ray plasma (dark blue) fills the entire $r\gta$\,30 kpc halo of the massive galaxy and becomes steeper at large radii because of the NFW potential. In the inner region, the X-ray and UV phase (light blue) reach similar electron density, with both temperatures drifting toward the $10^6$\,K regime. The X-ray phase (which includes the soft 0.1\,-\,0.2 keV plasma) has a slight decline in the nuclear region as dropout becomes more intense; in the hard X-ray ($>0.5$\,keV) it remains essentially flat.
There is 0.5 dex variance in the profiles driven by turbulent mixing.
CCA differs from hot models (Bondi, ADAF, etc.; \citealt{Yuan:2014}) which all predict a cuspy X-ray profile. Recent observations have detected flat or declining X-ray $T$ profiles in massive galaxies with resolved Bondi radius, such as M87 (\citealt{Russell:2015}), NGC 3115 (\citealt{Wong:2014}), NGC 4261 and 4472 (\citealt{Humphrey:2009}).
Remarkably, they also show a -1 slope in the inner X-ray density profile, consistently with CCA.
Regarding normalization, the inner observed {\it Chandra} bin in these 4 nearby ETGs reside at $r\approx$\,100\,pc with observed $n_{\rm e,x}\simeq0.3$\,-\,0.5\,cm$^{-3}$, well matching the values of the CCA model (Fig.~\ref{f:cca_prof}).
As further support of inner plasma condensation, many ETGs harbor within the central kpc low-entropy X-ray `coronae' prone to rapid cooling (\citealt{Sun:2009b}).
By integrating the X-ray emissivity profile up to 100\,pc, with inner density normalization $n_{\rm e,x}=1$\,-\,10\,cm$^{-3}$, we retrieve a nuclear X-ray luminosity (purely due to hot gas) in the range $10^{38}$\,-\,$10^{40}$\,erg\,s$^{-1}$, which does not exceed the point source X-ray luminosities observed in typical massive galaxies $<10^{43}$\,erg\,s$^{-1}$ (Fig.~13 in \citealt{Russell:2013}).

The optical/IR gas (red) has similar declining emission-weighted density profile but $\sim$\,1 dex higher than that of the X-ray phase.
We remark this is emission-weighted, not volume-weighted density; the warm/cold gas fills only a small fraction of the volume, thus the hot X-ray mass still dominates the mass content in the core.
The emission-weighted $T$ profile is very flat dominated by gas near $10^4$ K. As the condensing gas is not spherically converging toward the SMBH, compressional heating is minor and molecular clouds can condense while the warm filaments drift in the core.
The molecular phase (green) is volatile, with gas density varying by 1 dex within the inner 100 pc and reaching internal densities up to a few $10^{-21}$\,g\,cm$^{-3}$.
The relative contrast with the hot plasma can reach ratios up to 3 dex.\footnote{
Single molecular clouds commonly aggregate in larger giant molecular clouds and associations (GMAs). Observed GMAs can be thus perceived has having 1-2 dex lower volumetric density, but most of the molecular mass is contained in compact structures.}
The warm/cold gas cools rapidly in non-isobaric mode (toward the isochoric regime, as the related cooling time is much shorter than hot gas sound crossing time) and is further supported by turbulent pressure, thus preventing extreme contrast ratios $>$\,3 dex.
The emission-weighted $T$ of molecular clouds is varying in the core between 30\,-\,100\,K, with the lower values reached at later times. 
Any shock and compressional heating is quickly radiated away via rapid cooling ($\propto \rho^2$). 
The inner molecular phase variance is related to the major locus of cloud collisions (\S\ref{s:cca_dyn}), and augmented by the local intermittency in the turbulent plasma eddies. Albeit profiles are typically not presented, real data maps show that warm filaments and molecular clouds are linearly decreasing with radius (\citealt{Werner:2014,David:2014}), being truncated beyond 10 kpc where the cooling time rises above 10s $t_{\rm ff}$. Future ALMA radio data will be instrumental in unveiling the coldest part of the CCA rain, as shown in A2597 (\citealt{Tremblay:2016}), where the $30\,{\rm K}$ molecular clouds traced by CO(2-1) are condensing out of the warm filaments and falling inward toward the center.
We note not all ETGs are expected to reside in the CCA state; some ETGs may reside in the quiescent or disc state (as low-luminosity galaxies), others may be experiencing a major merger with substantial injection of external warm/cold gas (e.g., \citealt{Davis:2011}).

\begin{figure}
      \centering
      \subfigure{\includegraphics[scale=0.42]{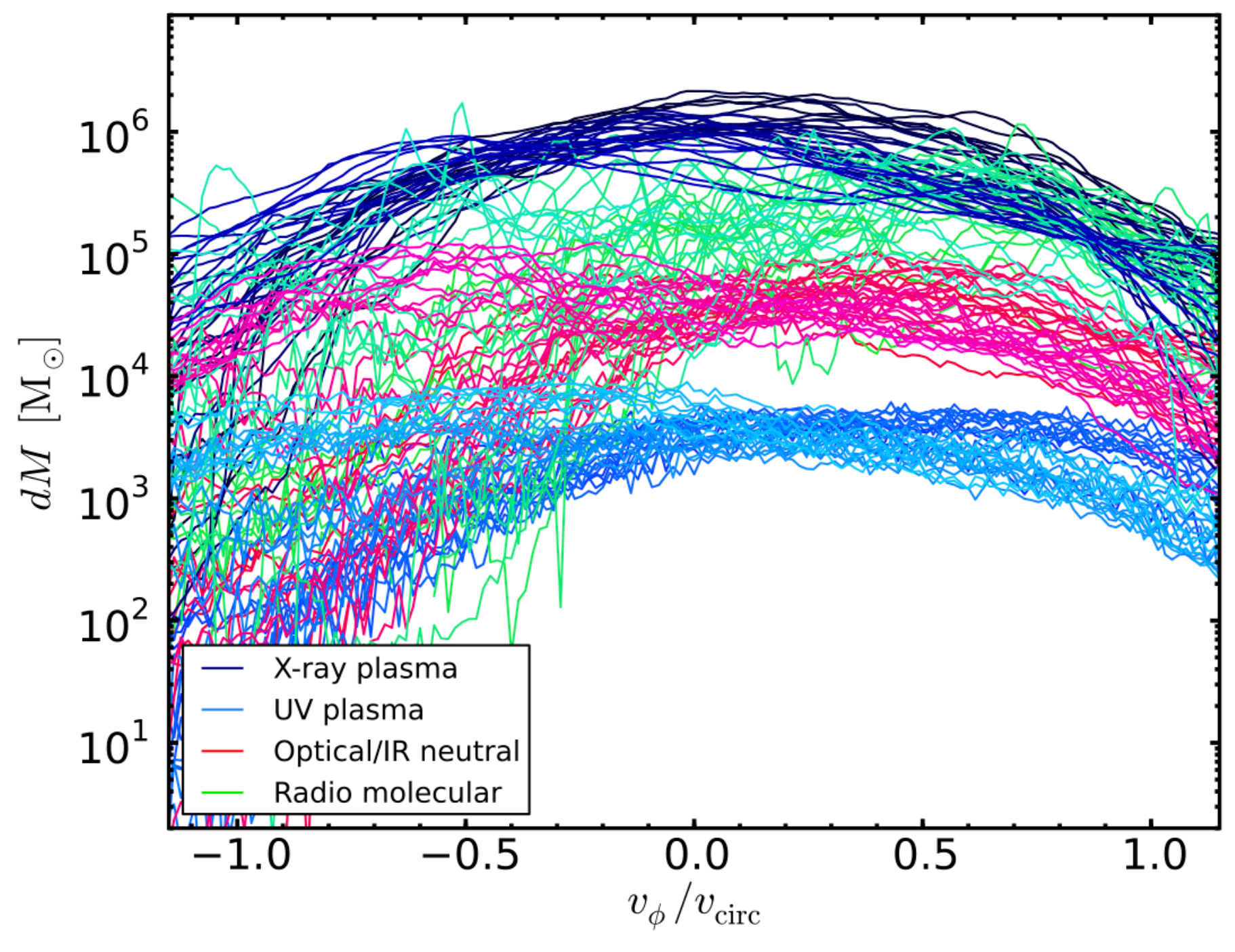}}
      \caption{Accretion with turbulence, cooling, AGN heating, and rotation: multiphase mass distribution per bin of rotational velocity normalized to circular velocity (cf. Fig.~\ref{f:cool_e03_vphi}), with extraction radius of 2 kpc.  
      The lines cover the last 60 Myr of evolution with 2 Myr step.
      Turbulence broadens the plasma angular momentum distribution creating a negative/retrograde tail.
      The multiphase clouds can often condense with low angular momentum. Clouds emerging out of the tails experience recurrent, chaotic collisions which mix and reduce angular momentum.
         }
      \label{f:cca_e03_vphi}    
\end{figure} 

The distribution of the rotational velocity normalized to the circular value among the main phases is shown in Fig.~\ref{f:cca_e03_vphi}.
At variance with the disc evolution (\S\ref{s:cool_e03_multi}), turbulence broadens the X-ray plasma angular momentum distribution creating a retrograde, negative tail. The neutral and molecular clouds can thus condense with low angular momentum. Similar to G15, those emerging out of the tails with large rotational velocity experience several collisions before being accreted. As inelastic collisions mix and partially cancel angular momentum (\S\ref{s:cca_dyn}), they induce rapid variations in the tails of the distribution, in particular the negative wing which is not supported by initial rotation. 
Shrinking of the distribution via collisions and regeneration via condensation compete throughout the evolution and 
are particularly manifest in the neutral and molecular phases.     
      
\begin{figure}
      \centering
      \subfigure{\includegraphics[scale=0.42]{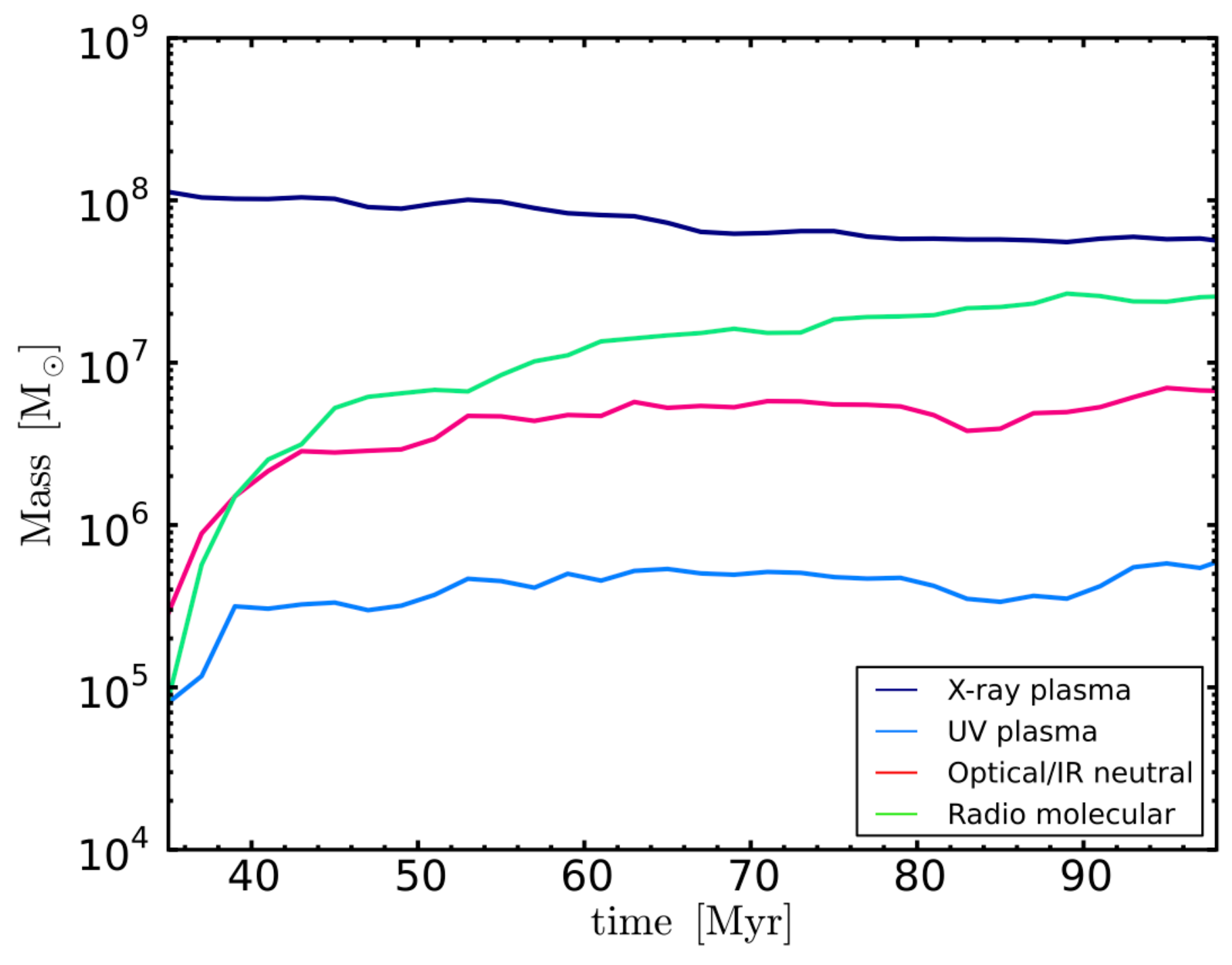}}
      \caption{Accretion with turbulence, cooling, AGN heating, and rotation: evolution of mass for each of the major multiphase components within $r\lesssim2\,$kpc (\S\ref{s:phases}) -- plasma (blue), warm neutral gas (red), and cold molecular gas (green).
      The molecular gas dominates the mass budget of the condensed CCA rain, with values comparable to what ALMA is detecting in massive ETGs (e.g., NGC 5044). A small fraction of the warm phase 
      exists in the form of a partially ionized skin which is traced by H$\alpha$ emission.
       }
       \label{f:cca_masses}      
\end{figure}

Fig.~\ref{f:cca_masses} shows the evolution of the multiphase masses within 2 kpc radius. The X-ray mass decreases by 40\% at the end of the run from the initial $10^8\,\msun$. While condensation renders the hot halo more diffuse, there is no steady cooling inflow as in the disc evolution and turbulent eddies can transport the plasma outwards from the core. The UV mass is relatively small, barely approaching $10^6\,\msun$.
The molecular, radio gas dominates the mass budget of the condensed gas, reaching nearly $3\times10^7\,\msun$ during 60 Myr of CCA rain.
Albeit with uncertain CO-to-H$_2$ conversion factor,
\citet{David:2014} estimate in NGC 5044 a molecular mass within 2.5 kpc of $\sim\,$$5\times10^7\,\msun$, in line with our findings. 
The presence of molecular gas may be corroborated by the [C$^+$] emission found in the core of massive ETGs (\citealt{Werner:2014}).
A longer and/or less quenched CCA will steadily increase the simulated amount of molecular gas up to several $10^7\,\msun$; we avoided parameter fine-turning to match a single data point, also considering the limitations of the present molecular modeling, as no explicit chemical network.
With respect to the whole hot halo, the cold gas mass fraction is $\lta 1$ percent, yet it has a profound impact on the SMBH growth and the galaxy self-regulation. In comparison, the Milky Way hosts $10^9\,\msun$ of H$_2$ gas (\citealt{Heyer:2015}), 20$\times$ higher molecular gas than a massive ETG. 

Star formation is secondary during the simulated evolution.
The average depletion time of the molecular gas due to star formation in massive galaxies is 1 Gyr (\citealt{ODea:2008}).
Thereby we expect a newly generated stellar mass $\approx2\times10^6\,\msun$ in the last 60 Myr dominated by molecular gas. The inferred star formation rate is $\dot M_{\ast}\approx0.04\ \msun\,{\rm yr^{-1}}$, not far from WISE estimate of $0.07\ \msun\,{\rm yr^{-1}}$ (\citealt{Werner:2014}) and over 1 dex lower than that of common disc galaxies. 
In the current CCA model, turbulent pressure is the dominant mechanism counterbalancing the self-gravitational collapse, thus helping to explain the low star formation efficiencies, $\varepsilon_\ast\lta5$\%, although magnetic fields and stellar feedback may play important role (\citealt{Federrath:2015}) and will be studied in the future.
Interestingly, in CCA the tiny SMBH is directly linked to the cold gas of the whole host ETG, thus to the stellar mass,
which is observationally known as Magorrian relation (\citealt{Kormendy:2013}); the long-term AGN feedback duty cycle and high-redshift starburst history are required to properly set the normalization of such linear scaling.

The optical/IR (and 21-cm), neutral gas forms the secondary component of the CCA rain, reaching masses $\simeq 7\times10^6\,\msun$ (Fig.~\ref{f:cca_masses}), significantly lower than the molecular gas mass. 
The typical occupied volume is instead $\gta$\,0.7 dex higher than that of the cold phase, presenting more filamentary morphology.
The average volume fraction for the combined warm and cold phase is a few percent within 1 kpc radius.
Unfortunately, neutral hydrogen maps are not commonly investigated in deep single-target ETGs. 
Albeit notoriously difficult to detect, The ATLAS$^{\rm 3D}$ survey finds that at least half ETGs host HI gas (\citealt{Serra:2012}), with massive galaxies populating the lower end at $10^7\,\msun$ (more discussions on neutral gas in \S\ref{s:cca_comp}). 
The most visible tracer of the filamentary phase is H$\alpha$ (and [N$^+$]) emission from ionized gas near $10^4$\,K (\S\ref{s:cca_comp}).
Converting NGC 5044 H$\alpha$ luminosity detected with SOAR (\citealt{Werner:2014}) to total ionized gas mass yields (\citealt{Macchetto:1996,Kulkarni:2014}) $M_{\rm H^+}\simeq1.1\times10^6\,(L_{\rm H\alpha}/10^{40.3}\,{\rm erg\,s^{-1}})(n_{\rm e}/40\,{\rm cm^{-3}})^{-1}\ \msun$. 
The condensed structures thus have a skin of ionized gas, which
should account for $\lta 5$\,percent of the total condensed mass (neutral and molecular).
Collisional ionization can partially explain such ionization fractions. 
By using our UV phase as rough indicator for the ionized gas layer, 
which well wraps the warm filaments and is dominated by gas at $10^4$\,-\,$10^5$\,K, 
the ionization fraction is of the order of a few percent.
Unresolved mixing with the hot plasma, cosmic rays, and radiation from the stars or the AGN -- which are not included in our runs -- can further increase the H$\alpha$, ionized layer at the expense of the neutral and molecular phase. 

\begin{figure}
     \centering
     \subfigure{\includegraphics[scale=0.5]{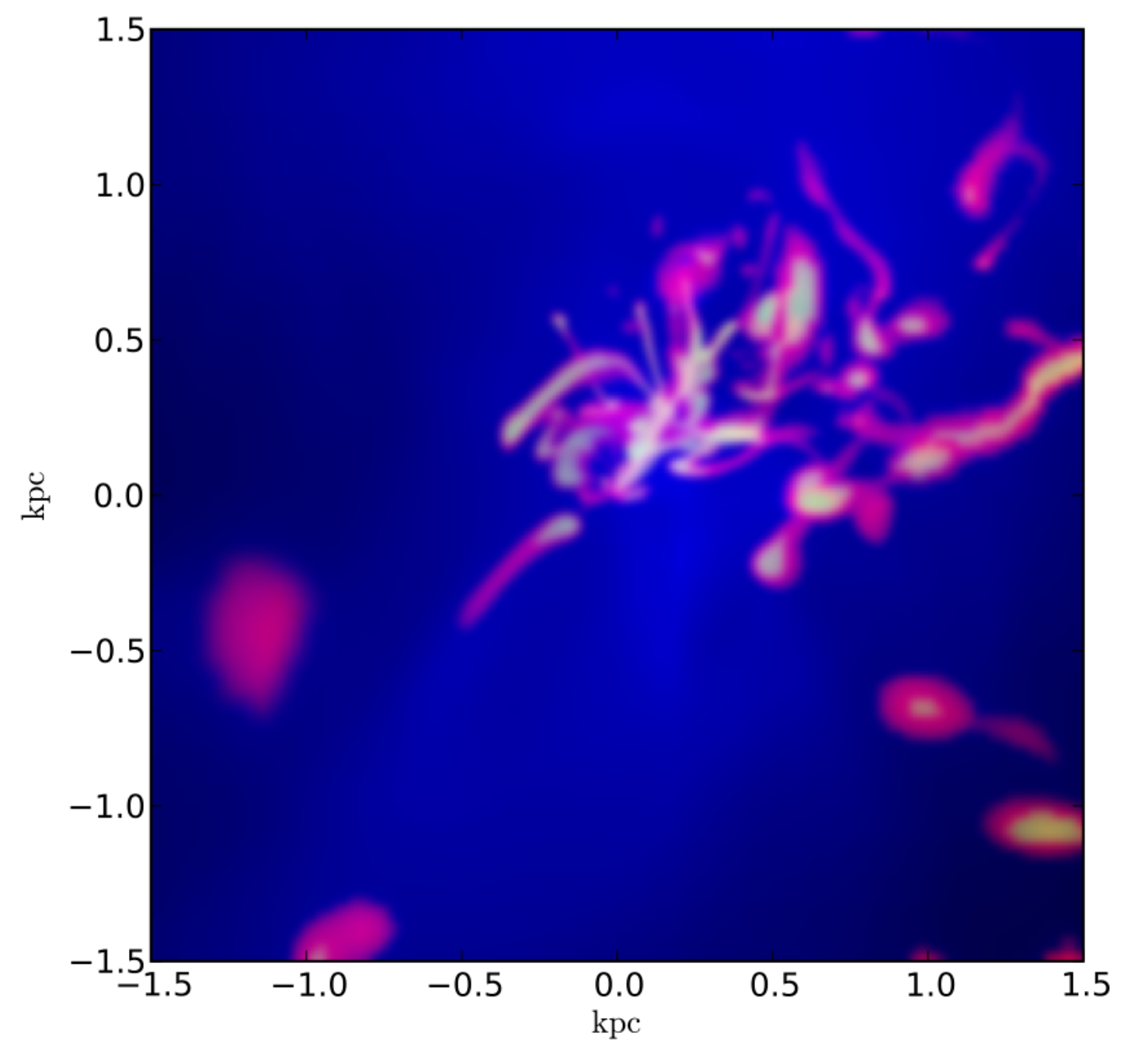}}  
     \caption{
     Accretion with turbulence, cooling, AGN heating, and rotation: composite RGB image of the surface density tracing the plasma (blue), warm gas (red), and cold gas (green), which shows the projected view of the developed CCA rain along a random inclination angle ($t=68$\,Myr). A Gaussian filter mimics the observational limited beam. 
     CCA is associated with a cospatial 3-phase medium.
     Warm and cold clouds coalesce in projection, forming a network of filaments and giant associations. 
     The interface layer between turbulent eddies provides the optimal locus for the multiphase condensation forming extended fractal structures.
        }
         \label{f:cca_RGB}
\end{figure} 

Fig.~\ref{f:cca_RGB} shows the composite RGB image of the projected surface density related to the 3 major phases during developed CCA rain in the central region (next \S\ref{s:cca_comp} tackles in depth the observational brightness images).
The typical, central surface densities of the hot, warm, and molecular gas are 10, 25, 100 $\msun$\,pc$^{-2}$, respectively.
Clearly, the multiphase condensation evolves in a stratified way, with the cold phase embedded in the warm filaments, which emerge out of the low-entropy plasma. Each major phase has further intermediate layers, as the ionized gas enveloping the neutral phase (\S\ref{s:cca_comp}).
The cospatiality attribute is thus a major signature of CCA and internal multiphase condensation, although a few clouds can be randomly comprised of purely one phase.
In projection, molecular clouds are seen as giant molecular structures or associations (GMAs). Similarly, the projected warm phase can form a complex web of filaments, although large-scale filaments can also naturally form at the interface layer between turbulent eddies, where the vortices compress nonlinearly.
In other words, the condensing gas tends to trace the high-density peaks during the turbulence cascade.
Warm filaments are thus much easier to identify in projected, observational images (e.g., \citealt{Fabian:2016}).

\subsection{Condensation criterium \& thermal instability}  \label{s:cca_TI}
 
In the central 500 pc, the clouds continue to collide inelastically, as previously discussed. On larger scales, the filaments tend to chaotically drift in the 3D turbulent field and are not necessarily precipitating in free fall. 
Moreover, the impact of AGN heating is to prevent the halo from catastrophic collapse (avoiding a large negative background $v_r$), again diminishing the sinking time. 
Thereby, the well-known radial $t_{\rm cool}(r)/t_{\rm ff}(r)\lta 10$ criterium (\S\ref{s:intro})
is a very crude estimate of the condensation radius with large scatter ($\gta0.7$ dex; Fig.~3 in \citealt{Gaspari:2012a} -- see also \citealt{Valentini:2015}), as the pure gravitational time is not a robust characteristic timescale. 
We suggest to track as much as possible the ratio {\it locally} and propose as more efficient estimator for the multiphase condensation the criterium $t_{\rm cool}(l')/t_{\rm eddy}(l')\equiv [l'/v_{\rm cool} (l')]/[l'/\sigma_v (l')]=\sigma_v(l')/v_{\rm cool}(l')\lta1$ over a chosen spatial scale $l'$, which may differ from radial distance. This implies that the cooling time is the dominant factor having the largest variations associated with plasma density. As the turbulent velocity is always sub-virial, in particular on small scales, the above threshold of 10 decreases toward a more natural value of order unity.

AGN outflow uplift -- not tested here -- is possibly important to further enhance the triggering of multiphase condensation (\citealt{Salome:2006,Brighenti:2015,McNamara:2016}). 
Recently, \citet{McNamara:2016} suggest that a strong directional input of kinetic energy via the inflated X-ray cavities can lift the low-entropy plasma, thus condensing at the altitude where the cooling time is lower than the infall time ($\gg t_{\rm ff}$). Our proposal can be seen as a generalization of this concept, i.e., the real dynamical timescale at the denominator is not the free-fall, but the current, local dynamical field of the gas, being it a chaotic large-scale eddy (containing most of the kinetic energy) or a more directional outflow uplift. Notice that a buoyant cavity quickly disrupts into smaller and smaller vortices (cf.~\citealt{Brighenti:2015}, Fig.~10) stimulating condensation and the raining down of cold clouds as in CCA. Thereby, the two are complementary processes.
Three-dimensional simulations are required to test in-depth the outburst stage.

Regardless of the perturbation driver (AGN outflows/bubbles, stellar feedback, mergers, galaxy motions, etc.), it appears evident that fluctuations in realistic hot halos do not start with tiny linear amplitude so the usual convection versus thermal instability analysis has limited applicability (\S\ref{s:hot}). 
Multiphase condensation occurs here in a region with significant entropy gradient, where slow buoyancy might in principle suppress a linearly cooling element. Instead, chaotic motions (either the turbulent eddies or the directional large-scale driver) generate significant variance in the plasma properties; the related plasma density peaks locally drop out if they have enough time to cool before being non-radially sinked or directly reheated by the next AGN outburst.

\vspace{-0.41cm}
\subsection{X-ray, UV, optical/IR, and radio images and comparison with observational data}  \label{s:cca_comp}
We compare now the projected images from the simulated NGC 5044 during the CCA rain 
with the best available observed images in the key bands tracing the 3 main phases: X-ray, optical/IR, and radio.
In the simulated images, we project the emissivity of the gas residing in a specific band and associated phase interval (\S\ref{s:phases}).
We did not attempt to fine tune the simulation parameters to match 1:1 the observational data because of time variance and the intrinsic telescope uncertainties (background noise, response, etc.), which are beyond the scope of this paper. 
The aim here is to show the key CCA predictions emerging in surface brightness
and how they can be probed in multiwavelength images, e.g., retrieved with {\it Chandra}, SOAR, or ALMA.     

\begin{figure}
      \centering
      \subfigure{\hspace{+0.2cm} \includegraphics[scale=0.47]{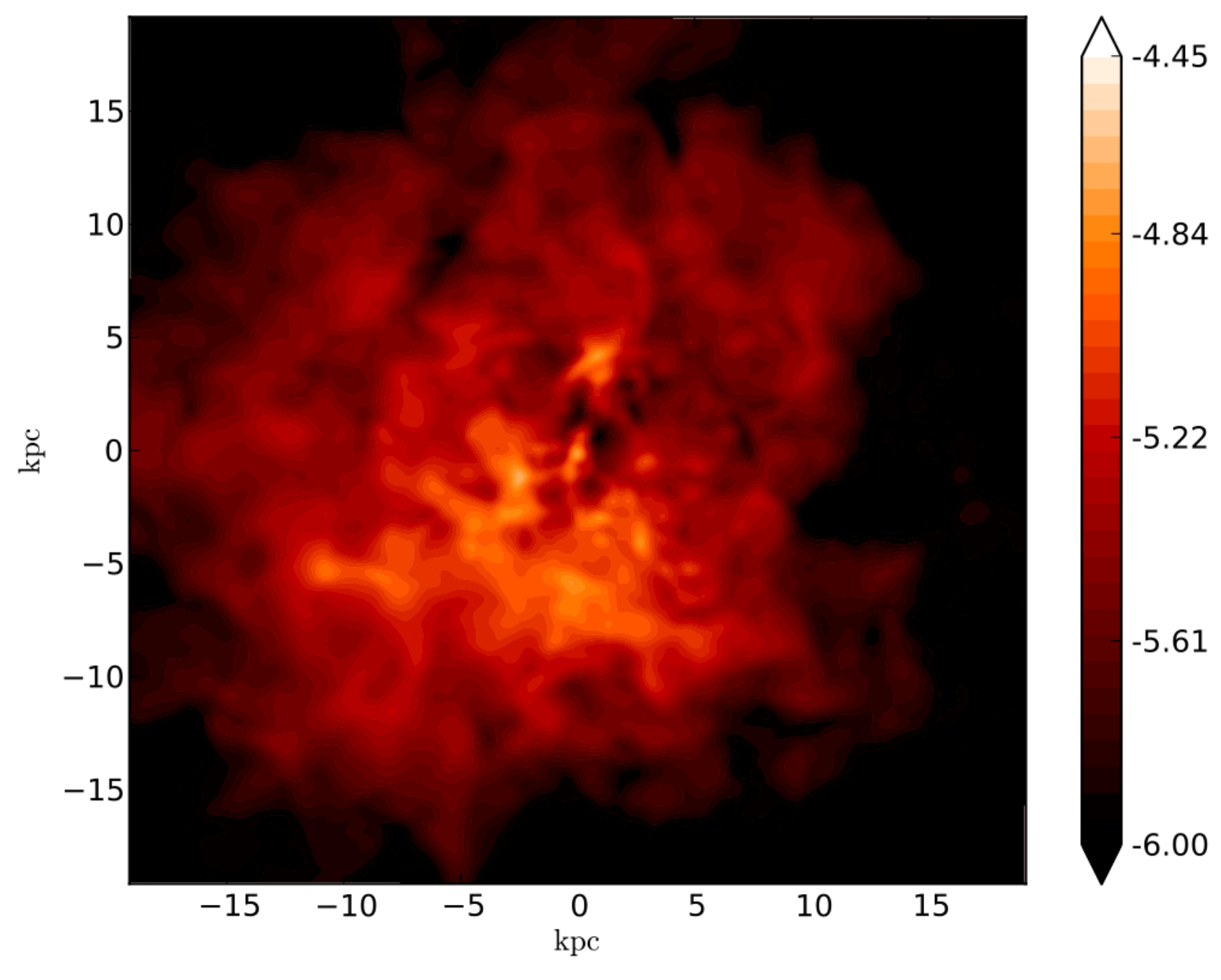}}
      \subfigure{\hspace{+0.2cm} \includegraphics[scale=0.47]{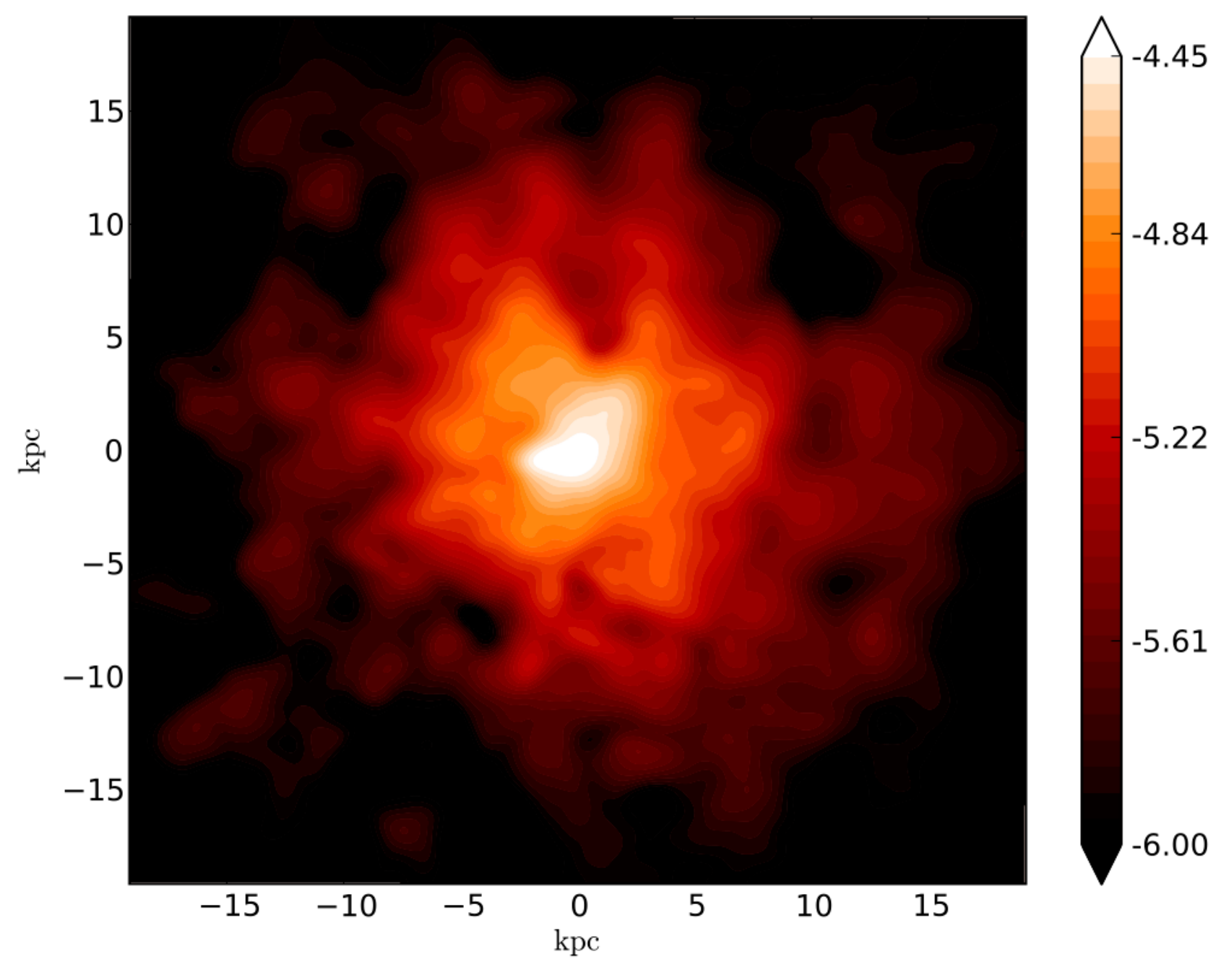}}      
      \subfigure{\includegraphics[scale=0.4]{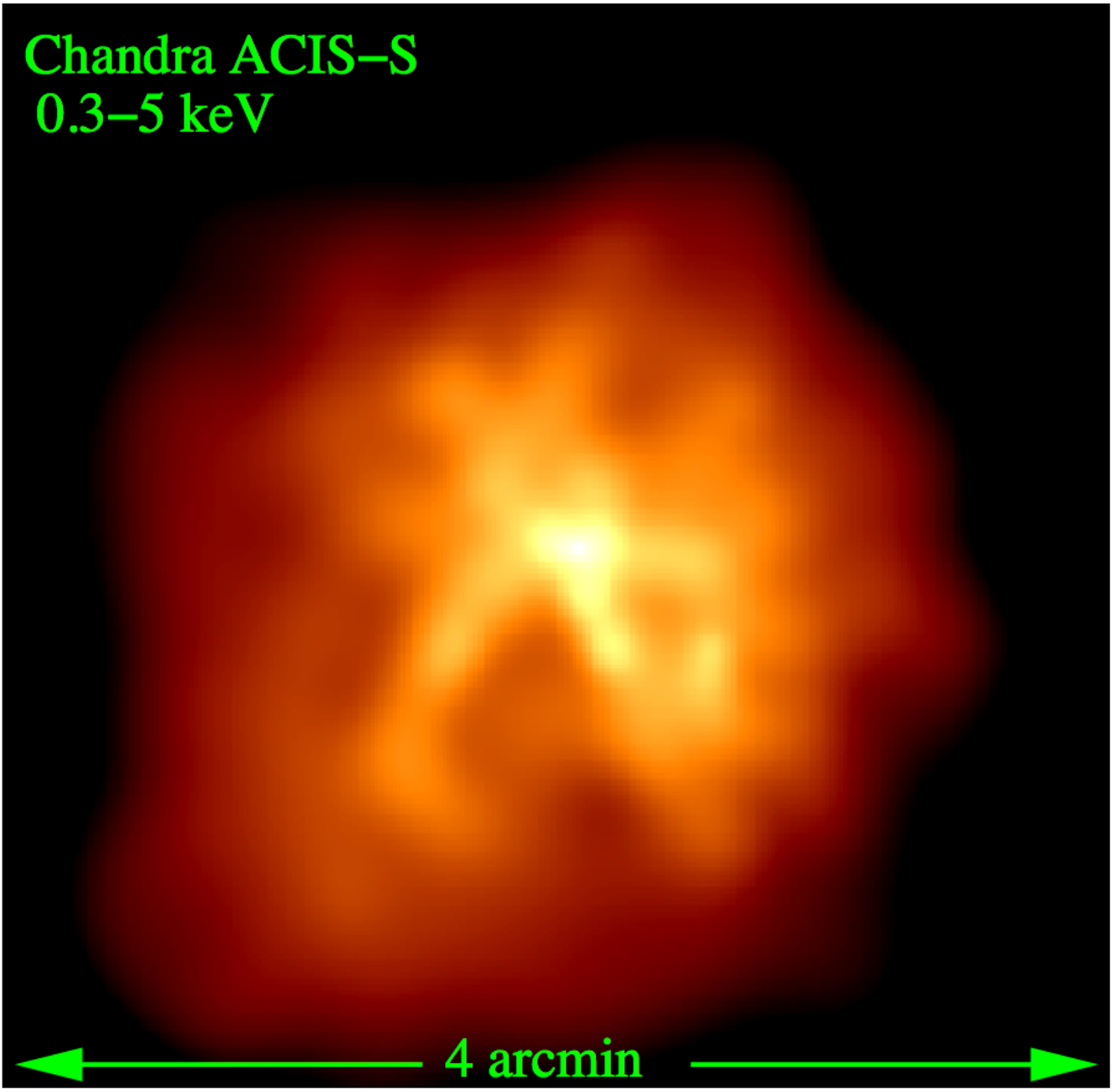}    }
      \caption{Top and middle: soft (0.1-0.5 keV) and hard ($> 0.5$\,keV) X-ray surface brightness (erg\,s$^{-1}$\,cm$^{-2}$\,sr$^{-1}$) of the simulated NGC 5044 during the CCA rain ($t=88$\,Myr). A gaussian filter is applied to mimic the observational beam smoothing.
        Bottom: {\it Chandra} ACIS-S X-ray image of NGC 5044 (from Fig.~1 in \citealt{Buote:2003} -- \textsuperscript{\textcopyright}AAS, reproduced with permission; \href{http://dx.doi.org/10.1086/377094}{DOI}); $1''=0.16$\,kpc). Given the {\it Chandra} drop in effective area at low energy, such image is most sensitive to photons > 0.5 keV, similar to the middle panel.        
        All images have 38.4\,kpc width. 
        The similarity between the synthetic and real image is remarkable.
        The imprints of the turbulent eddies are visible in the hard X-ray, while condensing filaments start to emerge in soft X-ray.
        }
      \label{f:comp_X}
\end{figure} 

Figure \ref{f:comp_X} shows the simulated X-ray surface brightness (SB$_{\rm x}$) map of NGC 5044 compared with the {\it Chandra} ACIS-S map from \citet{Buote:2003}. The top and middle synthetic maps separate the contribution of the hot plasma into the very soft (0.1\,-\,0.5\,keV) and hard X-ray ($>0.5$\,keV), respectively.
The hot plasma halo in hard X-ray fills the whole tens kpc galactic halo, emitting photons mostly within the central 5 kpc. 
The X-ray halo is quasi spherically symmetric.
The signature of mild, subsonic turbulence are clearly visible in the fractional asymmetries, inducing relative density perturbations and thus surface brightness fluctuations (see \citealt{Gaspari:2013_coma, Gaspari:2014_coma2} for more on this subject).
The synthetic map shows that depressions in surface brightness can be naturally created by turbulent eddies.
Some of the weaker cavities in hot halos could be thus not be associated with AGN jets or sloshing motions. 
We note a mild subsonic turbulence level ${\rm Ma_{3D}}\sim0.3$\,-\,0.4 -- as detected by recent {\it Hitomi} telescope (\S\ref{s:turb}) -- is fully sufficient to stimulate CCA; while in such regime dissipational heating is subdominant and slow, the variance induced by turbulence still plays a key role.
The hard X-ray halo is the original progenitor of the multiphase rain. 
In soft X-ray, the condensing filaments start to be visible and the hardness ratio (hard to soft X-ray emission) shows corresponding decrements, as found by observations (Fig.~5 in \citealt{Tamura:2003}).
In the central 5 kpc, the peaks in SB$_{\rm x}$ correlate with the locus of the warm filaments and enclosed radio, molecular clouds, forming a coherent dynamical structure from the 10s kpc down to the pc scale.
Despite the decreasing sensitivity below 0.5 keV and the adaptive smoothing, the {\it Chandra} image well captures the combined effect of 
the turbulent atmosphere and the plasma cooling features,
with remarkable similarity to the synthetic data.
The higher the angular resolution, the better the locus of condensation it will be resolved.
Future {\it Athena} mission (\citealt{Nandra:2013}) is in excellent position to spatially resolve such X-ray `clumpiness' and to directly constrain the initial stage of the {\it local} runaway condensation. 

\begin{figure} 
      \centering
      \subfigure{\hspace{+0.0cm} \includegraphics[scale=0.45]{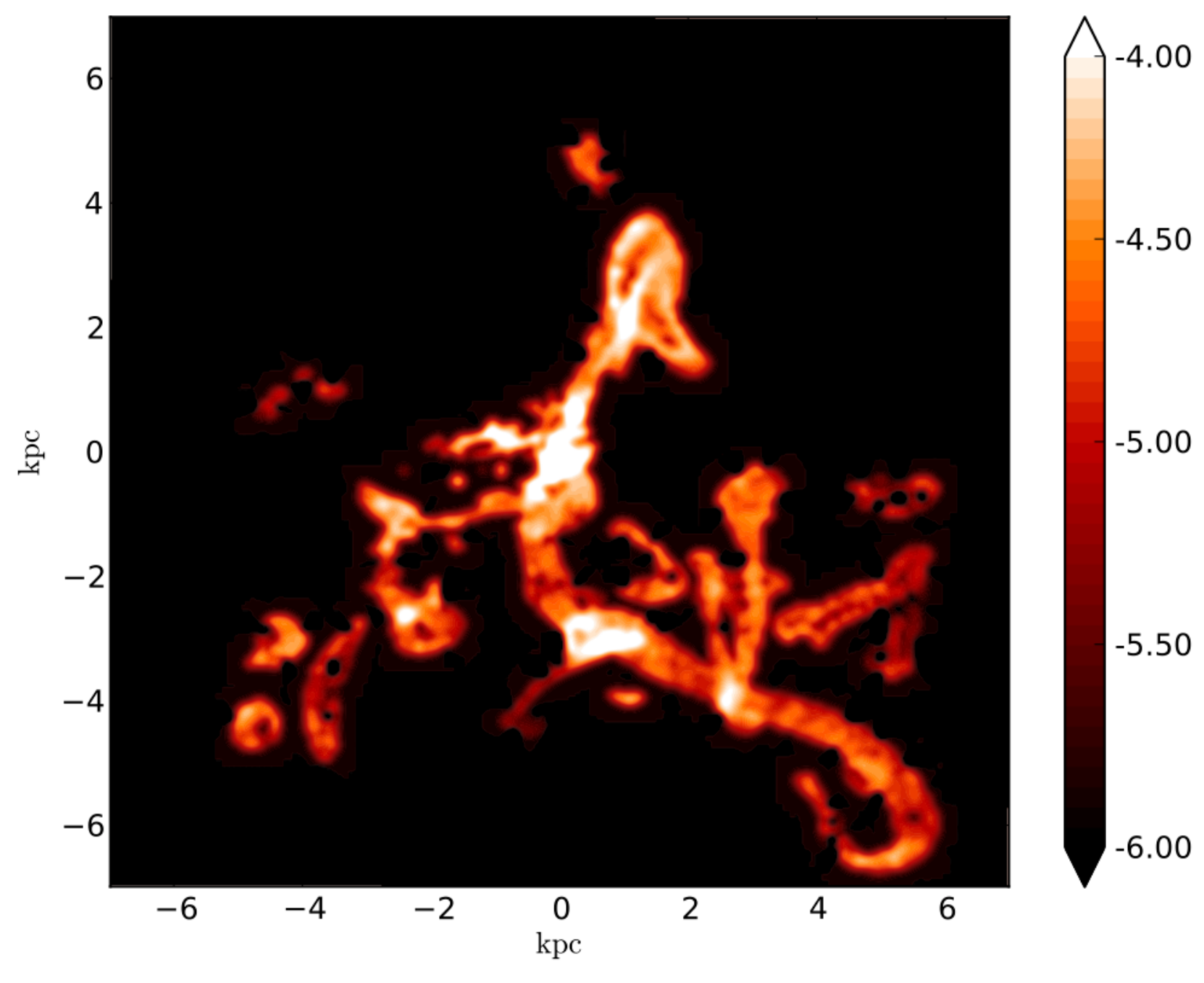}}  
      \subfigure{\hspace{+0.0cm} \includegraphics[scale=0.45]{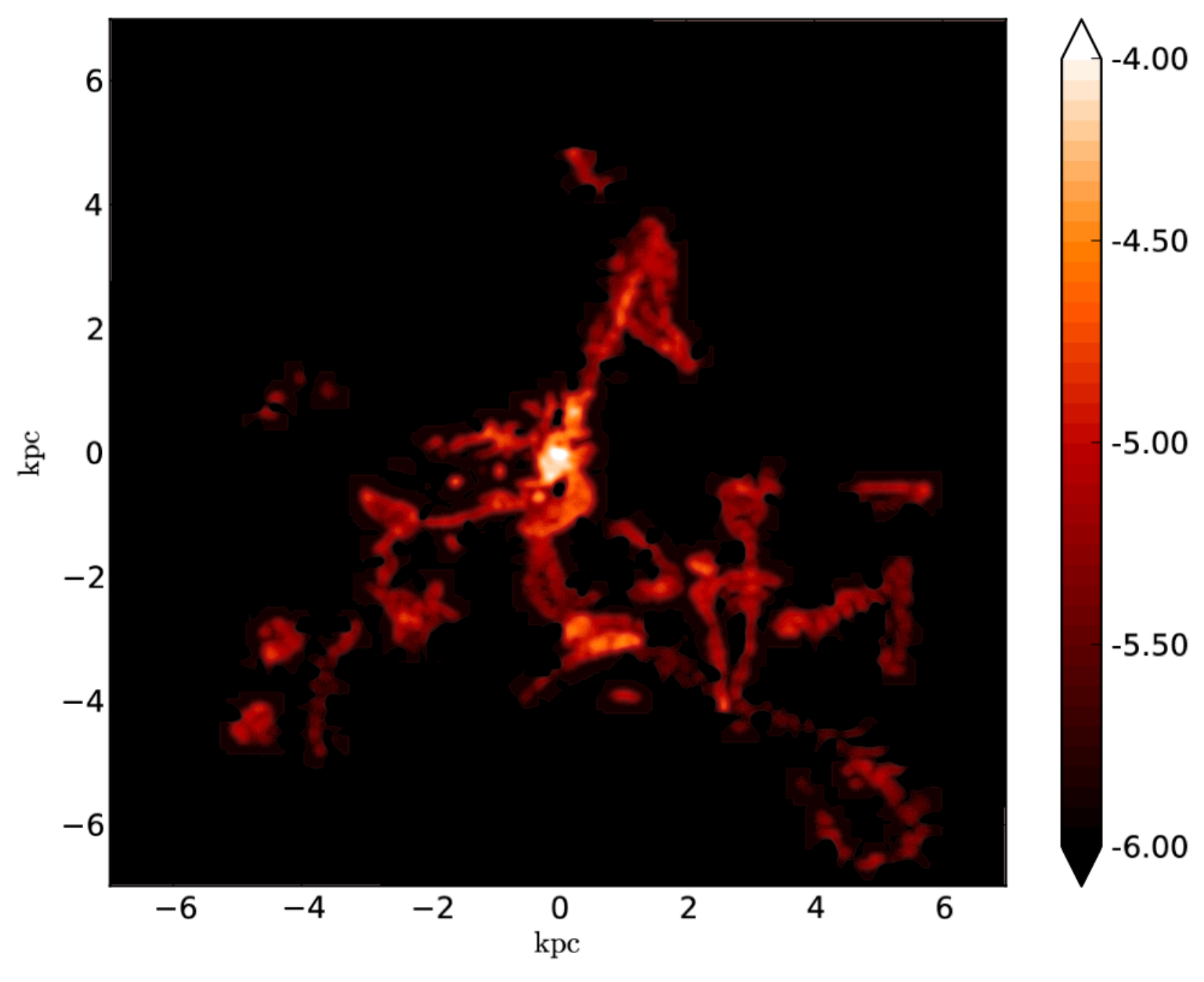}}
      \subfigure{\includegraphics[scale=0.365]{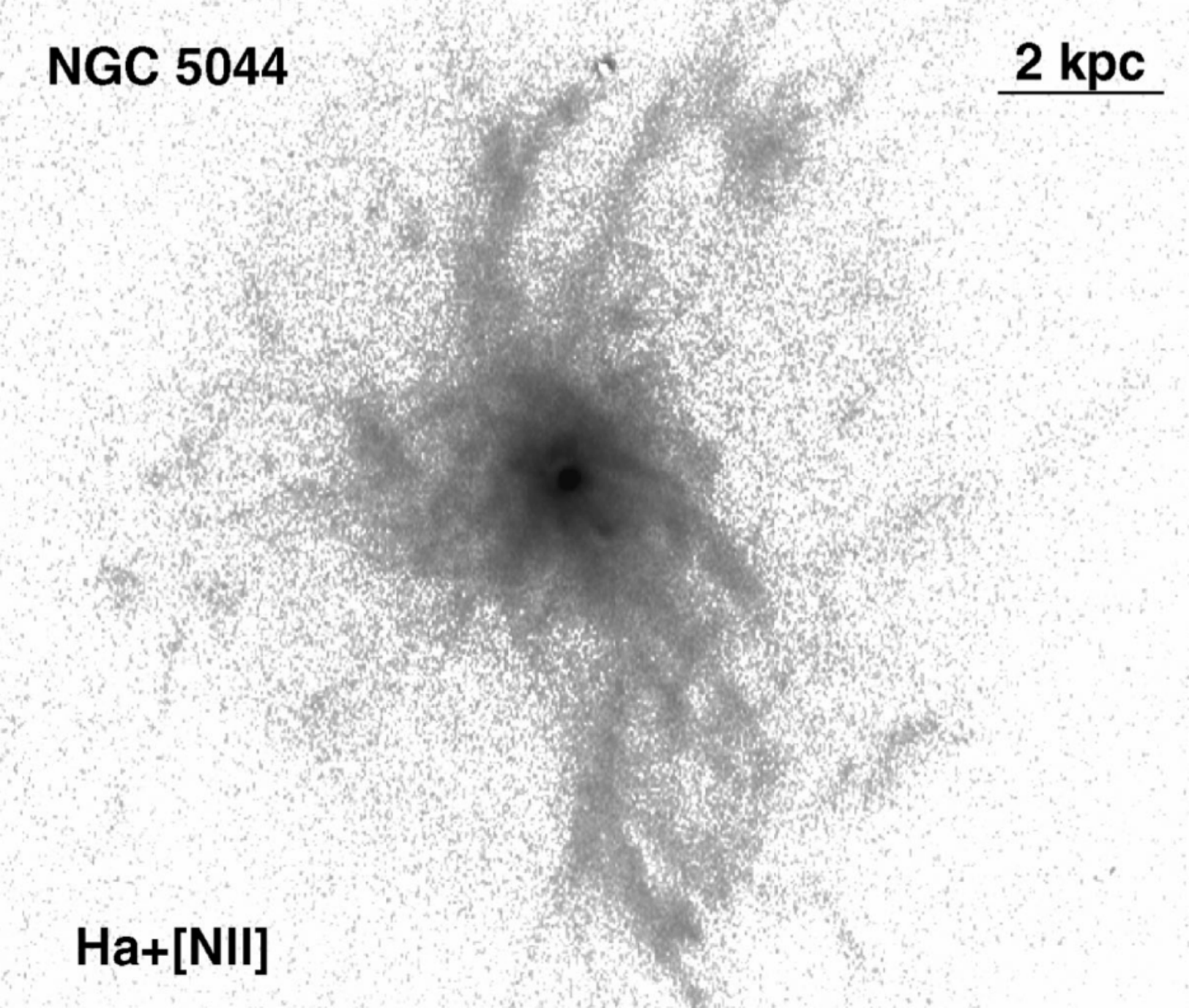}}
      \caption{Top: surface brightness of the ionized UV gas ($1.6\times10^4\,{\rm K}<T\le1.1\times10^6$\,K). Middle: surface brightness of the neutral gas ($200\,{\rm K}<T\le1.6\times10^4$\,K), mainly shining in the optical/IR band and 21-cm. The field of view is $7\times7$\,kpc$^2$ ($t=88$\,Myr).
      Bottom: NGC 5044 SOAR H$\alpha$+[N$^+$] emission tracing part of the ionized gas around the filaments (from Fig.~1 in \citealt{Werner:2014} -- reproduced as per Oxford Journals policy; \href{http://dx.doi.org/10.1093/mnras/stu006}{DOI}).
      The low-temperature ionized gas, having smaller masses yet substantial brightness, envelops the thinner neutral filaments and clumpy structures as a skin. Extended straight or curved filaments naturally form out of the interacting sheets between the large-scale turbulent eddies.
                   } 
      \label{f:comp_Halpha}
\end{figure}

Figure \ref{f:comp_Halpha} shows the typical condensed ionized and neutral structures emerging from the CCA rain within the central 6\,kpc, residing in the UV (top) and optical/IR or 21-cm regime (middle). 
Key feature is that filamentary structures can be naturally formed in the turbulent flow. Condensation preferentially occurs between the interface of two or more turbulent eddies, which compress the gas along sheets, thus creating extended filaments. This simple process does not require magnetic fields, anisotropic conduction, AGN jet entrainment, or kinetic plasma physics, albeit these may further enhance the filamentary morphology and properly set emission line ratios.
The multiple clouds also combine in projection forming extended filaments, in particular in the neutral phase which appears more clumpy. Compared to G15 with higher temperature floor (e.g., due to photoionization heating), 
the warm phase appear thinner and more fragmented, although current turbulent evolution is also longer.
Second key result is that the low-temperature ionized gas envelops the thinner neutral filaments and clouds as a skin.
Albeit the UV mass is much smaller than that of the neutral gas, its emissivity is significant as the cooling function 
has the strongest peak near $10^5$\,K (Fig.~\ref{f:lambda}).
Similar UV filaments and clumpy structures have been discovered in several massive galaxies, including A1795, A1068, A2597, and RXJ1504 (see the UV sample of 16 BCGs in \citealt{Tremblay:2015}).\footnote{
We attempted to subtract FUV-NUV to remove the stellar contribution in {\it GALEX} data of NGC 5044 (\citealt{Marino:2011}), but the maps are too shallow to constrain filamentary emission. A deep spectrum or IFU instrument would be ideal.}

The bottom panel in Fig.~\ref{f:comp_Halpha} shows the H$\alpha$+[N$^+$] as seen by SOAR (\citealt{Werner:2014}). 
Albeit in part contaminated by the stellar component,
several curved filaments and clump formations emerge up to 7 kpc radii, very similarly to our evolved galaxy during the developed CCA rain (and many other observed massive galaxies; \S\ref{s:intro}). 
Such radial condensation region is roughly enclosed by $t_{\rm cool}/t_{\rm ff}\lta8$; however, as proposed in \S\ref{s:cca_TI} a more accurate ratio results to be $\sigma_v/v_{\rm cool}\simeq1.3$ (at 7 kpc).
The inner kpc core dominates the emission, being the preferred locus of cloud accumulation during the rainfall (\S\ref{s:cca_dyn}).
The average velocity dispersion of the warm gas is 165\,km\,s$^{-1}$, which resides within the range observed by ESO 3.6m telescope in NGC 5044 (100\,-\,200\,km\,s$^{-1}$) and similarly in several other massive ETGs (\citealt{Caon:2000,Werner:2014,Hamer:2016}). 
The H$\alpha$+[N$^+$] emission traces the skin of ionized gas (its flux is only a fraction of the ionized gas shown in Fig.~\ref{f:comp_Halpha}; cf.~G15), while neutral filaments are predicted to be thinner, denser, and with feeble emission. 
Albeit high-resolution images are lacking and HI has been proven difficult to detect in emission in the past,
neutral gas is present in recent 21-cm ETG surveys (\citealt{Serra:2012,Kim:2016}). The presence of neutral oxygen detected by {\it Herschel} in NGC 5044, 5813, 6868, and 4636 (\citealt{Werner:2014}) further corroborates the incidence of neutral gas in giant ellipticals.
Fine-structure [C$^+$] can be also used to trace atomic gas;
by converting NGC 5044 [C$^+$] flux (\citealt{Werner:2014}) into atomic hydrogen masses (\citealt{Hamer:2014}) yields $M_{\rm HI}\simeq8.3\times10^6\,[D^2/({\rm 31.2\ Mpc})][F_{\rm [C^+]}/(3.2\times10^{-13}\ {\rm erg\,s^{-1}\,cm^{-2}})]\ \msun$, which matches our final neutral gas mass.
Overall, HI in ETGs remains an open question. Being always subdominant compared to the molecular phase, we expect to be challenging to detect it. Nevertheless, as ALMA is revolutionizing the molecular gas field in massive galaxies, we propose neutral gas as a key new angle to probe the multiphase condensation process during the intermediate stage of the cascade.

\begin{figure}
      \centering
      \subfigure{\hspace{+0.7cm} \includegraphics[scale=0.22]{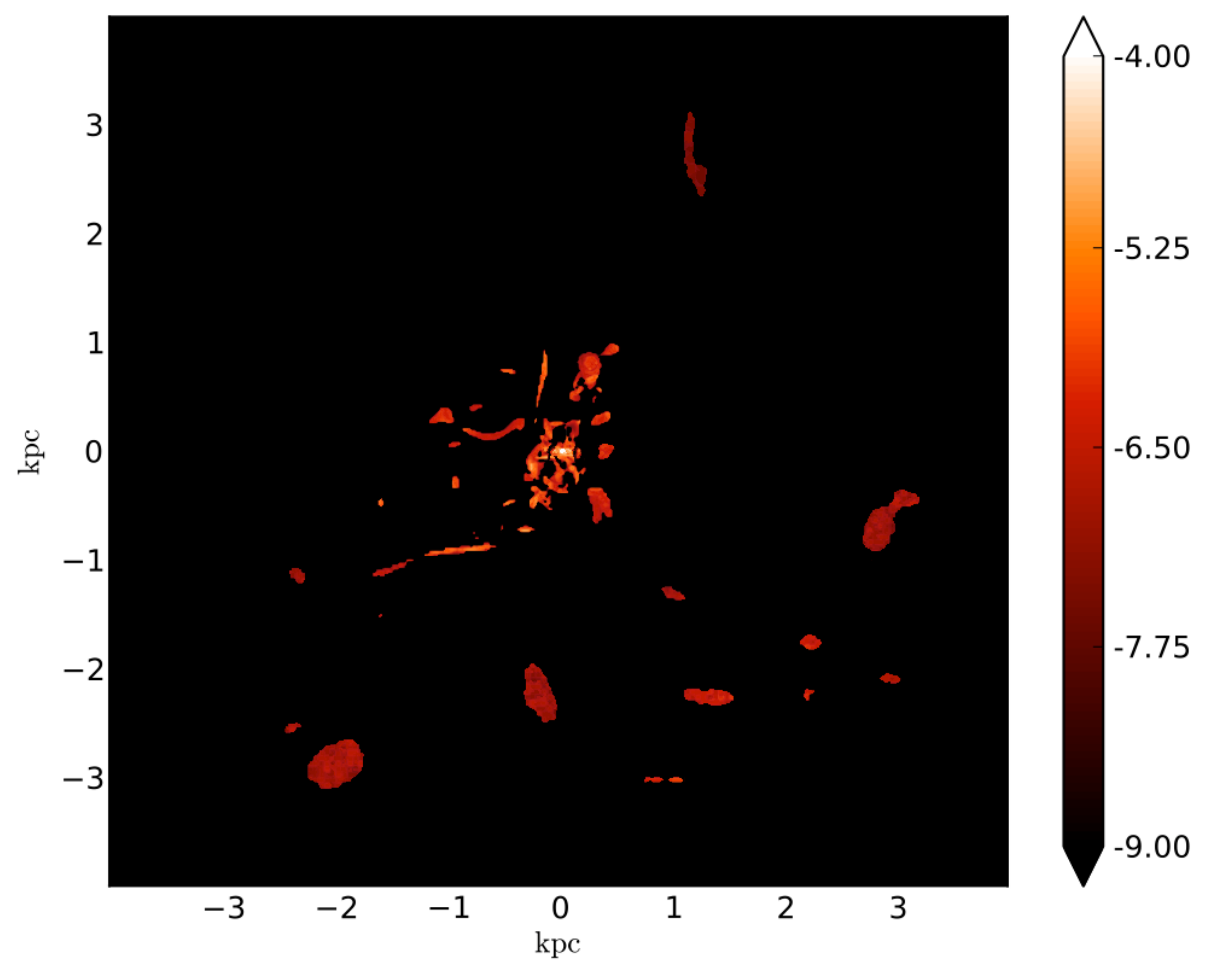}}
      \subfigure{\includegraphics[scale=0.29]{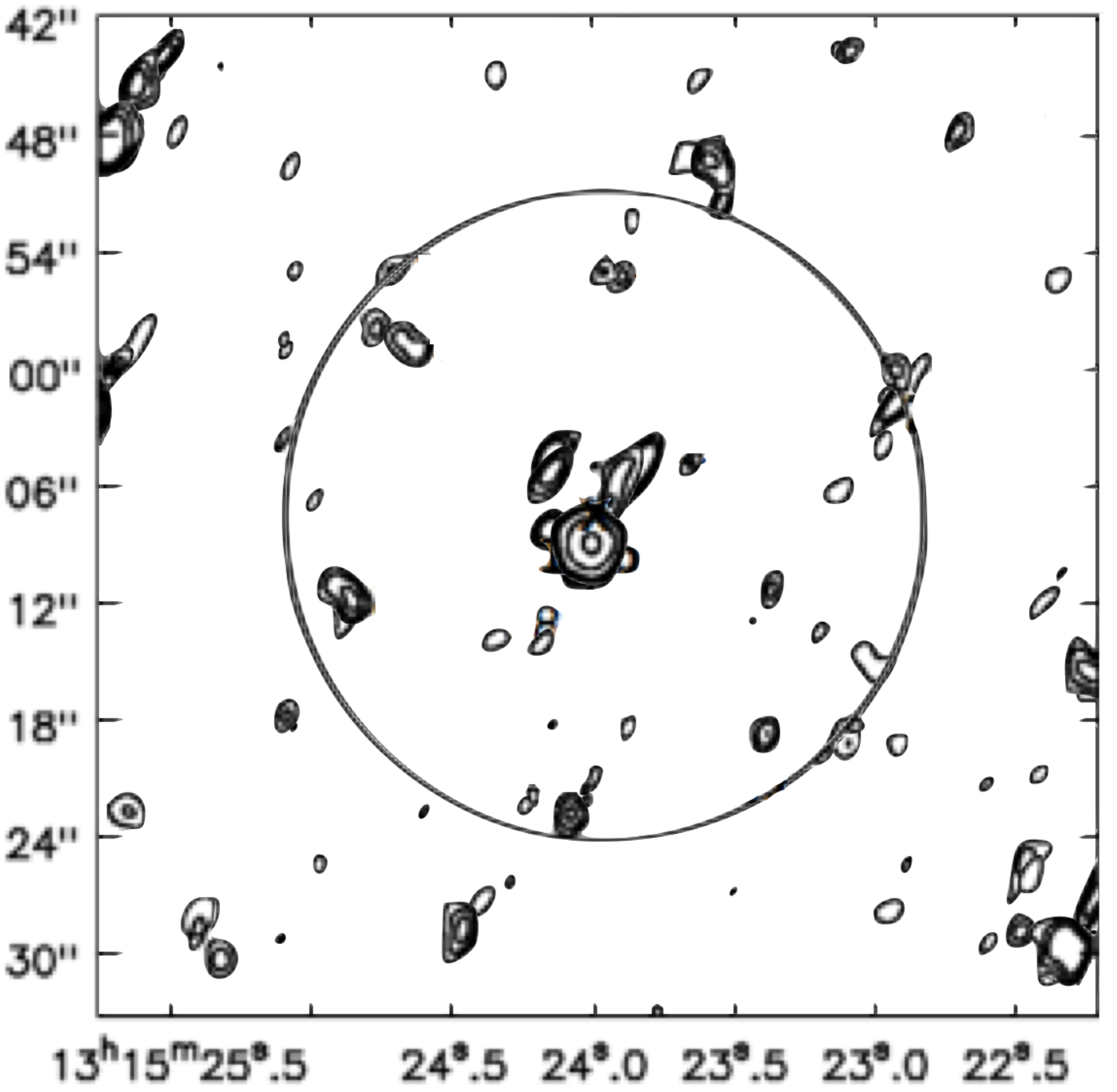}    } 
      \caption{Top: surface brightness of the radio, molecular phase ($20\,{\rm K}\le T\le 200$\,K). The field of view is $4\times4$\,kpc$^2$ ($t=88$\,Myr).
                   Bottom: superposed NGC 5044 ALMA contours of integrated CO(2-1) intensity over 4 line-of-sight velocity channels with centroid -75, -25, 25, 75 km\,s$^{-1}$ with 50 km\,s$^{-1}$ width (adapted from Fig.~2 in \citealt{David:2014} -- \textsuperscript{\textcopyright}AAS, reproduced with permission; \href{http://dx.doi.org/10.1088/0004-637X/792/2/94}{DOI}). 
                   The circle demarks 2.5 kpc; the thicker clouds are those with $4\sigma$ detection above noise. The radio clouds are more compact compared with the warm filaments, drifting in the turbulent velocity field at 150\,km\,s$^{-1}$.
                   Clouds mostly reside in the 0.5\,-\,1 kpc region, where collisions lead to the mixing of angular momentum. 
                   The clouds combine in giant molecular associations dominating emission in projection as shown by ALMA.
                   }
      \label{f:comp_CO}
\end{figure}

In Figure \ref{f:comp_CO} (top), we show the final stage of the simulated CCA rain, i.e., the molecular clouds shining in the radio band.
The molecular phase has more compact volume filling than the warm (neutral or ionized) component, albeit containing 3/4 of the condensed gas mass. The clouds aggregate in giant molecular associations (GMAs) which dominate the emission in projection.
The majority of the clouds reside in the inner few kpc, where they collide and will eventually sink radially onto the SMBH.
As molecular hydrogen is nearly invisible, CO emission is used as its most efficient tracer.
In the bottom panel, we show the composite ALMA CO(2-1) image of 4 velocity channels covering -125 to 125 km\,s$^{-1}$ (\citealt{David:2014}). Consistently with the CCA evolution, the cycle 0 ALMA data unveils the presence of 24 GMAs chaotically floating within 4 kpc radius and peaking in the central 500 pc, i.e., the region of strongest H$\alpha$+[N$^+$] emission and lowest X-ray entropy gas. 
The 3 better resolved GMAs have reported effective radii between 115\,-\,140 pc, with surface mass density ranging 53\,-\,220 $\msun$\,pc$^{-2}$, which are the typical values found for our larger molecular associations (and comparable to molecular cloud values found in spirals; \citealt{Federrath:2013_GMC,Miville:2016}). 
IRAM 30m detects additional diffuse emission: as shown by the simulation, several molecular clouds have smaller size which cannot be resolved by the ALMA beam. 
Diffuse cold gas emission is supported by the {\it Herschel} [C$^+$] maps too (\citealt{Werner:2014}). 
In spiral/star-forming galaxies, the [C$^+$] emission has been found to correlate with molecular gas (\citealt{Crawford:1985}); however, the incidence of photodissociation regions in ETGs is not clear, 
and the excellent spatial correspondence with H$\alpha$+[N$^+$] (Fig.~4 in \citealt{Werner:2014}) points [C$^+$] 
as better tracer of the cooling warm phase. As further support, by converting the [C$^+$] luminosity into ionized gas masses (\citealt{Guillard:2015}) yields $\sim10^6\,\msun$, consistent with the masses retrieved from H$\alpha$ gas.

The ensemble of the 24 ALMA GMAs have a 1D, line-of-sight velocity dispersion of 122\,km\,s$^{-1}$, which is consistent with a kinematics inherited from a plasma subsonic turbulence ${\rm Ma_{3D}}\approx0.4$ (becoming supersonic in the cooling medium). As found for other massive ETGs (\citealt{Werner:2014}) and BCGs (\citealt{Russell:2016}), the cloud kinematics on kpc scale remains typically below the stellar velocity dispersion. 
Relatively low velocities, 100\,-\,200\,km\,s$^{-1}$, arise naturally during CCA. 
The condensed multiphase structures acquire the characteristic chaotic velocity of the turbulent field $\sigma_v$ (or other possible large-scale motions not included here, as a trailing X-ray bubble). The filaments and clouds are not ballistic point-mass objects in free fall, but they are integral part of the hydrodynamical flow, continuously interacting via inelastic collisions and drifting via torques and ram pressure (\S\ref{s:cca_visc}).
The NGC 5044 GMA {\it internal} velocity dispersion (from the observed linewidth) is roughly an order of magnitude lower, which can be explained via the turbulence inertial cascade from the large to small scales. A prediction of CCA is indeed that the molecular clouds are dynamically supported, being internal thermal pressure $\gta2$ dex lower than turbulent pressure, in agreement with the NGC 5044 ALMA findings (lengthscale estimates related to sound speed are thus not relevant here). 
Spatially resolved giant molecular clouds in spiral galaxies similarly show substantial, internal turbulent support with line-of-sight velocity dispersions 5\,-\,20\,km\,s$^{-1}$ (\citealt{Donovan:2013}).
The molecular phase can not be thus easily squeezed, consistently with the low star formation rates (\S\ref{s:cca_multi}); in other words, virial parameter is $5\sigma_v^2\,r_{\rm c}/(GM_{\rm c})\gg1$ for the majority of the clouds, rendering self-gravity secondary.
If internal dynamical pressure slightly exceeds external plasma pressure, the cloud will re-expand to find equilibrium, unless being sinked inside $r_{\rm B}$.

In the nuclear region, after recurrent collisions, CCA predicts rapid funneling of some of the clouds with radial inflow at several 100\,km\,s$^{-1}$. In the presence of a bright AGN (e.g., strong 230 GHz continuum) behind the cloud line of sight, such infalling structures should appear as HI or CO absorption features.
By using such method, a recent cycle 1 ALMA observation (\citealt{Tremblay:2016}) has resolved three $10^5\ \msun$ molecular clouds falling at 300\,km\,s$^{-1}$ in A2597; the authors argue that they are confined within the central $\sim$\,100 pc radius, but uncertainties remain.
\citet{David:2014} also report a pronounced central absorption feature in NGC 5044, indicating an inner cloud falling radially at 260\,km\,s$^{-1}$. 

A limitation of the current setup is the absence of a strong directional uplift, as a buoyant X-ray bubble, which may further stimulate condensation along its path (\citealt{Brighenti:2015, McNamara:2016}) and crate an anisotropic molecular distribution as seen in a few available ALMA clusters as PKS 0745-191 (see \S\ref{s:intro}). The core mechanism is however not dissimilar, i.e., instead of a few chaotic large-scale eddies affecting the core (which contain the bulk of the kinetic energy), the kinetic driver of perturbations becomes a pair of directional cavities again promoting cooling by locally raising density and preventing a quick free fall (\S\ref{s:cca_TI}).
We note, in groups and ETGs, the cavity size are much smaller than in galaxy clusters, diameter $\lta1$\,kpc (\citealt{Shin:2016}), and strong directional entrainment is not seen as in massive clusters. The cold gas kinematics may thus drift toward more anisotropic from the gentle quasi-continuous group regime to the more powerful impulsive cluster environment, as happens for the AGN duty cycles (compare \citealt{Gaspari:2011a} with \citealt{Gaspari:2011b}).

The new, expanding ALMA interferometer will be instrumental in advancing our knowledge of CCA, multiphase condensation, and the different dynamical stages of the molecular gas (e.g., drifting, infalling, and colliding).
Our team is currently carrying out an observational campaign to unveil the formation and evolution of cold gas in massive ETGs; more exciting discoveries are expected to come. Preliminarily (Temi et al., in prep.), NGC 5846 and NGC 4636 display analogous chaotic molecular associations and properties (as velocity dispersion and bulk velocities) as seen in NGC 5044, even better correlated with H$\alpha$ and dust distribution.

\vspace{-0.41cm}
\section{Summary and future prospects}  \label{s:conc}
\noindent
By using 3D AMR hydrodynamical simulations, 
we continued the systematic investigation of cold accretion flows onto SMBHs
linking the galactic 50 kpc scale to the sub-pc scale under realistic astrophysical conditions.
Focus of this work has been the development and impact of neutral and molecular gas cooling.
At variance with the \citet{McKee:1977} {\it bottom-up} ISM model, in which the hot phase is created via stellar evolution
and driven by compressive SN turbulence, the CCA model proposes a {\it top-down} condensation rainfall from the 
hot halos born from the cosmological collapse.
This is particularly relevant for massive galaxies, groups, and clusters, which host high luminosity X-ray halos. Such diffuse, plasma atmospheres are continuously perturbed by subsonic large-scale turbulence, mainly via AGN feedback in the central 50 kpc region. 
Massive galaxies (mostly ETGs) are clearly not `red and dead'; as disc galaxies, they display a complex live multiphase structure, despite hosting 1-2 dex lower cold gas masses.
The morphology of the multiphase halo follows a `disc vs.~filament' dichotomy dictated by the current dynamical state of the system, i.e., `rotation vs.~turbulence' supported.
The key features of each multiphase flow are summarized as follows.

As rotation dominates over turbulent motions (${\rm Ta_t}\equiv v_{\rm rot}/\sigma_v>1$) in the massive galaxy,
the condensation occurs along a conical helix, spinning up the gas which finally settles onto the equatorial region in a {\bf multiphase disc} (\S\ref{s:cool_e03}).
The disc is cylindrically stratified: the ionized (UV) gas envelops the neutral (optical/IR) layer which englobes the very thin molecular (radio) structure,
dominating 95\% of the condensed mass. There is no major fragmentation observed, with the warm phase experiencing mild perturbations
via pressure torques induced by condensation.
The BHAR is {\bf suppressed} by almost 2 dex compared with the cooling rate as most gas has found circularization.
Accretion proceeds along a narrow polar {\bf funnel} ($\lta10^\circ$) or after reaching the inner equator through gas which condensed out of the plasma with low angular momentum.
After 50 Myr the disc has reached 1 kpc extension. Extended molecular and warm discs with similar surface density have been observed in $\sim$\,1/3 of observed massive, typically quiescent, galaxies (\citealt{Caon:2000,Ocana:2010,Young:2011,Werner:2014,Hamer:2016}).

When rotation and turbulence are both inhibited, the hot plasma condenses radially, forming a compact {\bf multiphase cloud} within the inner 100 pc (\S\ref{s:cool_e00}).
The inner multiphase density profiles are steep, following a free-fall slope ($\propto r^{-3/2}$ within the SMBH influence region).
The BHAR is almost identical to the cooling rate.
Adiabatic, {\bf compressional heating} inhibits the formation of the third, molecular phase. 
The warm 8000 K (optical/IR) gas dominates the condensed medium, albeit highly concentrated, with a very thin ionized UV layer.
Similarly, a significant number of quiescent massive galaxies display a spherically symmetric warm H$\alpha$ core (\citealt{Caon:2000,Werner:2012,Werner:2014}) and tend to lack cold gas, in particular when no AGN feedback or other major perturber are detected.

In the more frequent, intermediate state the hot halo experiences both rotation and subsonic turbulent motions ($\sim$\,160\,km\,s$^{-1}$) leading to {\bf chaotic cold accretion} (CCA; \S\ref{s:cca}). The multiphase condensation occurs in the high density peaks via local runaway cooling.
AGN heating prevents a global cooling inflow, facilitating the growth of multiphase structures.
Initially, warm filaments condense out of the low-entropy plasma. The denser, perturbed regions further condense into 
molecular clouds, following a {\bf top-down} rainfall.
The CCA angular momentum transport is driven by {\bf pressure torques} on $>1$\,kpc scale and dominated by chaotic {\bf inelastic collisions} in the core (ram-pressure drag is secondary).
The accretion rate can be modeled via {\bf quasi-spherical viscous accretion} (\S\ref{s:cca_dyn}), $\dot M_\bullet \propto \nu_{\rm c}$ with an effective clump collisional viscosity $\nu_{\rm c}\equiv\sigma_v\,\lambda_{\rm c}$; the mean free path is $\lambda_{\rm c}\approx100$\,pc with large scatter. The core cloud volume filling is a few percent, rapidly declining beyond 1 kpc. As ensemble, the multiphase clouds inherit the large-scale velocity dispersion of the turbulent field, $\sigma_v\approx100$\,-\,200\,km\,s$^{-1}$, although the internal dispersion are over 1 dex lower. The condensed gas is {\bf dynamically}, and not thermally, supported.

Turbulence broadens the angular momentum distribution of the plasma allowing clouds with low angular momentum to condense out. Clouds emerging out of the tails require multiple collisions to mix or cancel angular momentum. This can boost the BHAR up to 2 dex higher than in the disc evolution. The CCA BHAR follows a quasi lognormal distribution with typical mean and 3-sigma $\log(\dot M_\bullet/\dot M_{\rm cool})\simeq-0.7\pm1.0$, which is sub-Eddington at low redshift. The relative BHAR variations $\Delta \dot M_\bullet/\dot M_\bullet$ have analogous distribution, unveiling a self-similar {\bf peak within peak} time structure. 
In Fourier space, the BHAR frequency power spectrum shows {\bf pink/flicker noise} slope, $P_f\propto f^{-1}$ (constant variance per log bin), which is characteristic of {\bf chaotic and fractal phenomena} found across numerous fields (e.g., technology, biology, economics). 
The lognormal and pink noise behavior is consequence of the turbulence imprint, as also seen in Earth weather data series, which share strong similarities with the CCA rain.
The BHAR self-similar fluctuations driven by CCA can solve the problem of large and {\bf rapid variability} observed in compact objects, as AGN, quasars, and high-mass X-ray binaries (\S\ref{s:time_domain}; exemplary one is 3C273).

The multiphase medium arising from CCA is comprised of {\bf 3 main phases} which are highly {\bf cospatial}: {\bf hot plasma} (0.1\,-\,1 keV), {\bf warm neutral gas} ($\sim$\,9000\,K), and {\bf cold molecular gas} ($<$\,50\,K).
After 60 Myr the total molecular mass is $3\times10^7\,\msun$ (similar to recent ALMA detection in NGC 5044), dominating the condensed gas mass, being at least $5\times$ larger than the warm mass.
The lowest entropy, ionized phase forms a thin layer around the condensed gas, with masses 2 dex lower than that of the cold gas.
CCA predicts emission-weighted $\rho$ and $T$ {\bf radial profiles} of all phases which tend to follow a {\bf -1 and 0} logarithmic slope, 
as observed in M87, NGC 3115, NGC 4261 and NGC 4472.
ETGs which do not display a molecular phase are likely experiencing strong photoionization heating, preventing the formation of the third phase;  the 2-phase CCA model presented in G15 is more suited for such galaxies, albeit the core CCA mechanism remains intact.

We propose as improved `TI', or better, {\bf condensation criterium} $\sigma_v/v_{\rm cool}\lta 1$ instead of $t_{\rm cool}/t_{\rm ff}\lta 10$, preferably in local non-radial sectors, as filaments tend to {\bf drift} and collide with sub-virial velocities in the turbulent field, not necessarily precipitating in free-fall. AGN outflow uplift -- not tackled here -- is a particular directional and supersonic case of such delayed infall and is expected to help stimulating condensation (e.g., \citealt{Brighenti:2015, McNamara:2016}).
We warn the commonly used `thermal instability' term can be misleading, since initial tiny, linear fluctuations are severely idealized in realistic hot halos. Linear thermal stability analysis has thus limited applicability here.
What matters is that chaotic motions, cascading top-down via turbulence, create significant variance in relative density, which leads to the top-down {\bf nonlinear multiphase condensation}.

CCA can self-consistently explain key observations in massive galaxies.
As shown by {\it Chandra}, the hard SB$_{\rm x}$ is diffuse and perturbed. While globally spherically-symmetric,
the {\bf X-ray halo} displays increments and depressions which are driven by the chaotic turbulent eddies 
(and may be confused with weak cavities/edges tied to AGN jets/cold fronts).
The required turbulence to trigger CCA is mild, of the order of 150\,km\,s$^{-1}$, which has been directly detected by {\it Hitomi} via line broadening (\citealt{Hitomi:2016}).
The first stage of the multiphase condensation appears in the very soft X-ray ($\lta0.5$\,keV).
As traced by H$\alpha$+[N$^+$] SOAR images (\citealt{Werner:2014}), {\bf extended ionized (UV/optical) filaments} emerge up to a radius of $\sim$\,10\,kpc, 
enveloping thinner, clumpier {\bf neutral (optical/IR/21-cm) structures}.
Large-scale filaments can be produced between the interface of the turbulent eddies, without invoking magnetic fields or anisotropic plasma physics.

Consistently with ALMA detections (\citealt{David:2014}), 
{\bf molecular, CO radio-emitting clouds} condense up to several kpc and typically reside in the inner 500 pc radius.
Projection along line of sight and limited beam size show visual aggregation of small elements in more extended structures, such as giant molecular associations (GMAs) with size up to 100 pc, mass of several $10^6\,\msun$, and surface density up to a few 100\,$\msun$\,pc$^{-2}$.
Considering the extended emission of [C$^+$] coincident with H$\alpha$, [C$^+$] may be a better tracer of neutral gas, with estimated HI mass (\citealt{Hamer:2014}) matching our final warm mass. Neutral gas has been detected throughout ETG surveys (\citealt{Serra:2012}) and could be a new key angle to probe multiphase condensation.
Within the Bondi radius, molecular clouds are efficiently funneled toward the SMBH with several 100\,km\,s$^{-1}$ velocity, as recently discovered in A2597 with ALMA (\citealt{Tremblay:2016}).
The clumpy nature of CCA -- increasing toward smaller radii -- can explain the presence of the BLR/NLR and the obscuration required in AGN Unification models (\citealt{Netzer:2015}).
In the presence of a bright AGN backlight, we frequently expect CO and HI absorption features during CCA mode while the cloud and SMBH align along line of sight, as found in Centaurus A, NGC 5044, and A2597 (\citealt{Espada:2010,David:2014,Tremblay:2016}).

The updated, long-term {\bf AGN feedback loop} now includes a third key phase, the cold molecular gas, and works as follows.
As long as turbulence dominates over rotation, extended warm ionized filaments 
condense out of the diffuse X-ray plasma, englobing neutral filaments, which envelope molecular clouds and associations. 
The inner chaotic and inelastic collisions funnel the central multiphase clouds boosting the accretion rate, thus the AGN outflows/jets, 
which then thermalize in the core of the massive galaxy, group, or cluster, increasing the plasma core entropy
(e.g., \citealt{Gaspari:2012a}).
The CCA rain thus subsides and turbulence decays; the accretion rate decreases toward the low-level disc evolution.
This enables entropy and cooling time of the gas to decrease again, reigniting another phase of plasma perturbations, multiphase filaments and cloud condensation -- another CCA rainfall.
We remark CCA is a cyclical, rather than persistent, state and does not exclude some ETGs may be acquiring external warm/cold gas from ongoing mergers. 

Molecular CCA opens up {\bf new windows} into modeling and studying accretion physics in more realistic way,
e.g., in cosmological simulations and SMBH (semi)analytic studies, and in the interpretation of {\bf multiwavelength observations} of gaseous halos in galaxies, groups, and clusters.
In future works, we continue our theoretical investigations on how CCA behaves with additional physics
and in different environments.
The CCA rain is expected to be more frequent in massive hot halos due to the larger cooling rate and the velocity dispersion dominated regime; 
however, all galaxies are expected to host a diffuse X-ray halo, including our Milky Way, 
thus experiencing phases of (self-similar) CCA rain.
From the observational perspective, 
the new and expanding telescopes, such as ALMA, JWST, and {\it Athena},
supported by ongoing major missions ({\it Chandra}, XMM, CARMA, SOFIA, IRAM)
will be instrumental in unveiling the role of warm (neutral and ionized) and cold (molecular) gas in the multiphase condensation
of hot halos. It will be key to resolve molecular clouds, improving the short ALMA exposures limited to GMAs, and to understand the different dynamical stages of the cold phase in a much larger sample, both in groups and clusters.
Major advancements will be possible via multiwavelength studies and a continuous joint effort between simulations and observations.

\vspace{-0.41cm}
\section*{Acknowledgments}
\noindent
M.G. is supported by NASA through Einstein Postdoctoral Fellowship Award Number PF-160137 issued by the Chandra X-ray Observatory Center, which is operated by SAO for and on behalf of NASA under contract NAS8-03060. 
FLASH code was in part developed by the DOE NNSA-ASC OASCR Flash center at the University of Chicago. 
HPC resources were provided by the NASA/Ames HEC Program (SMD-15-6208, SMD-15-6209, SMD-16-6718). 
We thank M.~Sun, B.~McNamara, M.~Voit, J.~Stone, N.~Werner, G.~Tremblay, P.~Green, P.~Kretschmar, F.~Gastaldello, D.~Eckert, S.~Molendi, A.~Goulding, and M.~McDonald for useful discussions and insights.
\vspace{-0.6cm}
{\footnotesize
\bibliographystyle{mnras}
\bibliography{biblio}
}

\vspace{-0.5cm}
\begin{appendix} 
\section{Quasi-spherical viscous accretion}\label{app:visc}
The fluid conservation equations of mass and angular momentum affected by a viscous shear force
 reduce in spherical symmetry to
\begin{equation}\label{e:sph_M}
\frac{\partial \rho}{\partial t} + \frac{1}{r^2}\frac{\partial}{\partial r}(r^2\rho v_r)=0
\end{equation}
\vspace{-0.41cm}
\begin{equation}\label{e:sph_l}
\frac{\partial \rho l}{\partial t} + \frac{1}{r^2} \frac{\partial}{\partial r}(r^2 \rho l v_r) = \frac{1}{r^2}\frac{\partial}{\partial r}\left[\nu\rho r^4\frac{\partial}{\partial r}\left(\frac{l}{r^2}\right)\right].
\end{equation}
Notice that the viscous shear stress (per unit solid angle) is $\nu\rho\,r\,d\Omega/dr$, where
$\Omega=l/r^2$ is the angular velocity on the plane perpendicular to the specific angular momentum $l$. Such 1D approximation is valid within small solid angles. Assuming steady state and expanding the angular momentum equation, Eq.~\ref{e:sph_M} and \ref{e:sph_l} simplify to
\begin{equation}\label{e:sph_M2}
r^2\rho v_r={\rm const} \equiv -\dot M 
\end{equation}
\vspace{-0.41cm}
\begin{equation}\label{e:sph_l2}
\dot M \frac{\partial l}{\partial r} = \frac{\partial}{\partial r}\left[\nu\rho r^4\frac{\partial}{\partial r}\left(\frac{l}{r^2}\right)\right],
\end{equation}
that is the angular momentum advection term is due to the viscous torque gradient. Without loss of generality, we can rearrange the last equations as
\begin{equation}\label{e:sph_vr}
v_r = \frac{\nu}{r}\,\frac{\frac{\partial}{\partial \ln r}[\nu\, \rho l r \,(\xi-2)]}{\nu\,\rho l r\, \xi} = -\frac{\nu}{r}\,(2-\xi),
\end{equation}
where $\xi\equiv \partial \ln l/\partial \ln r$.
The last equality arises by
using the continuity equation in logarithmic form and assuming power-law dependence of the flow variables. 
The accretion rate (per unit solid angle) is thus
\begin{equation}\label{e:sph_Mdot}
\dot M = r\rho\nu\,(2-\xi)
\end{equation}
For Keplerian angular momentum $\xi = 1/2$, while a dark matter distribution has $\xi = 1$, therefore the radial velocity magnitude is the kinematic coefficient divided by radius within order unity, leading to
\begin{equation}\label{e:sph_vr_fin}
v_r \simeq -\frac{\nu}{r},
\end{equation}
\vspace{-0.41cm}
\begin{equation}\label{e:sph_Mdot_fin}
\dot M \simeq r\rho\nu
\end{equation}
Interestingly, recasting the equations in cylindrical coordinates and changing the proper operators, the radial velocity equation \ref{e:sph_vr_fin} remains the same, while $\dot M \simeq \Sigma\nu$, as customarily used in accretion disk studies (e.g., \citealt{Balbus:1999}). The $\rho r$ term in Eq.~\ref{e:sph_Mdot_fin} can be thus interpreted as an effective surface density $\Sigma$. Finally, Eq.~\ref{e:sph_l} is exact only on the rotational plane, while accretion progressively reduces toward the poles, thereby the total accretion rate is not the maximal $4\pi\,\dot M$, but of order $\pi\,\dot M$. A more precise normalization would require a 2D analysis, which is beyond the purpose of our analytic estimate and keeping in mind the large temporal variations in the collisional mean free path of the chaotic clouds which introduce considerable scatter.

\end{appendix}

\label{lastpage}

\end{document}